\documentclass{book}
\usepackage{subfiles}
\usepackage[margin=1in]{geometry}
\usepackage{amssymb,amsmath}
\usepackage{amsmath,amsfonts,amssymb}
\usepackage{t1enc}
\usepackage[OT2,OT4]{fontenc}
\usepackage[cp1250]{inputenc}
\usepackage{amsmath}
\usepackage{amsmath,amsfonts,amssymb}
\usepackage{graphicx}
\usepackage{lipsum}%% a garbage package you don't need except to create examples.
\usepackage{fancyhdr}
\usepackage{wrapfig}
\usepackage{color}
\usepackage{url}
\usepackage[small,nohug,heads=vee]{diagrams}
\diagramstyle[labelstyle=\scriptstyle]

\newtheorem{theorem}{Theorem}
\newtheorem{definition}{Definition}

\usepackage{listings}
\newcommand*{\email}[1]{\href{mailto:#1}{\nolinkurl{#1}} }

\usepackage{cmbright}
\usepackage[T1]{fontenc}

\usepackage[noend]{algorithmic}
\usepackage{algorithm}

\graphicspath{{/home/pawel/Dropbox/tutorials/paris/images/}}

\newcommand{\card}[1]{\ensuremath{\left|#1\right|}}

\begin{document}
\thispagestyle{empty}
\title{Computational and applied topology, tutorial.}
\author{Pawe{\l} D{\l}otko\\ Swansea University\\
(Version 3.14, and slowly converging.)}
\date{\today}
\maketitle

\chapter{Introduction.}
If you are reading this, it probably means that you got interested in applied and computational topology and perhaps you want to use it to solve your problems. Or perhaps you are searching for inspiration to your lectures. In any case, I hope you will find this resource useful. Some of the examples and ideas have been taken from my friends in the field. In each of those cases I tried to make it clear in the text. Please note that the script is under constant development and it is not (nor has it ever been) intended to be a complete, rigorous book. It is rather a collection of evolving ideas, anecdotes and examples to illustrate what \emph{applied and computational topology} is. Or at least my view on it. I hope you will enjoy it!

Completion of this lecture will require you to get your hands dirty with data and implementation. Most of the concepts are illustrated with python (or C++) examples you may play with. They are all based on \emph{Gudhi} library. Please follow the compilation instructions available in this webpage: \url{http://gudhi.gforge.inria.fr/}. There are at least a few ways you can install Gudhi - you can go through the whole compilation (requires cmake, Boost, Eigen3, Cgal,...) or use an anaconda package installation. If you have any problems please email me as well as the Gudhi mailing list \emph{gudhi-users@lists.gforge.inria.fr}

The archive with exercises accompanying this script can be downloaded from here \url{https://www.dropbox.com/s/n5mwv3bntslwpsw/code.zip?dl=0}

So far I have presented some version of this material at the one week long lecture series I gave at EPFL in February 2015 and a day tutorial in Topological Data Analysis in Paris region in June 2018. This document is designed as supporting material for the tutorial. You will find various comments clearly indicating it throughout the text. You will miss all the presentation part, but I strongly encourage you to go ahead and play with this tutorial. If you have problems that no one else can help with, and if you can find me, then perhaps I will be able to help. 
\vskip 1 cm
Enjoy!
\vskip 1 cm
Pawe{\l} D{\l}otko, Swansea University, July 2018.
\vskip 1 cm
I would like to thank Nicolas Scoville for his great help in improving the text. Thanks Nick!
%export PYTHONPATH="$PYTHONPATH:/home/pawel/Dropbox/gudhi/persistence_representation_cytonization/persistence_representation_cythonization/build/src/cython/"

%\tableofcontents
\chapter{Why not topology?}

Topology studies global structure of spaces. It integrates local information and provides concise summaries of data. Non local, provable properties are very important in contemporary data analysis, since \emph{local} methods are very popular. Analysing data locally reassembles a process of reading done by a child who has only just learn how to read. In this process a child reads letter bit-by-bit and tries to put-all-of-them-together into words so that something that makes sense is obtained. On the other hand, a person who is more experienced in reading does not look at the single letters, but at more \emph{global} linguistic structures. Especially if you are fluent with skim reading. What is the structure we are referring to? Please read quickly the text below proposed by Matt Davies from MRC Cognition and Brain Sciences Unit at Cambridge.

\begin{center}
\textit{Aoccdrnig to a rscheearch at Cmabrigde Uinervtisy, it deosn't mttaer
in waht oredr the ltteers in a wrod are, the olny iprmoetnt tihng is
taht the frist and lsat ltteer be at the rghit pclae.  The rset can be
a total mses and you can sitll raed it wouthit porbelm.  Tihs is
bcuseae the huamn mnid deos not raed ervey lteter by istlef, but the
wrod as a wlohe.}
\end{center}
Despite some local errors, the message is easy to understand. But when you try to read it as a child, by combining single letters together, you will fail. The reason why you were able to read and understand the text is because the \emph{shape} of words is preserved. Permutations of letters correspond to some, quite considerable, noise. Analysing shape of words is the basis of skim reading. That being said, we get to our credo: shape matters! Even in the presence of noise, shape determines the meaning. Through the lecture we will give you more and more examples to support this and methods to analyse shape.

Related to the previous example Have you noticed that most legal agreement and licences ARE WRITTEN WITH CAPITALS? WHY IS THAT? WELL, NOW THE WORDS DO NOT HAVE SHAPE, SO YOU CANNOT SKIM READ THEM. AND THIS IS WHAT THEY WANT: THEY TYPICALLY DO WANT TO DISCOURAGE YOU FROM READING MANY PAGES OF A TEXT LIKE THIS. Apparently even the legal people are aware that shape matters.

This is one of the simplest examples of a situation when global structure allows us to understand the data. But any local analysis will, most likely, fail. So, here we are, in the world of topology. But, what is topology? Let me give you a few examples.

\vskip 0.5cm
Topology is about aggregating local information to global output. Think about a great number of cheap, gps-free sensors which are distributed in a vast area that are to monitor and detect a fire. The boundary of this area is fixed with a chain of sensors that can communicate with the neighbouring ones. Each sensor can communicate with other sensors lying not further than a radius $r$ away. It also can detect a fire if only it breaks not further away than $\frac{\sqrt{3}}{2}r$ from that sensor. In this setting, can the sensor network determine if the whole area or interest is observed? It may seems very unlikely, since sensors do not have localization information. They do not have metric information (apart from binary knowledge if a sensor that they can 'hear' is in the distance smaller than $r$, or not). But, the answer is: yes, it can, check~\cite{rob_sensors} for details! And it is all thanks to topology and its ability to integrate local connectivity information to global data.
\begin{figure}
  \begin{center}
    \includegraphics[scale=0.07]{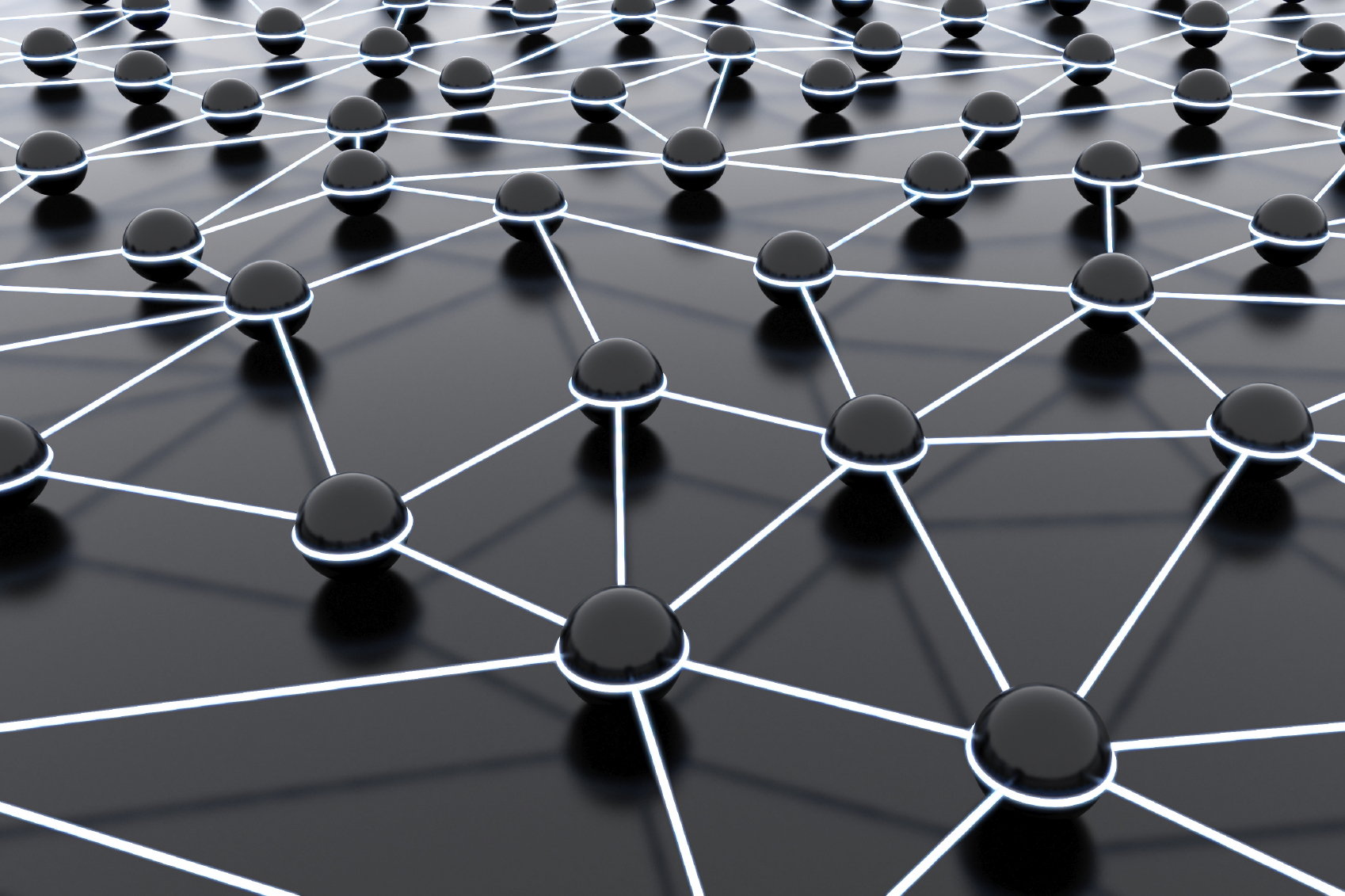}
  \end{center}
  \label{fig:rob}
  \caption{By Robert Ghrist.}
\end{figure}

\begin{figure}[!h]
  \begin{center}
    \includegraphics[scale=0.2]{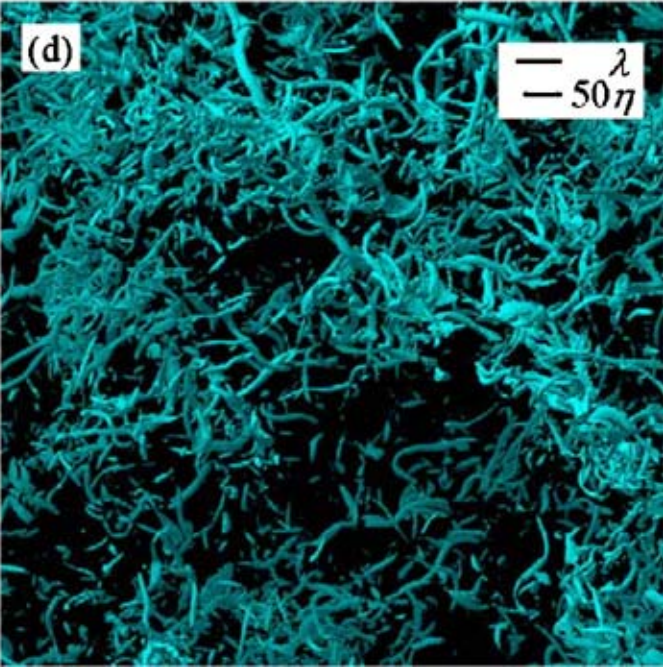}
  \end{center}
  \caption{By Takashi Ishihara.}
  \label{fig:turbulence}
\end{figure}

\vskip 1 cm
Topology gives a rigorous dimension reduction technique. Think about a massive turbulent flow dynamics simulation the dynamics of which we want to characterize. See Figure~\ref{fig:turbulence}. For the simulation the domain is divided into a $12000^3$ grid, a $2048^3$ sample of which is depicted in the right. An important factor is how \emph{turbulent} the flow is. Typical measures of turbulence is the \emph{enstophy}. In the picture, blue regions denote high and black low enstrophy. 

A typical characterization of the flow is an \emph{average  enstrophy}. That is, a single scalar value to characterize this large and complicated dynamic. If the number is high (or locally high) then the simulation, or its region is consider highly turbulent while low in the other case. But we can look at this picture through the lens of topology. Now we see a cubical domain in $\mathbb{R}^3$ with a scalar value function defined on it. In this course we will discover tools to characterize it better than with a single scalar value. 
 
\vskip 1cm
Topology gives us information about obstacles sometimes called \emph{cuts}. Suppose we are given a non directed graph $G$ consisting of a set of vertices $V$ and edges $E$ (we will have a more precise definition of a graph in one of the next sections). Let $P$ be a set of all paths in $G$. We want to define a function $f : V \rightarrow \mathbb{R}$ (often called a potential) such that for every path $p \in P$ and for two constitutive vertices $v_1, v_2$ in this path, $f(v_1) > f(v2)$. It is easy to find out that such a function exists if and only if $G$ does not contain a \emph{cycle}.

\begin{figure}
  \begin{center}
    \includegraphics[scale=0.3]{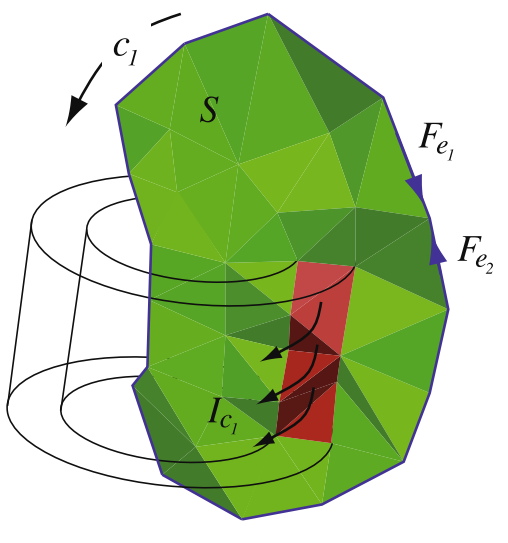}
  \end{center}
  \caption{By Ruben Specogna.}
  \label{fig:cuts}
\end{figure}

\vskip 1cm
Quite often in engineering we face less trivial scenarios when a similar situation happens. Let me focus here on one specific example dealing with solutions of Maxwell's equations. There are various ways of solving them, but there is one particularly interesting due to its speed and low memory consumption.
It is called the $T-\Omega$ approach to Maxwell's equations. It aims in defining a discrete counterpart of Maxwell's law given a three dimensional \emph{mesh} $\mathcal{K}$ of a considered electric circuit. The mesh $\mathcal{K}$ decomposes into sub meshes $\mathcal{K}_a$ and $\mathcal{K}_c$, representing insulator and a conductor. One of the intermediate steps requires defining a potential--like function in the insulator $\mathcal{K}_a$. As in the case of graphs, this is not possible if $\mathcal{K}_a$ has some sort of cycle -- the one that does not bound anything. We will deal with those cycles a lot when talking about homology groups. In the case of $\mathcal{K}_a$ being non trivial, some topological correction has to be added to the system. In engineering we refer to them as \emph{cuts} and they are helpful to solve the problem better, faster and with less memory resources. See Figure~\ref{fig:cuts}.  

\chapter{From graphs to complexes.}

Graphs are object that, often independently, appeared in different disciplines of science. They are typically used as a abstraction of a set-up of some real-word problem. One of the first problems formulated in the language of graphs was thes \emph{Seven Bridges of Königsberg} problem formulated by Leonhard Euler in 1735. This problem, according to many scientists, give a foundation for graph theory and subsequently topology. Let us have a look at it.

\begin{figure}[h!tb]
\centering
\includegraphics[scale=0.9]{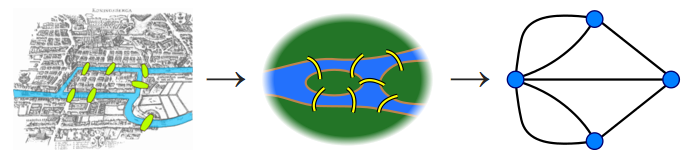}
\caption{From Wikipedia.}
\end{figure}

The problem is to find a walk through the city of Königsberg. A person has to cross each bridge exactly once and come back to the starting point. Euler's contribution was to show that such a walk is not possible. But, the solution to this particular was not important, but a technique he used to reach it. Euler has abstracted away all the unnecessary information about the map of the city. He used only the important parts, which were:
\begin{enumerate}
\item The regions of the city that can be reached without crossing any bridge. Those were contracted to single vertices, giving rise to so called vertices in Euler's graph representing the problem.
\item The bridges, which connected the regions from the point (1). They give rise to the edges connecting vertices from point (1).
\end{enumerate}
This reduction gave a very concise way of representing the problem. It is also a very good example of a process of \emph{abstraction} which allows us to give up all unnecessary details and construct a simplistic model that inherits all the properties required to solve the problem. 

So why does such a walk not exist? I encourage you to experiment with this representation by yourself and see why. Really, this is just simple counting. You can find the short description of the reason in the comments to this section.

Let us define graphs formally. An \emph{abstract undirected graph} $G$ is a pair $(V,E)$ such that $V$ is a set of vertices and $E$ is a set of edges, each of which is a pair of vertices.

A $n-$\emph{clique} in a graph is a set of vertices $v_1,\ldots,v_n$ such that there is an edge from $v_i$ to $v_j$ for every $i,j \in \{1,\ldots,n\}$, $i \neq j$.

A \emph{path} between two vertices $u,v \in V$ is a sequence of vertices $u = v_1, \ldots , v_n = v$ such that there is an edge between $v_i$ and $v_{i+1}$ for every $i \in \{1,\ldots,n-1\}$.

Having this set of definitions we can discuss the first topological concept of this lecture - a concept of \emph{connected components}. A graph is connected if there exist a path between any pair of its vertices. If there exist two vertices for which such a path do not exist, then the graphs is \emph{not connected}. We can define a relation of connectivity between two vertices: two vertices are in the relation if a path between them exists. So called \emph{classes of abstraction} of that relation are the connected components of a graph. 
The concept of a connected component is classical in graph theory and in topology. But there are more common concepts in the intersection between the two. Let us talk about \emph{cycles}. A path having the same beginning and end point $u = v_1, \ldots , v_n = u$ is called a \emph{cycle}. For simplicity we will add an additional assumption that there are no repetitions among vertices $v_1,\ldots,v_n$ in a cycle. Simply speaking, a cycle is a closed path in a graph. Not all graphs admit cycles. A graph that is connected and does not have a cycle is a \emph{tree} (if it has unique connected component), or a forest if it is disconnected and does not have a cycle.

\begin{figure}[h!tb]
\centering
\includegraphics[scale=0.9]{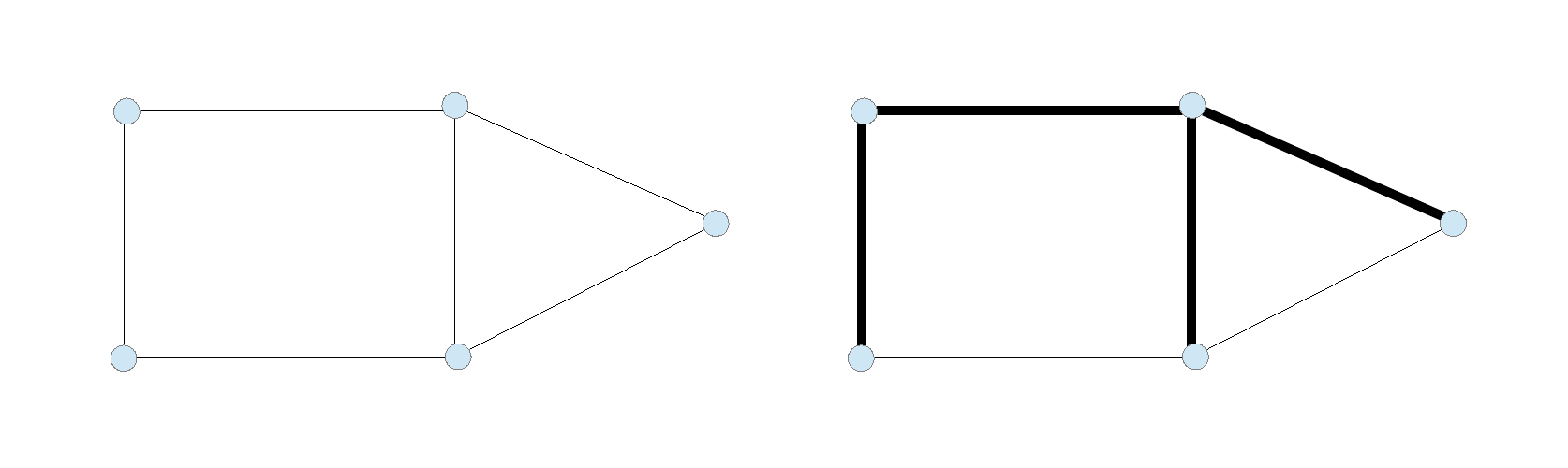}
\caption{On the left, the initial graph $G$, on the right, with bold, the edges that belong to a spanning tree $T$ of $G$.}
\label{fig:spanningTree}
\end{figure}

Given a connected graph $G$, we can construct a maximal acyclic subgraph of $G$. Any such subgraph is often refereed to as a \emph{spanning tree}. Note that for a given graph, there are many possible spanning trees. For a given graph $G$, let $T$ be such a spanning tree. See an example in Figure~\ref{fig:spanningTree}.
It is clear that we cannot add any more edges to $T$ without introducing a cycle. Let me call every edge that is not a tree edge a \emph{cotree} edge. Every cotree edge gives rise to a cycle. The collection of all the cycles obtained from all cotree edges forms a \emph{cycle basis of a graph}. A cycle is yet another concept in common between graph theory and topology. One can interpret cycles as one dimensional holes in the structure of the graph.

\begin{figure}[h!tb]
\centering
\includegraphics[scale=0.9]{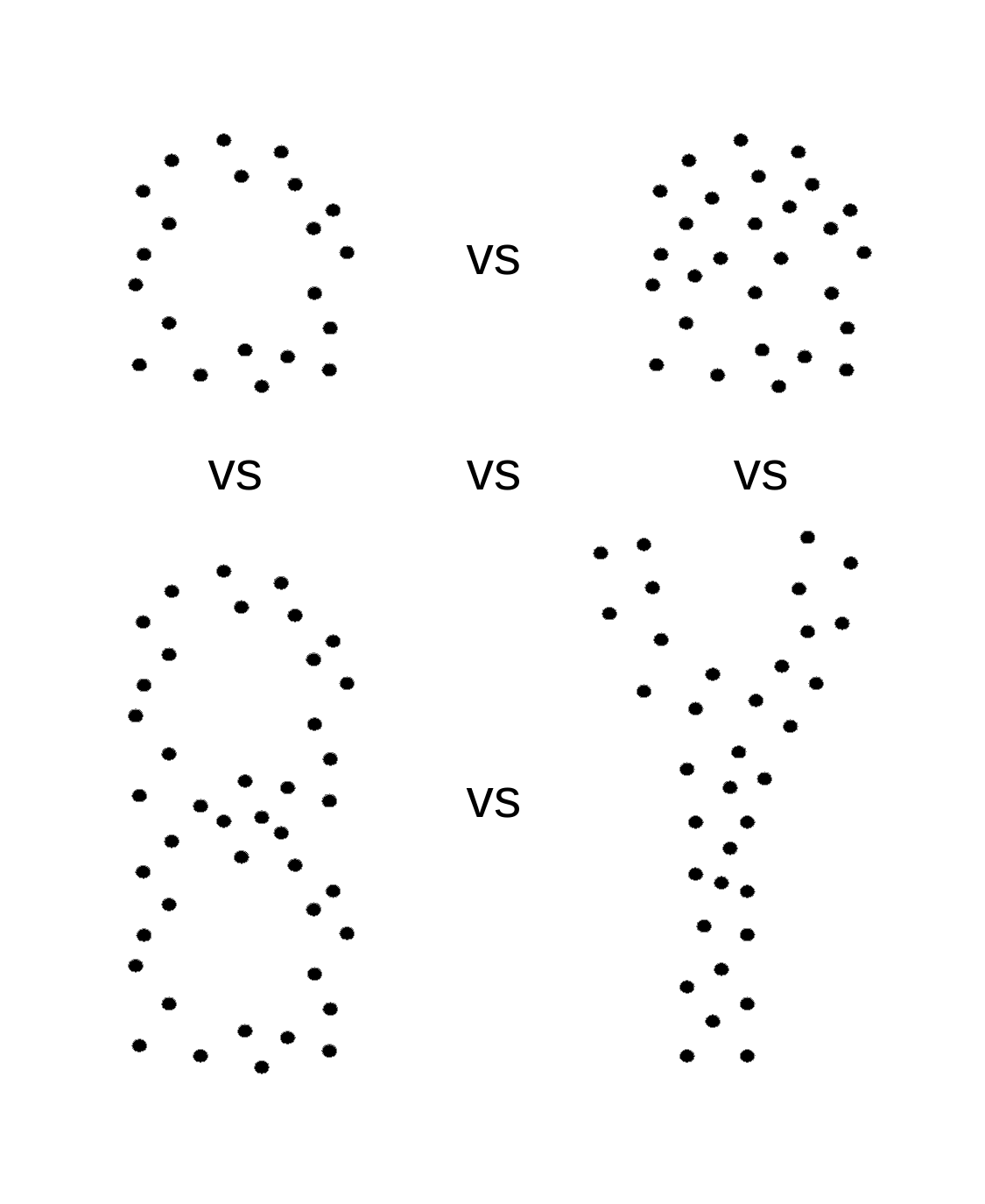}
\caption{The same connected components have different meanings. One of toy examples showing why we may need higher order topological information.}
\label{fig:beyond_connected_components}
\end{figure}

The basic topological description in dimension zero is that of connected components. Think of them as holes in the space. Various clustering methods can be used to capture connected components. The question you may be asking right now is: do we really need more? It turns out that in some cases we do. Have a look at the Figure~\ref{fig:beyond_connected_components}. All sets have a different number of connected components, but they are very different. Topology allows to discriminate between them. 

Given a cycle, we can \emph{grasp} such a graph using that cycle to cage it. This cycle is like a 1-dimensional hole. 

This raises a question: are there two, or even higher dimensional holes? The answer is, yes, there are. Not in graphs, but in more general objects called complexes that are generalizations of graphs. We will study those objects in the next section. To talk about them, we need to leave the safe harbour of graph theory into the uncharted sea of algebraic topology. But we shall always remember where we started from: we should remember that complexes and homology theory which we will discuss, is a generalization of a well know concept in graph theory and we can get a lot (but not all!) motivation from graph theory. Coming back to this basic intuition turns out to be very useful.

Before we start our voyage to the second star from the right, let us have a look at yet another nice intersection of classical graph theory and topology. Let us talk about \emph{flows and cuts} and max flow min cut \emph{duality}. To introduce this concept, we need \emph{directed graphs}. They are almost like standard graphs. The only difference is that the set of edges $E \subset V \times V$ is formed by ordered pairs i.e. the edge $(v_1,v_2)$ and the edge $(v_2,v_1)$ are two different edges (pointing in opposite directions).

Often for various problems we are given some functions defined on vertices and edges of the considered graph. For the flow problem let us consider a \emph{capacity} function on the graph's edges $c : E \rightarrow \mathbb{R}$. The capacity of an edge $e$ tells us how much stuff can be pushed through $e$ in a unit of time. 

A \emph{flow} is a map $f : E \rightarrow \mathbb{R}$ subject to the following two constraints:
\begin{enumerate}
\item Capacity constraint: for every $(u,v) \in E$, $f( u,v ) \leq c(u,v)$. Simply speaking, we cannot push through an edge more stuff than its capacity.
\item Conservation of a flow: for every vertex $v \in V$ the amount of flow that enters $v$, exits from $v$. This is one of Kirchoff's laws from circuit theory.
\end{enumerate}

Let us now pick two vertices $s,t \in V$. The $s$ stands for \emph{source} while $t$ stands for \emph{target}. A typical question in graph theory is: how much flow can we send from a source to target?
\begin{figure}[h!tb]
\centering
\includegraphics[scale=0.8]{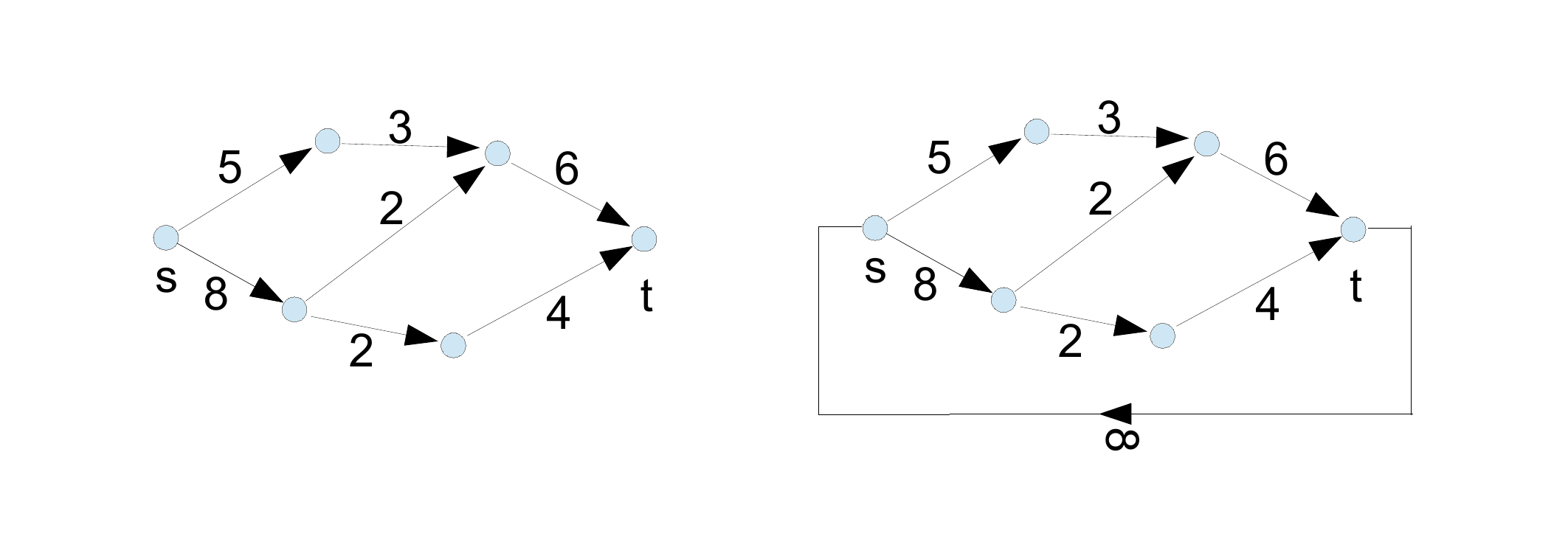}
\caption{On the left, the initial directed graph with source and target and capacity function. On the right, the same graph with an extra feedback edge of an infinite capacity.}
\label{fig:mfmc}
\end{figure} 
An example of a weighed graph with a capacity function is presented in the Figure~\ref{fig:mfmc}. This flow problem can be easily changed to a circulation problem by adding a directed feedback edge, let's call it $F$, from target to source of an infinite capacity. Now, I would like to consider only the cycles in a graph that involve the edge $F$. One can think that all the \emph{local} cycles in the graph are trivial/non important. 

A celebrated result in computer science says that maximal flow (or circulation) is equal to the minimal cut. Flow lines are cycles crossing $F$. \emph{Cuts} are the edges whose removal kills all flow. Mathematicians may already see an analogy to a well known theorem in topology. Flows are like the first homology generators. Cuts are the dual generators of the first cohomology group. Max flow min cut theorem is nothing more than a special version of the well known Poincare duality theorem. The cuts we have been talking about are just like the cuts in the graphs. The only difference is that they are embedded in a higher dimensional structures. We will be discussing this in the next section.

\section{Comments}
\begin{enumerate}
\item When talking about Euler's problem I have promised you an explanation why a described walk in Königsberg does not exist. Have a look at the graph. Note that there are an odd numbed of edges incidental to every vertex. Suppose by contrary that such a walk exists. If it exist, we can assume that we start at any region we like, call it $A$. Then in order to move to the next region, we need to cross a bridge. Pick any bridge and cross. Once the bridge was crossed, it cannot be crossed any more from the formulation of the problem. Now, there are an even number of bridges we can use when going through $A$. When getting in and getting out, we will make two of the remaining even number not usable. Therefore we can never get back to $A$ having used all the attached bridges. 
In his work Euler gave a deep analysis and proved theorems about existence of a path, which we nowadays call an \emph{Euler path} in graphs. 
\item Moore's law is an observation saying that the processing power of our computers double every $18$ months. That means that now for 1000\$ we can buy twice as powerful computer as we could 18 months ago. There is a huge amount of money put by the processor manufactures to maintain such a speed. But far fewer people are aware that there is a Moore's law present in algorithms and in general in science. In particular, the development of algorithms to compute maximal flows and minimal cuts extends Moore's law even though the money invested in developing algorithms are small fractions of money invested to build chips. So	 if you have a DeLorean which allows you to travel back in time and you want to make a lot of money by doing max flow min cut calculations 20 years ago, but can take either software or hardware, you should definitely take software with you.  
\item The genesis on work on max flow and min cut is a cold war analysis of transport capability of Soviet Union and the dependent countries\footnote{See USAF unclassified report available here: \url{http://www.dtic.mil/dtic/tr/fulltext/u2/093458.pdf}.}. The very urgent question for the US Air Forces was -- how much supply can be brought daily to the front lines and where to cut the rail network of the enemy with the minimal cost so that a supply for the front lines will be broken. Below is an original map of the railway network graph with weights on the edges:
\begin{figure}[h!tb]
\centering
\includegraphics[scale=0.4]{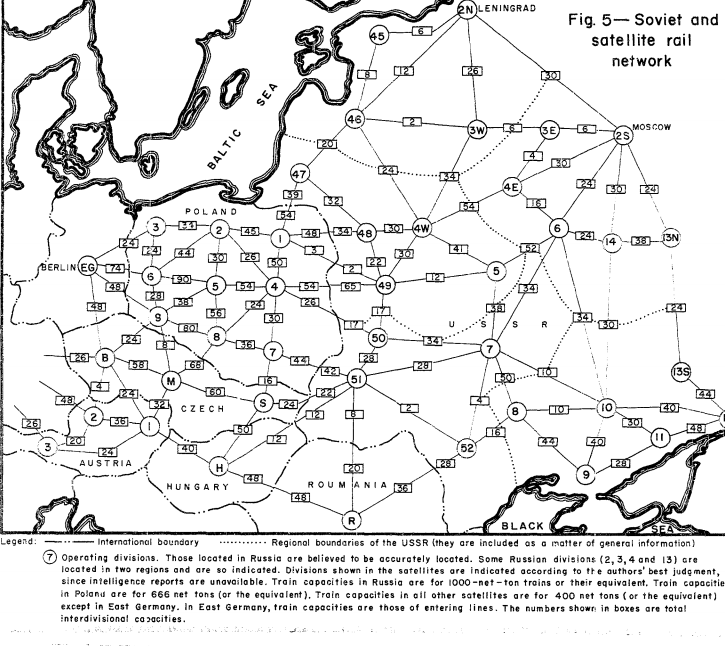}
\caption{USAF Project Rand, research memorandum.}
\end{figure}
\item A classical algorithm to compute maximal flow and minimal cut is due to Ford and Fulkerson. The idea is very simple:
\begin{enumerate}
\item Try to find a path $p$ from source to target such that the flow does not use all available capacity of that path.
\item Increase the flow through $p$ so that the maximal capacity of $p$ is reached.
\item Repeat the previous steps as long as such a path $p$ can be found.
\end{enumerate}

In order to find a minimal cut, one needs to start from the source and annotate all the vertices that can be reached with the edges $e$ such that $f(e) < c(e)$. A minimal cut consists of all the edges which has one endpoint annotated and the other one not annotated.
\end{enumerate}
\chapter{Why topology?}

Topology is about \emph{integrating local properties into global information}. Think of a set of beacons which can be heard from some distance. Imagine, that we have only binary information: can I, or can I not hear a given beacon? Note that very local information typically will not change under a \emph{continuous} deformation. An extreme example of such a continuous deformation is an equivalence of a coffee mug and a donut, one you can see in the picture below. 

\begin{center}
\includegraphics[scale=1]{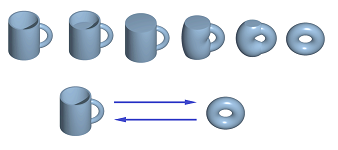}
\end{center}

In contract,	 a metric is about precise distances. In metric spaces we can determine exactly how far we are from a given object. How far we are from the Eiffel tower? How far it is from here to the nearest coffee shop? A metric gives much more precise information, but at a much greater cost. If we forget about metrics, we can have invariants that are very general - to the extent where they are the same for any pair of shapes that can be continuously deformed to each other, like donut and coffee mug above. 

Given this exposition you may wonder, what is topology good for? Is there any realistic scenario when it can be used? The answer is positive, and we will have an example of this situation in this section. 

Think of a collection of beacons. You do not know where they are. You do not know how far away from you they are. What you can measure is if you can hear it or not (and in some cases the strength of the signal). This is precisely the notion of proximity available via topology. 
There are some examples of beacons that use Bluetooth technology - you can find them in some shops, and they allow you finding various goods you are looking for (and perhaps even want to buy). 

In various campuses and urbanized areas there are different types of beacons that we can use.
If you happen to have osx or a linux machine you will be able to take part in the exercise. Unfortunately at the moment I do not have a code that works on windows machines (sorry...)
If you are a linux user please download this code:\\
\url{https://www.dropbox.com/s/qpxocdk57x9cmaf/ssidcollect_linux.py?dl=0}

If you are an osx user, please download this code:\\
\url{https://www.dropbox.com/s/bkf2yj6u1eza82w/ssidcollect_osx.py?dl=0}

Both are designed to recode ssid of wifi networks. Run them by using:
\begin{lstlisting}
python ssidcollect_linux.py
\end{lstlisting}
Now go for a walk. You may go around some obstacles (ideally of a considerable diameter). When you are back, stop the execution (you can do it by hitting CTRL+c or Command+C). 

In the folder you have run the code, the file \emph{output} is produced. Have a look at it! The first column is a time (number of seconds since 1 January 1970), second, the ssid of the router, and third: the signal strength. We will turn this data into a collection of points in $\mathbb{R}^N$. The number of dimensions, $N$, is equal to the number of distinct ssids we have in the file. Each sample in the time will correspond to a point.
You can either implement this simple code, or get it from here: \url{https://www.dropbox.com/s/2euv2zahn2dyirs/ssidaccumulate.py?dl=0}
To get the point cloud simply run:
\begin{lstlisting}
python ssidaccumulate.py
\end{lstlisting}

By doing so we will obtain a file \emph{file\_with\_points.csv}
Let us run the Principal Component Analysis on our data:

\begin{lstlisting}
import matplotlib.pyplot as plt
import numpy
import csv
from sklearn import decomposition

X = numpy.genfromtxt('file_with_points.csv', delimiter=',')
fig = plt.figure()

ax = plt.axes( projection='3d' )

pca = decomposition.PCA()
pca.fit(X)
X = pca.transform(X)

size = [200 for n in range(len(X[:,0]))]
color = [100*n for n in range(len(X[:,0]))]
ax.scatter(X[:, 0], X[:, 1], X[:, 2],s=size,c=color)#
plt.show()
\end{lstlisting}

I have done this experiment when walking close my house in Swansea. Without revealing too much of my private data, I have done one of the circles from the map below:
\begin{center}
\includegraphics[scale=0.5]{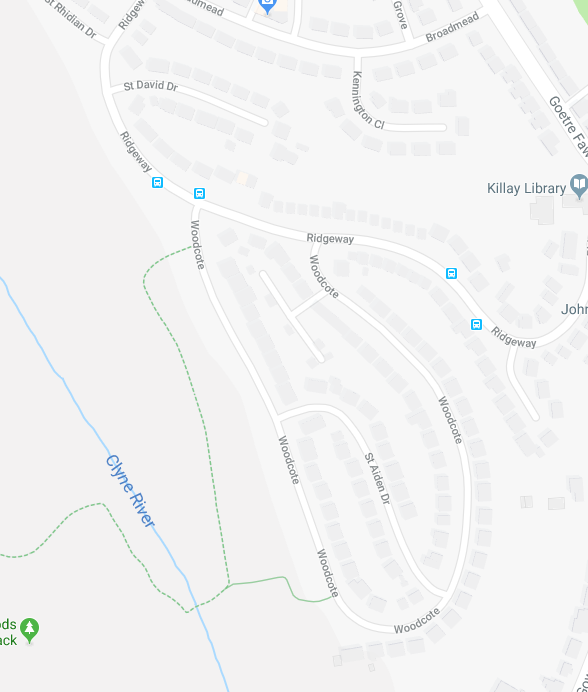}
\end{center}
Which results in the following:
\begin{center}
\includegraphics[scale=0.7]{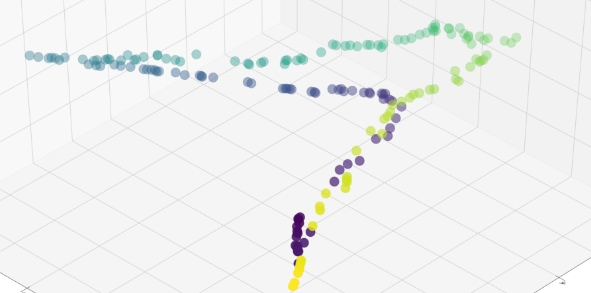}
\end{center}

As you see, no metric features are preserved here, but we can notice that I have made a loop. Colouring indicates time. All this is based on detecting proximity to wifi routers. Later on in this lecture we will learn the ways of quantifying that there indeed is a cycle. 

A historical note. I have done this exercise first attending a summer school in Topological Data Analysis organized by Gunnar Carlson, Rob Ghrist and Ben Mann in Snowbird in 2011. The majority of the code we have used here is due to Mikael Vejdemo-Johansson (also known as MVJ). Big thanks to him!
\include{complexes}
\include{complexesOfRelation}
\chapter{Cubical complexes}
In the previous section we have discussed the idea of simplicial complexes. They are very important in computational topology. But there are other ways of representing topological spaces which are also highly useful. One of them is the cubical complex. As far as I know cubical complexes were introduced and used by Jean-Paul Serre~\cite{j_p_serre}.
%J.P. Serre, Homologie singuli`ere des espaces fibr´es: Applications, Annals Math. 54(3) (1951), 425-505.

I came across cubical complexes when studding rigorous dynamics and computer assisted proofs. They are closely related with \emph{interval arithmetic} -- a computational tool that allows one to make calculations on a computer rigorous~\cite{moore_intervals}. The idea of interval arithmetic is to put all errors of the computations into a representation of a number. To do so, interval arithmetic defines all the elementary operations (together with elementary functions) on an interval instead of single numbers. More precisely, let $\mathbb{D}$ denote a set of \emph{double} numbers represented in a computer. Let us take $x, y \in \mathbb{D}$. By $low(x \diamond y)$ let us denote a largest number in $\mathbb{D}$ that is smaller than the result of the operation $x \diamond y$. By $high(x \diamond y)$ let us denote a smaller number in $\mathbb{D}$ that is greater than the result of the operation $x \diamond y$. We can obtain those numbers, since a processor gets the result with higher precision than the one available in double numbers. Those operations guarantee that the true value of the operation can be always enclosed in the resulting interval. Let us take two intervals $[a,b]$ and $[c,d]$ such that $a,b,c,d$ are \emph{representable} doubles. Then:
\begin{enumerate}
\item $[a,b] + [c,d] \subset [low(a+c),high(b,d)]$.
\item $[a,b] - [c,d] \subset [low(a-d),high(b-c)]$.
\item $[a,b] * [c,d] \subset [min\{ac,ad,bc,bd\} , max\{ac,ad,bc,bd\}]$.
\item I am skipping the formula for division. It is analogous to the previous ones. One has to make sure that zero is not in the interval in the denominator.
\end{enumerate}
Note that in the computations above the operations on the left hand side are rigorous mathematical operations defined on the intervals. The operations on the right hand side are made on computer representation of double numbers.

By using a basic Taylor expansion, we can also have an interval version of elementary functions. Given this, we can build mathematically rigorous numerical algorithms. This gives rise to a big field of computational mathematics called rigorous dynamics.

In dynamical systems, we often deal with maps acting from $\mathbb{R}^n$ to $\mathbb{R}^n$. To do a rigorous computation in this $n-$dimensional setting, we are using $n-$dimensional cubes. This is the rigorous dynamics genesis of \emph{cubical complexes}.

\begin{figure}[h!tb]
\centering
\includegraphics[scale=0.8]{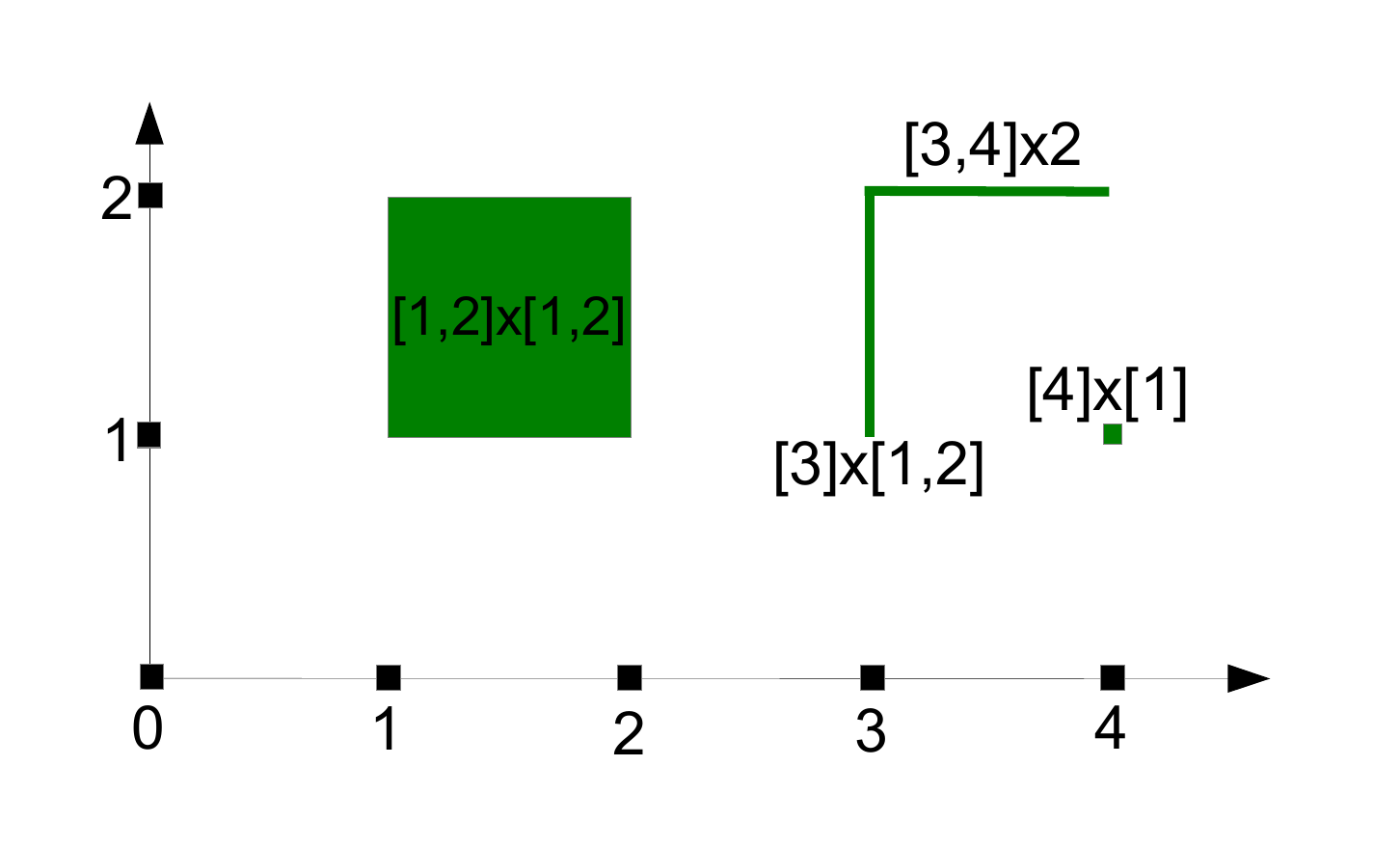}
\caption{A cubical complex.}
\label{fig:cubicalComplexes}
\end{figure}

For now, we restrict only to cubical complexes for which cubes of a fixed dimension have the same size. By doing a rescaling we can assume that the endpoints of the involved intervals have integer coordinates. By an \emph{elementary interval} we mean an interval $[n,n+1]$ or an interval $[n,n]$. The first one is non degenerated, the second one is a degenerated one. An \emph{elementary cube} is a Cartesian product of elementary intervals (degenerated of not). A few examples of elementary cubes are presented in the Figure~\ref{fig:cubicalComplexes}.

\begin{figure}[h!tb]
\centering
\includegraphics[scale=0.8]{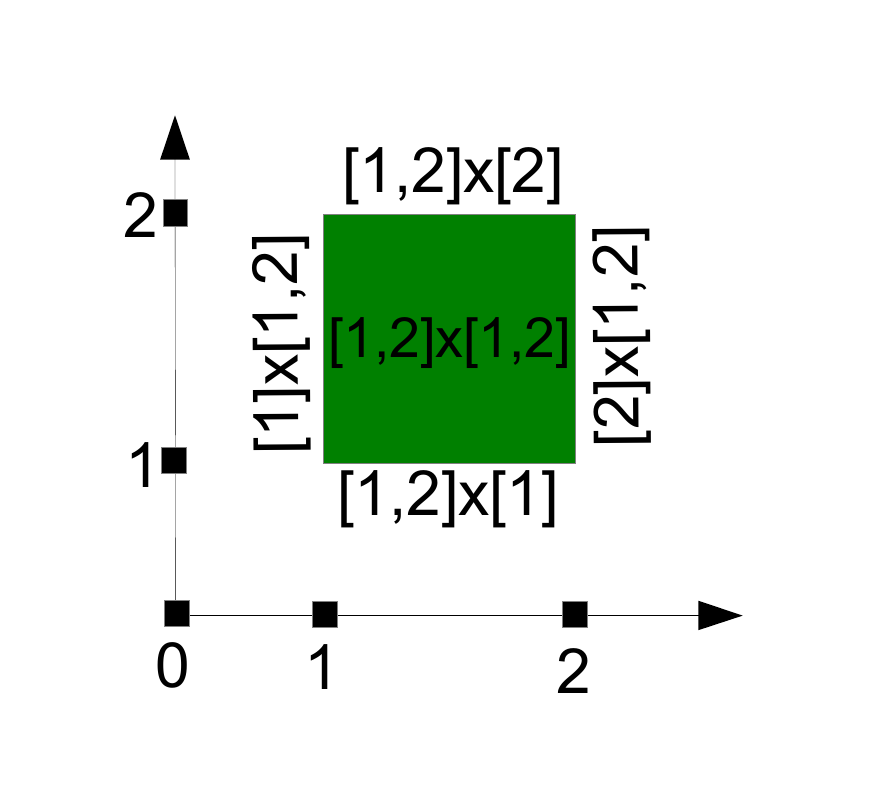}
\caption{Boundary of an elementary cube.}
\label{fig:boundaryOfACube}
\end{figure}

Elementary cubes are building blocks of cubical complexes. Let us take an n-dimensional cube $C = [x_0,y_0] \times [x_1,y_1] \times \ldots \times [x_n,y_n]$. At the beginning let us assume that all the intervals involved in $C$ are not degenerated. Then the boundary of $C$ is a collection of cubes $[x_0,y_0] \times [x_1,y_1] \times \ldots \times [x_i,x_i] \times \ldots \times [x_n,y_n]$ and $[x_0,y_0] \times [x_1,y_1] \times \ldots \times [y_i,y_i] \times \ldots \times [x_n,y_n]$ for every $i \in \{0,\ldots,n\}$. If some intervals in the product are degenerated, they are skipped in this procedure. Note that every element in the boundary of $C$ is a cube of a dimension equal to the dimension of $C$ minus $1$. See Figure~\ref{fig:boundaryOfACube} for an example.

For a typical kind of input there is one technical advantage of cubical complexes with respect to simplicial ones. They can be very efficiently stored in the computer memory. In the best case, we only need one bit (bit, not byte!) of information per cube. We do it by using so called \emph{bitmap} representation. Instead of defining it formally, let us have a look at an example. In the first case we will store only top dimensional cubes in the bitmap. In the second one we will show how to store the cubes of all dimensions.

\begin{figure}[h!tb]
\centering
\includegraphics[scale=0.8]{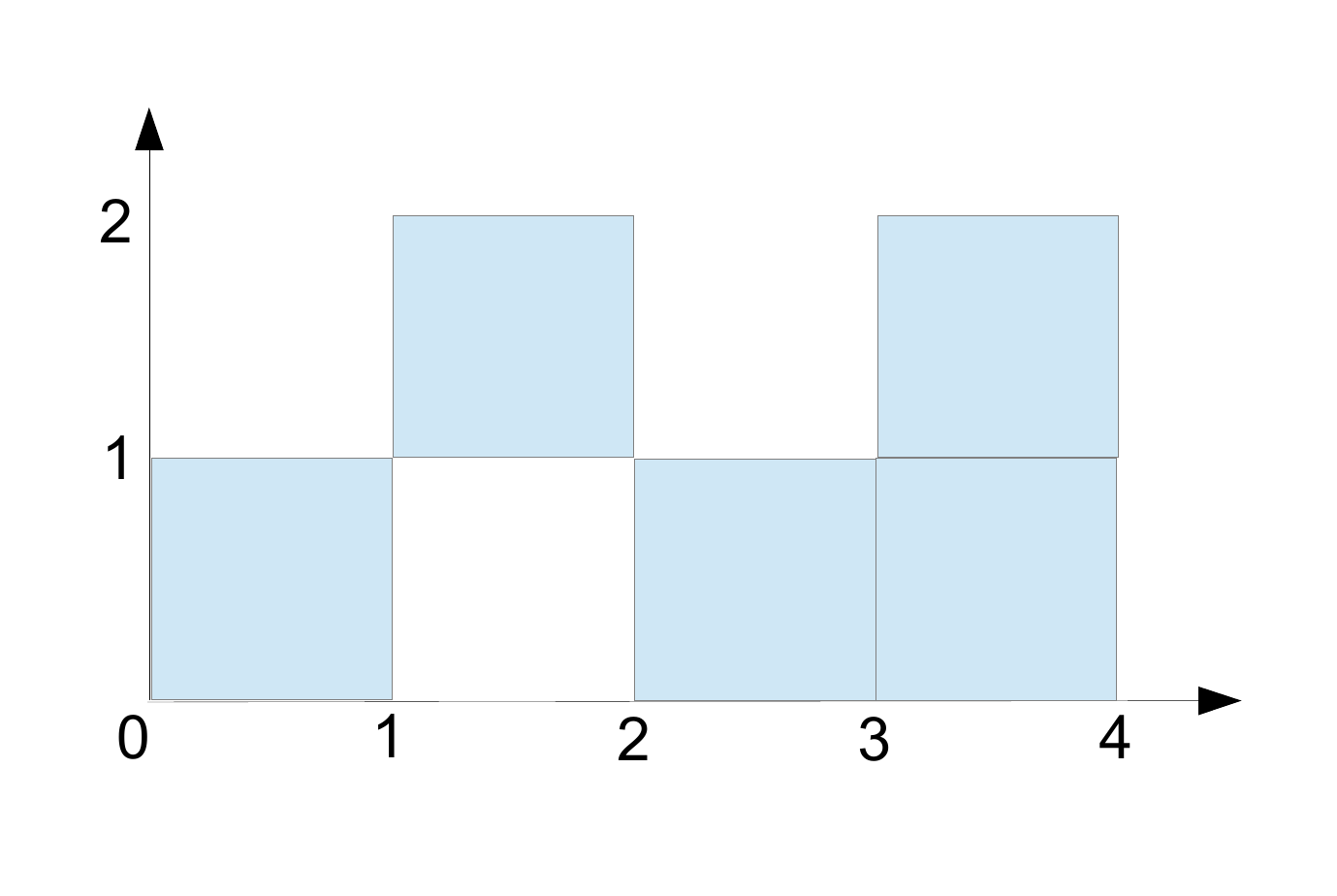}
\caption{Bitmap of a fixed dimension.}
\label{fig:bitmapFIxedDimension}
\end{figure}

In Figure~\ref{fig:bitmapFIxedDimension} an idea of a bitmap used to store cubes of fixed dimension is presented.
In this complex, the cubes $[0,1] \times [0,1]$, $[2,3] \times [0,1]$, $[3,4] \times [0,1]$, $[1,2] \times [1,2]$, $[3,4] \times [1,2]$ are present. All cubes have unit length, and the range of the out complex is from $0$ to $4$ in the x direction and from $0$ to $2$ in the y direction. We can therefore enumerate all the possible cubes in that range and get the following $2$ dimensional array:

\[
\begin{array}{|c|c|c|c|}
\hline
 [0,1]\times[1,2]      & [1,2]\times[1,2]    & [2,3]\times[1,2]     & [3,4]\times[1,2] \\
 \hline
 [0,1]\times[0,1]      & [1,2]\times[0,1]    & [2,3]\times[0,1]     & [3,4]\times[0,1] \\
 \hline
\end{array}
\]

To encode the complex, it suffices to encode which of the above cubes are present and which are not. Therefore, here is a representation of the complex:
\[
\begin{array}{cccc}
0      & 1    & 0     & 1\\
1      & 0    & 1     & 1\\
\end{array}
\]
This array, given the width and the height of a bitmap, can be encoded in a one dimensional array of bits\footnote{This is why we call it a bitmap.}.

\begin{figure}[h!tb]
\centering
\includegraphics[scale=0.8]{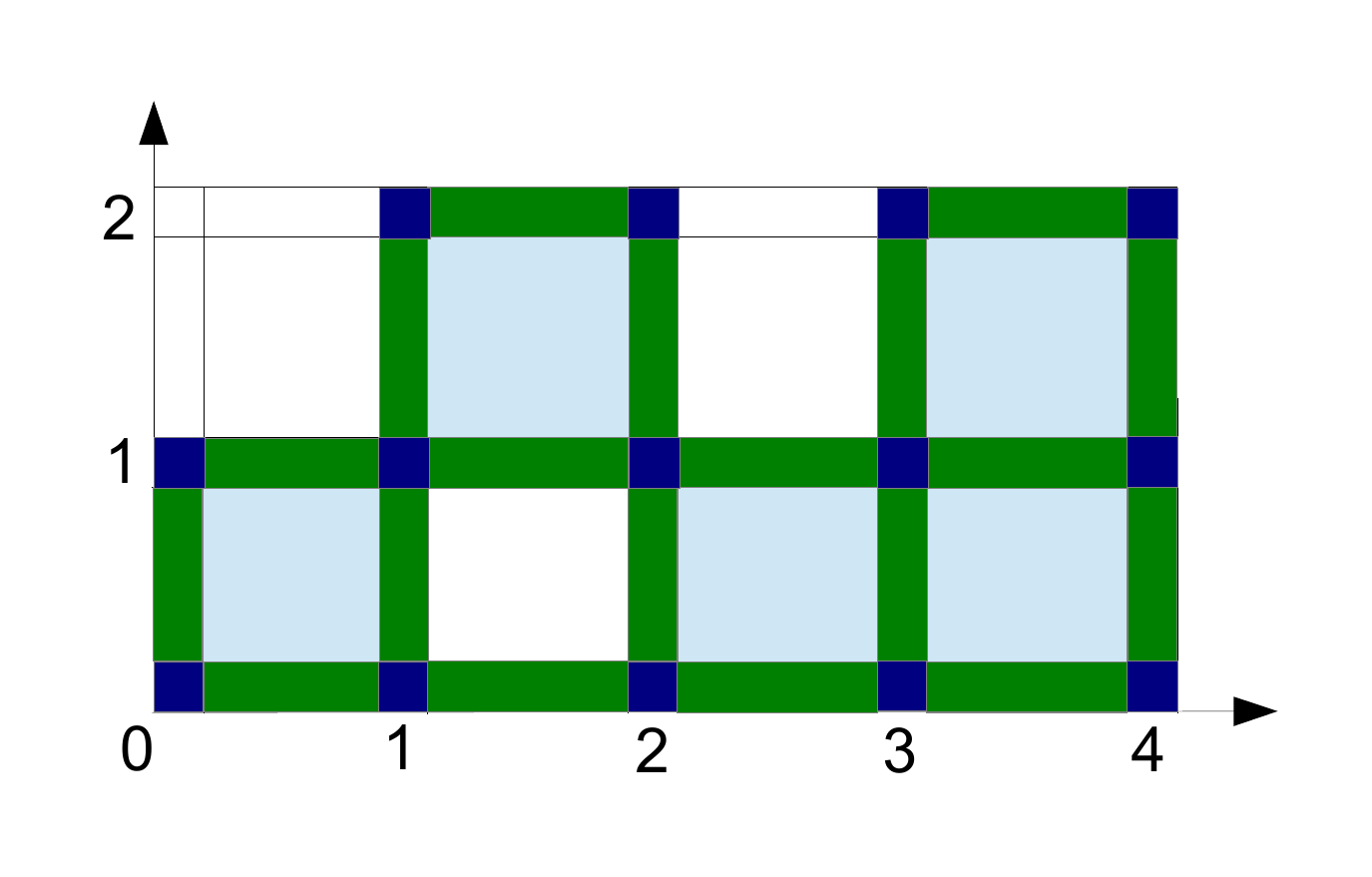}
\caption{Bitmap containing cubes of all dimensions.}
\label{fig:bitmapAllDimensionCubes}
\end{figure}

The idea of encoding the cells of all dimensions is very similar. Please consult the Figure~\ref{fig:bitmapAllDimensionCubes}. The obvious details are left for the reader.

\begin{figure}[h!tb]
\centering
\includegraphics[scale=0.8]{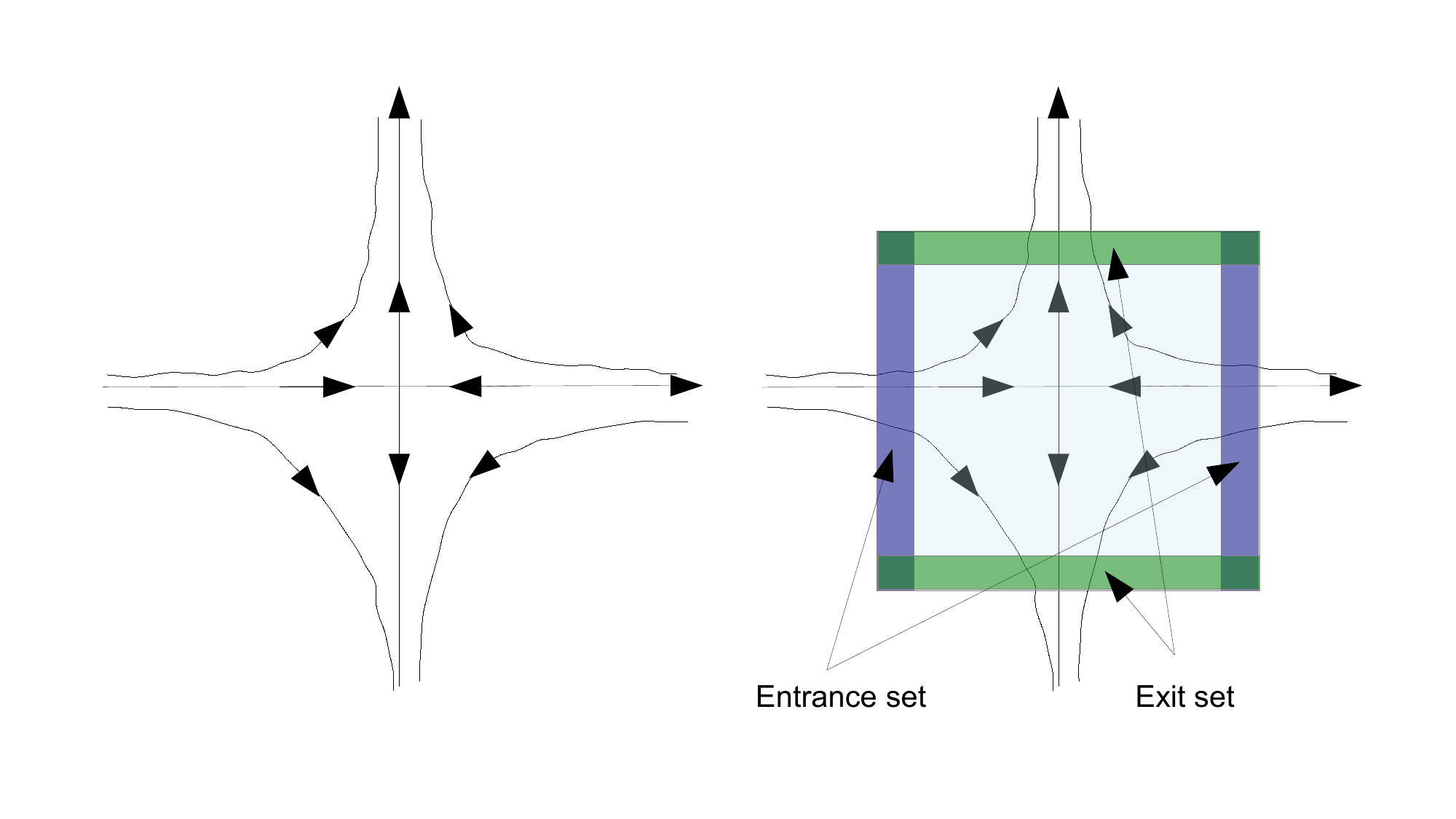}
\caption{Idea of a Conley index, for further details please
consult~\cite{conley_index}. Let us perform a simple phase space presented on the left.
There is contraction in a horizontal direction and expansion in the vertical one. Let us
suppose we build a so called \emph{insulating neighbourhood} $B$ as the box in the 
picture on the right. We assume that there are no fixed points in the boundary of 
$B$. We decompose the boundary of $B$ into two sets: \emph{exit set} is a set of all 
points that get out of $B$ for any sufficiently short time (marked with green) and 
\emph{entry set} which consists of all points in the boundary of $B$ that enters $B$ 
for any sufficiently small positive time (marked with blue). Now suppose we construct 
a so called \emph{quotient space}, where we contract the exit set into a point. To 
imagine the construction let us imagine first that the two horizontal stripes are 
contracted to a point. Later those two points have to be brought together. By dong so 
we will get a set having the homology of a circle. A celebrated theorem by Conley says 
that if there is an invariant set inside $B$, then topology of such a quotient space 
is not trivial. This is indeed the case in the picture above.}
\label{fig:conleyIndexIdea}
\end{figure}

Let me discuss one of the application of computational topology in rigorous dynamics. In this case, computational topology is used to compute a Conley index. The Conley index, if nontrivial, indicate the existence of an invariant part of a dynamical system (fix point, periodic orbit, etc.). An informal idea is presented in the Figure~\ref{fig:conleyIndexIdea}. 

Let us make a simple exercise which will lead us to the topic of the next section - filtrations. Let us define a two dimensional array of size $2N \times 2N$. Let us assume that this grid covers the area $[-2,2]^2$. On each grid element let us define a value of a function which is a distance to the unit circle. 
Note that for this very specific function, the distance to the unit circle is simply a norm of the grid element. Let us display it and create a cubical complex based on it. 
\begin{lstlisting}
import numpy as np
import math
import gudhi as gd
from PIL import Image


N = 100
array = np.zeros((2*N+1,2*N+1))
xExtrem = 2;
yExtrem = 2;

bitmap = []
for i in range(0,2*N+1):
	for j in range (0,2*N+1):
		x = i/(2*float(N)+1)*2*float(xExtrem)-xExtrem
		y = j/(2*float(N)+1)*2*float(xExtrem)-xExtrem
		norm = math.sqrt( x*x + y*y )
		norm =  math.fabs(norm-1)
		array[i][j] = norm
		bitmap.append(norm)

	
#Here we will display our creation:		
plt.imshow(array, cmap='gray', vmin=np.amin(array),vmax=np.amax(array))
#plt.savefig('circle.png')
plt.show()
				
#Given the input data we can buld a Gudhi btmap cubical complex:
bcc = gd.CubicalComplex(top_dimensional_cells = bitmap, dimensions=[2*N+1,2*N+1])
bcc.persistence()
\end{lstlisting}

Here is a an example of a picture you may get:
\begin{center}
\includegraphics[scale=0.6]{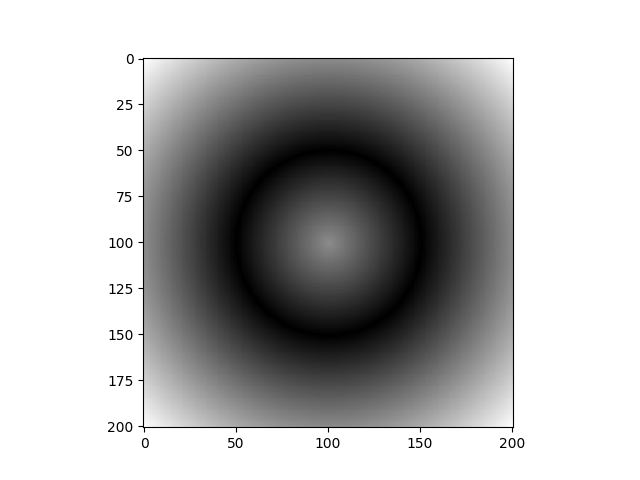}
\end{center}

Let us have a look at a similar example with a bit different way of obtaining a filtration on the maximal cubes (using Kernel Density Estimator). 

A crater dataset (for cubical filtration)
%http://bertrand.michel.perso.math.cnrs.fr/Enseignements/TDA/Tuto-Part1.html
This example was created and made available by Bertrand Michel. A	 thanks to Bertrand! Please visit his webpage for more interesting tutorials \url{http://bertrand.michel.perso.math.cnrs.fr/Enseignements.html}
The idea we want to illustrate is the following one. We start from a collection of points, and based on them, we create a continuous function: a kernel density estimator is built based on those ponts. Having the continuous function, we discretise it on a grid, and compute persistent homology of that discretization. For that purpose, we will be using cubical complexes. Currently we use python native mechanisms to get this filtration. Gudhi--dedicated tools for that will be soon available. Please download the data from that location~\url{http://bertrand.michel.perso.math.cnrs.fr/Enseignements/TDA/crater_tuto}

\begin{lstlisting}
import numpy as np
import pandas as pd
import pickle as pickle
import gudhi as gd
from pylab import *
import seaborn as sns
from mpl_toolkits.mplot3d import Axes3D
from IPython.display import Image
from sklearn.neighbors.kde import KernelDensity

f = open("crater_tuto")
#For python 3
#crater = pickle.load(f,encoding='latin1')
#For python 2
crater = pickle.load(f)
f.close()

plt.scatter(crater[:,0],crater[:,1],s=0.1)
plt.show()

#create 10 by 10 cubical complex:
xval = np.arange(0,10,0.05)
yval = np.arange(0,10,0.05)
nx = len(xval)
ny = len(yval)


#Now we compute the values of the kernel density estimator on the 
#center of each point of our grid. 
#The values will be stored in the array scores.
kde  =  KernelDensity(kernel='gaussian',  bandwidth=0.3).fit(crater)
positions = np.array([[u,v] for u in xval for v in yval ])
scores =  -np.exp(kde.score_samples(X= positions))

#And subsequently construct a cubical complex based on the scores.
cc_density_crater= gd.CubicalComplex(dimensions= 
[nx ,ny],top_dimensional_cells = scores)
# OPTIONAL, persistent homology computations:
pers_density_crater  =cc_density_crater.persistence()
plt = gd.plot_persistence_diagram(pers_density_crater).show()
\end{lstlisting}

\section{Filtration on complexes}
\begin{figure}[h!tb]
\centering
\includegraphics[scale=0.8]{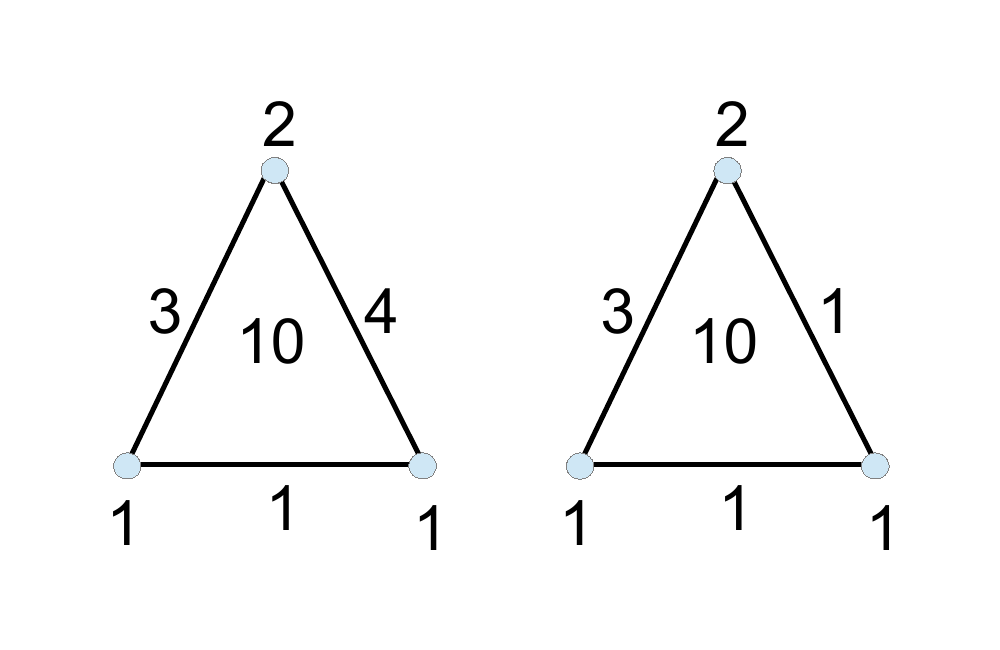}
\caption{On the left, a filtering function on a simplicial complex. The function on the right is not a filtering function, as some simplices appears before their boundaries.}
\label{fig:filteringFunction}
\end{figure}

In this section we will define a filtration of a simplicial or cubical complex. Traditionally in mathematics, a filtration is defined as an increasing sequence of subcomplexes. To make it as simple as possible, I prefer to define it in an equivalent way, by using a so called \emph{filtering function} defined on a finite complex. Let $\mathcal{K}$ be a finite simplicial or cubical complex. Then $f : \mathcal{K} \rightarrow A$, where $A$ is a set with a total order, is a filtering function if for every $a \in \mathcal{K}$ and for every $b$ in the boundary of $a$, $f(a) \geq f(b)$. In computational topology we often say that the boundary of a cell has to enter the filtration before the cell. In the Figure~\ref{fig:filteringFunction} please find a positive and negative example of a filtering function.

There are a few standard ways of defining a filtration in a complex $\mathcal{K}$. Let us have a look at them:
\begin{enumerate}
\item Given a filtering function $f$ defined on vertices. This may happen for instance when a scalar function is provided along with geometric data. Then, given a simplex or a cube $S$ which has vertices $v_0,\ldots,v_n$ we set the filtration value of $S$ to $max_{i \in \{0,\ldots,n\}}f(v_i)$.
\item Given a filtering function $f$ defined on top dimensional cells of a complex $\mathcal{K}$. This situation happens for instance when a $2d$ or a $3d$ image is given. The gray scale  level of the R, G or B from RGB levels on pixels or voxels provide such a function. Then the filtering function on a cell $S \in \mathcal{K}$ is defined as the minimum of filtering function values of maximal cells that have $S$ in their boundary.
\item Suppose that we are constructing a Rips complex. Then a natural filtration is induced by the length of the edges of a graph. A filtration value of vertices is set to $0$, filtration values of edges is their length and the filtration values of higher dimensional simplices is the maximal value of the length of edges in the boundary of that simplex.
\end{enumerate}
It is an easy exercise to show that those are filtering functions on a complex $\mathcal{K}$.
\chapter{Cycles and homology.}

\emph{Algebra is an offer made by the devil to mathematician. The devil says ''I will give you this powerful machine, it will answer any question you like. All you need to do is to give me your soul: give up geometry and you will have this marvellous machine''}
\\
Sir Michael Atiyah, 2002
\\

In order to start talking about applied \emph{algebraic} topology, we do need to dive a little bit deeper into algebra. Firstly we will define homology groups, which almost exactly fit the definition of the devil's offer mentioned above. However later on we will explore persistent homology, and by doing so we will regain our soul, as we will regain some geometric information.

The core idea of algebraic topology is to assign numbers to geometrical elements. Taking a collection of $n-$dimensional simplices, by assigning numbers to them we obtain $n-$dimensional chains. They form the structure of a group\footnote{A group in algebra means a structure with associative operation that poses unit element and such that every element has its inverse.} by a simple addition of elements component-wise (like polynomials). In this lecture we will restrict ourselves to the case of $\mathbb{Z}_2$, i.e. the elements of the group are $0$ and $1$, and the operations are defined \emph{modulo 2}. In this case it is important to remember that:
\[0+0=0\]
\[1+0=0+1=1\]
\[1+1=0\]
A \emph{group of n-dimensional chains} of a complex $\mathcal{K}$ will be denoted by $C_n(\mathcal{K})$. This group, for $n < 0$ or $n$ greater than the dimension of the complex $\mathcal{K}$, is a trivial group, as there are no simplices to support the chains. 

Right now we have a graded structure of groups in different dimensions. To relate them, we need to define a \emph{boundary operator}. Let us define it formally. In the lecture about complexes we have already encountered  boundaries of simplices. Now we have a formal framework of chains, so we can define the boundary of a simplex as the following chain:
%\begin{enumerate}
%\item 
Let us start from the case of simplices. Given a simplex $S = [v_0,v_1,\ldots,v_n]$, its boundary, $\partial S = \sum_{i=0}^n [v_0,v_1,\ldots,v_{i-1},v_{i+1},\ldots,v_n]$.
%
%\item For a cube $C = [x_0,y_0]\times \ldots \times [x_n,y_n]$ one need to find all the non degenerated intervals in the representation. The boundary of a cube $C$ is then defined as:
%\[ 
%\partial C = \sum_{[x_i,y_i]| x_i \neq y+i} 
%[x_0,y_0]\times \ldots [x_{i-1},y_{i-1}] \times [x_i,x_i] \times [x_{i+1},y_{i+1}] \times [x_n,y_n] 
%+
%\]
%\[
%[x_0,y_0]\times \ldots [x_{i-1},y_{i-1}] \times [y_i,y_i] \times [x_{i+1},y_{i+1}] \times [x_n,y_n] 
%\]
%\end{enumerate}
Now, given a chain $c = \sum \alpha_i s_i$, where $\alpha_i$'s are scalars and $c_i$'s are simplices, we can define a boundary of $c$ as $\partial c = \partial \sum \alpha_i s_i = \sum \alpha_i \partial c_i$. In this way, the boundary operator extends to chains maps $C_n(\mathcal{K})$ onto $C_{n-1}(\mathcal{K})$. A \emph{chain complex} is a sequence of chain groups joined by the boundary operator. It is a crucial property of the boundary operator that $\partial \partial = 0$. It suffices to prove this property for every simplex -- we leave it as a simple exercise.

In order to define homology groups, we need two subgroups of a groups of chains -- groups of cycles and group of boundaries:
\begin{enumerate}
\item Cycles in homology theory are a generalization of cycles we know from graph theory. A cycle in a graph is a closed path consisting of edges\footnote{Let us assume for simplicity that no edge is repeated in the cycle.}. We can ''pick up'' the edges belonging to the cycle by defining a chain $c$ having values $1$ for the edges in the cycle, and $0$ otherwise. The chain $c$ has a special property: every vertex of the cycle is incidental to an even number of edges in the cycle. Consequently $\partial c = 0$, so a cycle, treated as a chain $c$, has empty boundary. That is a definition that generalizes for higher dimensional complexes. An $n-$chain $c$ is a cycle if $\partial c = 0$. From graph theory we have an intuition that a $1-$cycle is a closed path. What about a $2-$cycle? Well, think about a closed surface. The boundary of a tetrahedron of a cube for a start. In that case, every edge in the tetrahedron is incident to two 2-dimensional cells, and therefore it vanishes in the boundary.

The intuition of a 2-cycle is a closed surface. Note that this surface does not have to be a sphere. Think for instance of the boundary of a torus we have explored so far. This is again a 2-chain. With the sufficient stretch of our imagination this analogy can be carried on higher up with the dimension. Cycles form a subgroup of a group of chains. The group of $n-$dimensional cycles of a complex $\mathcal{K}$ is denoted by $Z_n(\mathcal{K})$.

It is good to think about cycles as a way to encapsulate information. In homology theory we are particularly interested in information about holes. Given a complex with a hole, like the one in the Figure~\ref{fig:complexWithAHole} we can pick a cycle to be the bound the hole. Many other cycles will encapsulate the same hole. There are also cycles which do not encapsulate anything but a solid region of the complex. In order to distinguish them, we introduce the concept of a \emph{boundary}.
\begin{figure}[h!tb]
\centering
\includegraphics[scale=0.8]{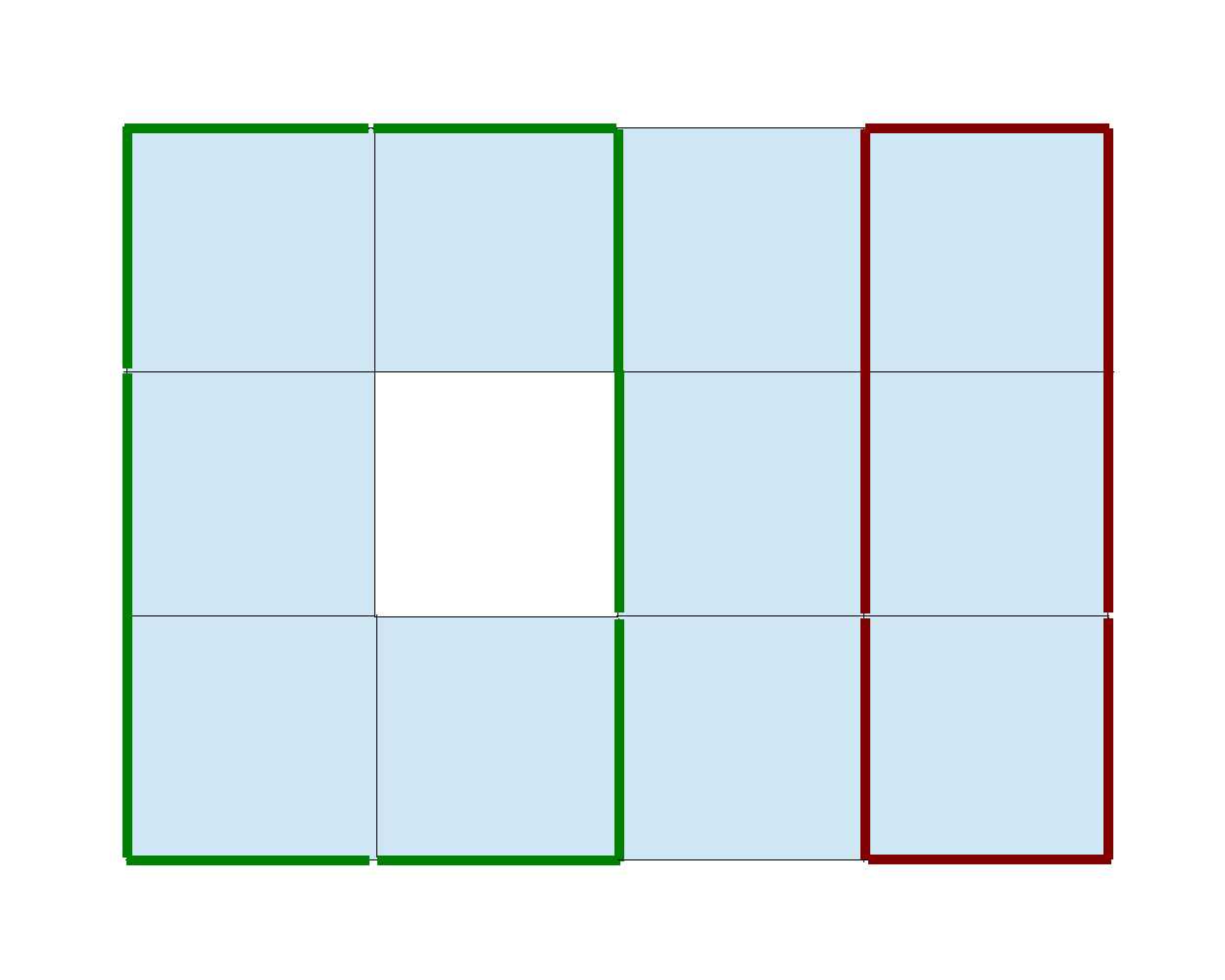}
\caption{Examples of cycles that bounds and that do not bounds a hole.}
\label{fig:complexWithAHole}
\end{figure}

\item \emph{Boundaries} are special kinds of cycles that do not bound a hole. Let us have a look at Figure~\ref{fig:complexWithAHole} at the red cycle (the one on the right). This cycle is a boundary of a 2-chain consisting of all cubes surrounded by the cycle. Why? Well, the edges in between cubes cancel out, so only the external ones remain. Formally a \emph{boundary} is an $n-$chain $c$ such that there exists an $(n+1)-$chain d such that $\partial d = c$.
Boundaries form a subgroup of cycles. The group of $n-$dimensional boundaries of a complex $\mathcal{K}$ is denoted by $B_n(\mathcal{K})$.
\end{enumerate}

One of the fundamental properties of the cycles and boundaries is the fact that every boundary is a cycle. There is a standard proof for simplices and cubes which can be figured out with paper and pencil in a short time. We will skip this proof here. 

\begin{figure}[h!tb]
\centering
\includegraphics[scale=0.8]{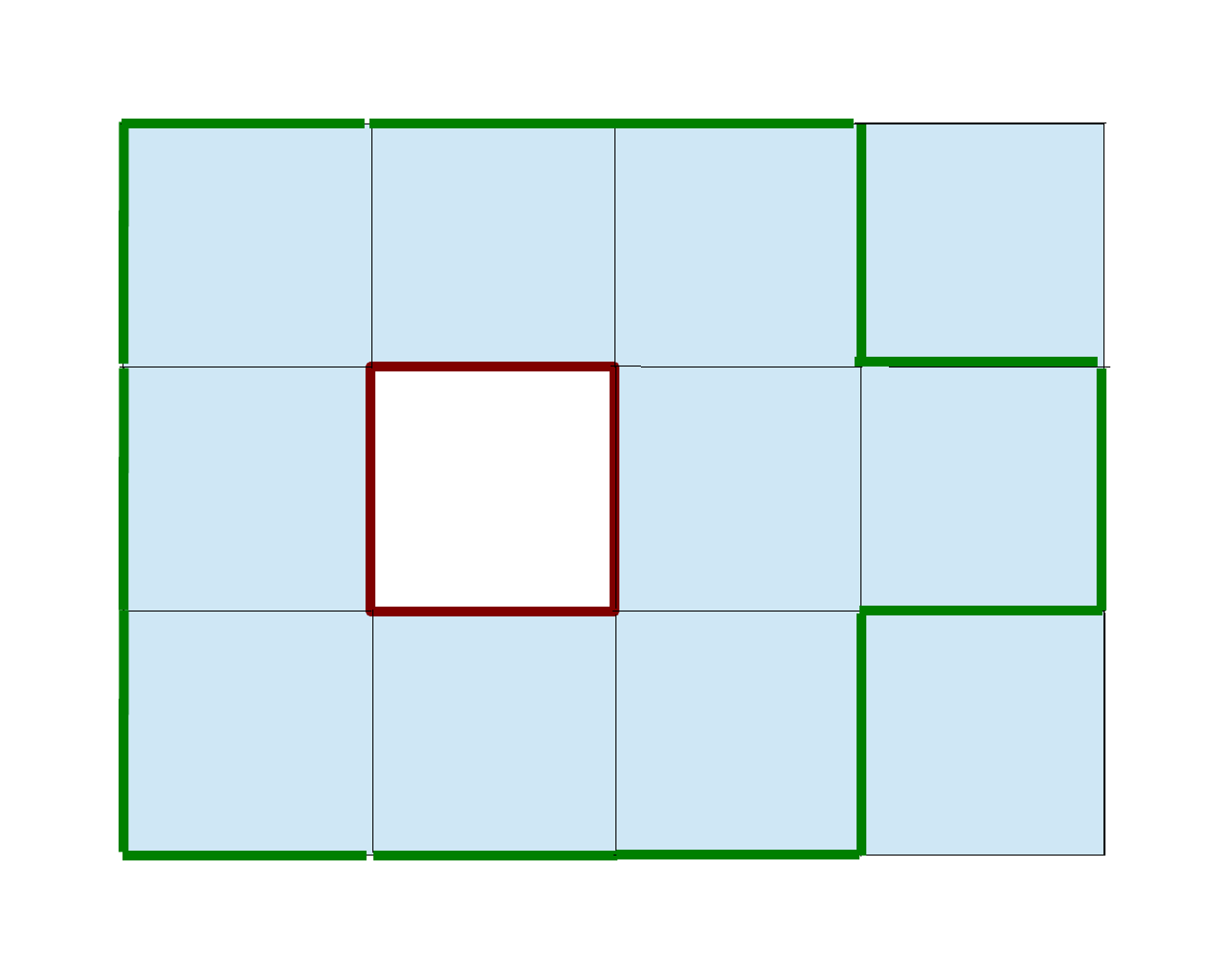}
\caption{Two equivalent (homologous) cycles.}
\label{fig:homologusCycles}
\end{figure}

It is typical in algebraic topology to consider the cycles which bounds a hole important, and those which do not, not important. Moreover two cycles $c_1$ and $c_2$ such that $c_1+c_2$ is a boundary are considered equivalent. See Figure~\ref{fig:homologusCycles} for an illustration. 

That intuition leads us to the definition of homology. Let us make it formal. Firstly we consider all the cycles that are boundaries as trivial. Secondly, all the cycles that bound the same hole are considered equivalent. Since $\partial \partial = 0$, every boundary is a cycle, and the group of boundaries is a subgroup of the group of cycles. Given this, when we quotient the group of cycles by the subgroup of	 boundaries, we will get information about the holes in the complex, one cycle per hole. Those are given by so called homology groups. More formally:
\begin{definition}
The p-th homology group is the p-th cycle group modulo the p-th boundary group. $H_p(\mathcal{K}) = Z_p(\mathcal{K}) / B_p(\mathcal{K})$
\end{definition}

Let me stress the main idea which is to count only the cycles that surround a hole. Therefore, one can interpret a homology group as a tool to count holes in the space. Typically integer homology is considered, which requires bringing a matrix into so called \emph{Smith Normal Form}. In our simplified scenario we compute homology over $\mathbb{Z}_2$, what requires a simplified matrix reduction algorithm. We will present it in a version suitable for computations of persistent homology and re-use it in the next section.

\begin{figure}[h!tb]
\centering
\includegraphics[scale=0.8]{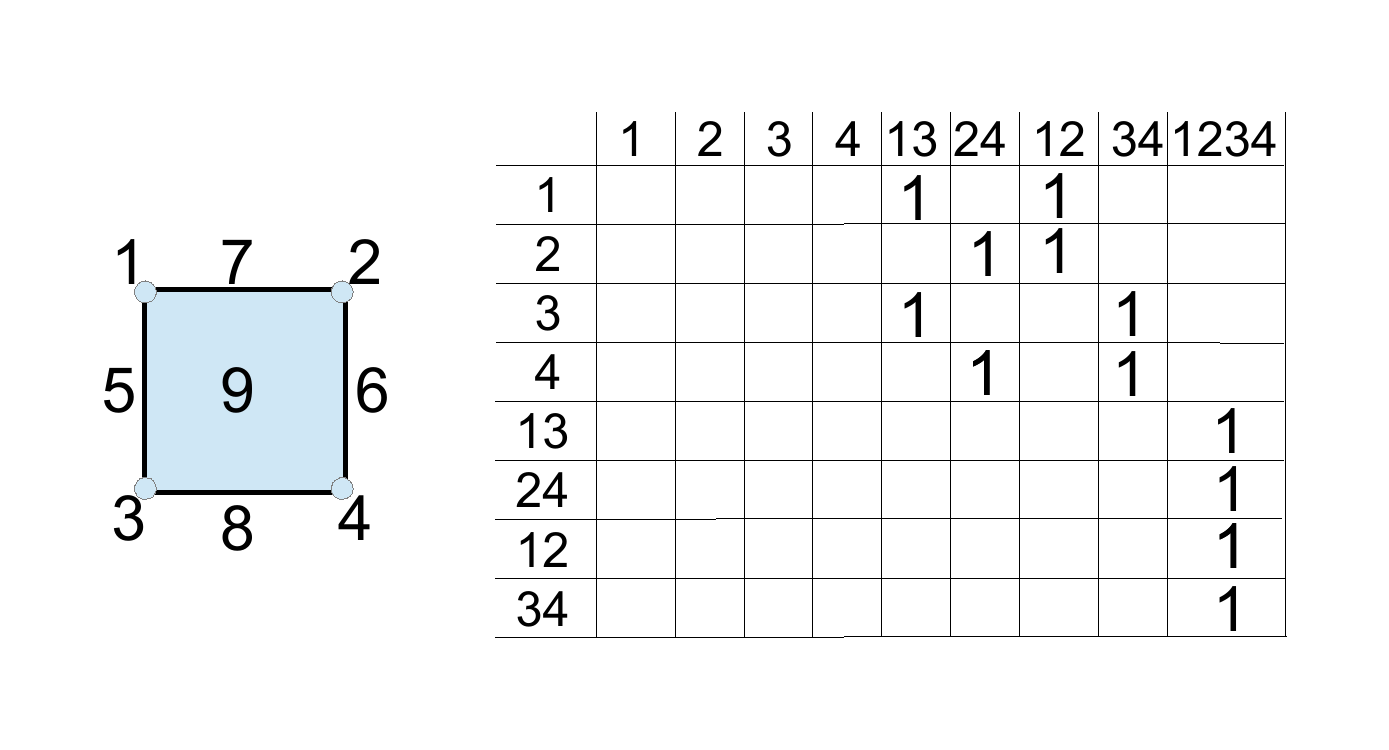}
\caption{Example of a complex with a filtering function on it that will be used in $\mathbb{Z}_2$ homology computations. Empty spaces denotes zero entries.}
\label{fig:matrixReduction}
\end{figure}

\begin{figure}[h!tb]
\centering
\includegraphics[scale=0.8]{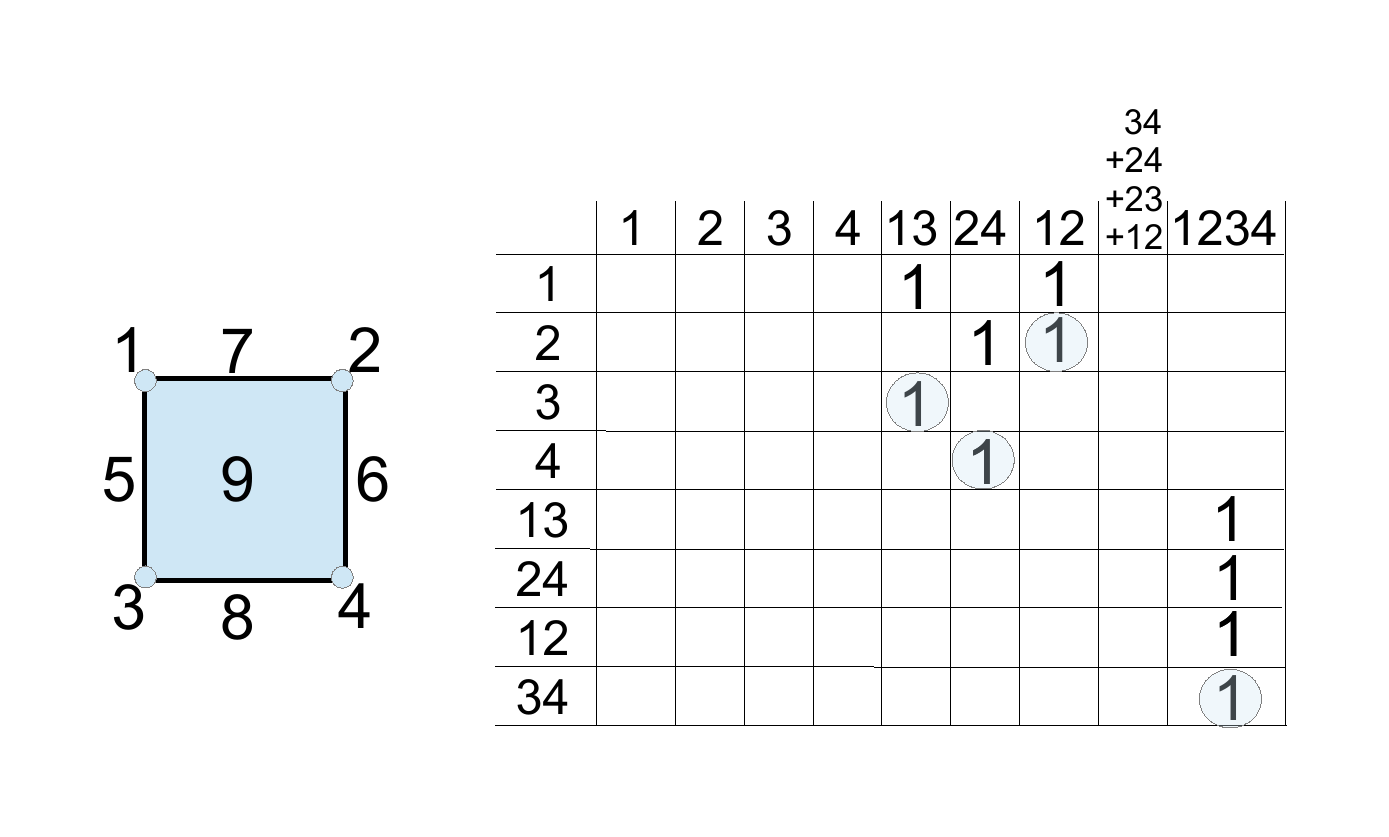}
\caption{Reduced boundary matrix. Lowest ones (which are unique after the algorithm terminates) are marked with ovals.}
\label{fig:matrixReduction2}
\end{figure}

In order to work with this algorithm we need to introduce a filtration (a.k.a. an ordering of elements) in the complex. For the sake of homology computations any filtration/ordering that preserves the order of dimension works. Let us consider an example presented in Figure~\ref{fig:matrixReduction}. It consist of a complex being a two dimensional cube. The filtration we have chosen is the following:
\[ [1] , [2] , [3] , [4] , [1,3] , [2,4] , [1,2] , [3,4] , [1,2,3,4] \]
Now, given the \emph{sorted boundary matrix} we want to perform the following algorithm:
\begin{algorithm}[H]
  \small
  \caption{Matrix reduction.}
  \begin{algorithmic}
	\REQUIRE Sorted binary matrix $M$ of a size $m \times m$;
	\
    \FOR { $i$ = 1 to m }
		\WHILE {there exist $j<i$ such that low(i) = low(j)}
			\STATE Add column $j$ to column $i$
		\ENDWHILE
    \ENDFOR
  \end{algorithmic}
\end{algorithm}
Here $low(i)$ is the lowest nonzero index in the $i-$th column. The idea of this algorithm is the following: We consider the columns of the matrix from left to right. If only the lowest one of the considered columns is equal to the lowest one of one of the previous columns, we add the previous column to the considered one. We check the previous columns from left to right. The reduced boundary matrix can be found on the Figure~\ref{fig:matrixReduction2}.

\begin{figure}[h!tb]
\centering
\includegraphics[scale=0.8]{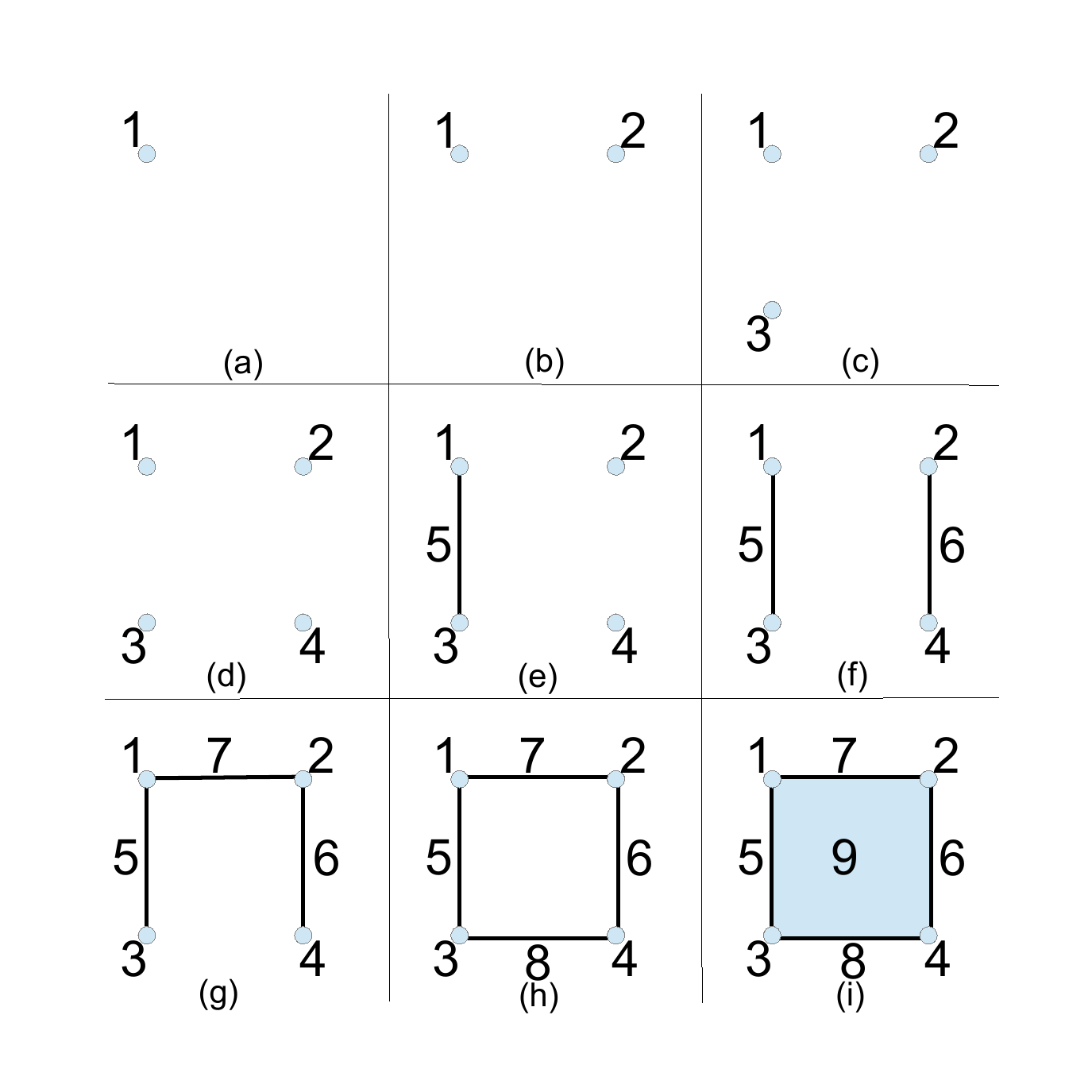}
\caption{Filtration of a complex.}
\label{fig:filtration}
\end{figure}

Let us now talk about how to interpret the reduced matrix. I will use a vocabulary of persistent homology even though it is not formally defined yet. To get the idea it is important to follow both the matrix in Figure~\ref{fig:matrixReduction2} and the evolution of the filtration presented in  Figure~\ref{fig:filtration}.

Take the first column of the matrix (indexed by a vertex [1]). It is empty. Empty columns mean that a homology class of a dimension equal to the dimension of the corresponding cell (in this case $0$) has been created. This is exactly the one connected component on Figure~\ref{fig:filtration}(a). The same holds for columns denoted by $2$, $3$ and $4$. The corresponding vertices, introducing new connected components, are on Figure~\ref{fig:filtration}(b-d). Then we have column $13$. It is nozero. In that case, we need to look at the lowest one. It is in row $3$. It means, that whatever was \emph{created} in the column indexed by $3$ is \emph{killed} in this column. Let us have a look at Figure~\ref{fig:filtration}(e). An edge $13$ appears and the number of connected components went down. So, indeed, a homology class died. The same situation takes place in column $24$. The lowest one is in row $4$, and therefore whatever was created in the column created by $4$ dies in this column. Again, when we look at Figure~\ref{fig:filtration}(f), we will see that adding edge $24$ decrease the number of connected components. An analogous argument hold for the column $12$. The next column, denoted as $34+24+23+12$, is zero. That means that a homology class is created there. Please note that $34+24+23+12$ explicitly gives us a representation of the nontrivial cycle that is created. The last column, $1234$ has a lowest one in $34$. This lowest one means that whatever was created in column $34$ (now denoted as $34+24+23+12$) is killed by $1234$. Look at Figure~\ref{fig:filtration}(i). Adding the two dimensional cell turns a homologically nontrivial cycle $34+24+23+12$ to a trivial one. 

Okay, so how do we get homology out of it? Simply check which classes were not killed. In this case, the only class which was not killed is the one in dimension zero, started by the vertex $1$. If we do not add a two dimensional cell $1234$, we would also have a one dimensional class $34+24+23+12$. It should be also clear, that we can stop the reduction at any stage, and with this procedure, we will get homology up to this stage.

\begin{figure}[h!tb]
\centering
\includegraphics[scale=0.8]{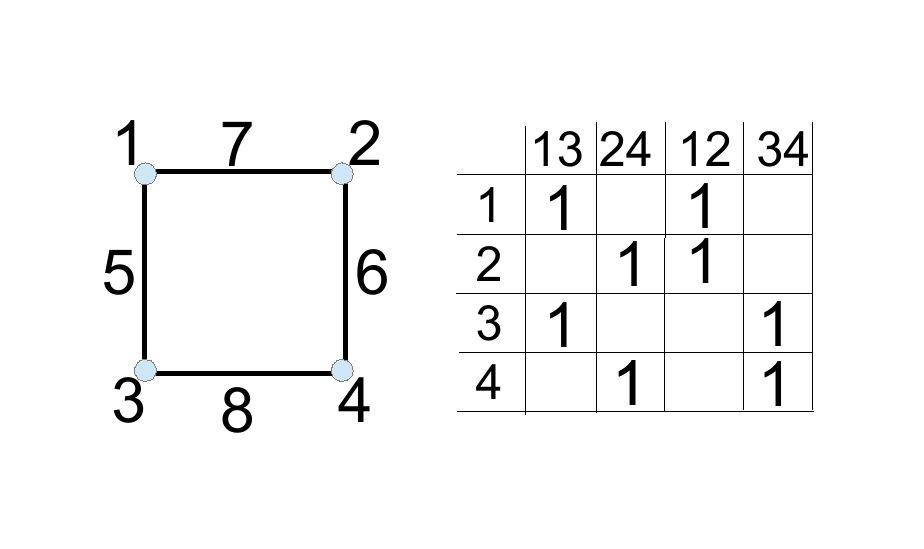}
\caption{Filtration of a complex.}
\label{fig:whyMatrixAlgorithmWork1}
\end{figure}

Okay, great. Magic. But the obvious question is -- why does it work?
Let's consider a few basic examples. The first one will be just a shorter and simplified version of the situation from Figure~\ref{fig:matrixReduction}. See Figure~\ref{fig:whyMatrixAlgorithmWork1}. There is no conflict until we get to the last column. 

Lets make a couple of observations. Firstly, every column in the matrix is either empty (all entries are zero), or it has lowest one. 
In the first case the boundary of the cell corresponding to this column can be generated from the cells that are already present in the complex. One may think about this as adding the final cell that closes a topologically nontrivial cycle. This is the moment of creation.

In the second case the lowest one in the $n$th column of the reduced matrix appears if the boundary of a  cell corresponding to that column creates a cycle that is not a linear combination of the cycles that have been generated so far. Since the boundary of a cell is a cycle itself, it means that this cycle cannot be generated as a linear combination of already available cycles, i.e. it is a nontrivial cycle. Addition of a cell make it trivial. So, this is the moment of destruction.

\begin{figure}[h!tb]
\centering
\includegraphics[scale=0.8]{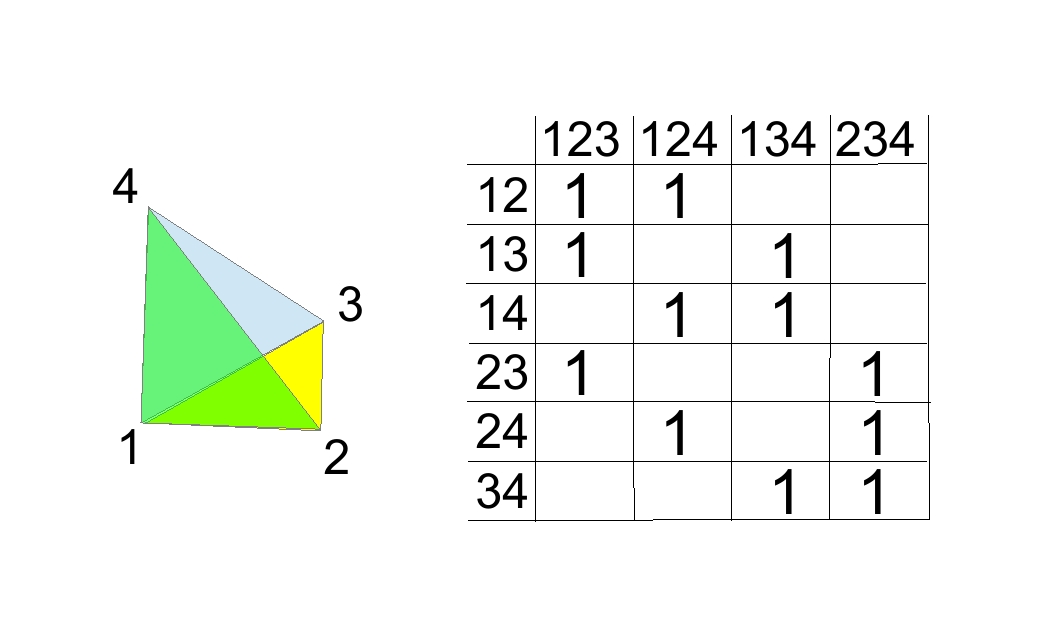}
\caption{Filtration of a complex.}
\label{fig:whyMatrixAlgorithmWork2}
\end{figure}

I am not going to give a proof here. Let us convince ourselves that this is the case by looking at the next example presented in Figure~\ref{fig:whyMatrixAlgorithmWork2}. In this example we have exactly the same situation, but one dimension higher. A detailed analysis is left to the reader.

Let us summarize what we have learned. In order to get a dimension of $i-$ dimensional homology group, one need to take a number of zero columns corresponding to the cells in dimension $i$ and subtract the number of nonzero columns corresponding to simplices in dimension $i+1$. The first collection of columns corresponds to the group of cycles, the second -- to the group of boundaries.

\section{Comments}
\begin{enumerate}
\item There exists a theory dual to homology theory called cohomology theory. If there is time, it will be intuitively introduced at the end of this lecture. Computational cohomology is also useful in applied science. Generating \emph{cocyles}, a concept dual to generating cycles in homology, is explicitly needed in some discrete formulations of Maxwell's equations. 
In a lot of cases homology and cohomology theory are dual. Do you remember the max-flow-min-cut duality we talked about in the first section? If there are no so called \emph{torsions} present in the homology groups\footnote{Torsions are generators of a finite order in the homology group.} then homology and cohomology groups are \emph{dual} (see the Universal Coefficient Theorem for Cohomology~\cite{hatcher}). There is a simple interpretation to this available in the picture of the annulus below. In red a representative of a cohomology generator is given. Formally it is so called \emph{cochain}. Think about it as a fence that has to be crossed by any homology generator. This is the basic interpretation of homology-cohomology duality.
\begin{center}
\includegraphics[scale=1]{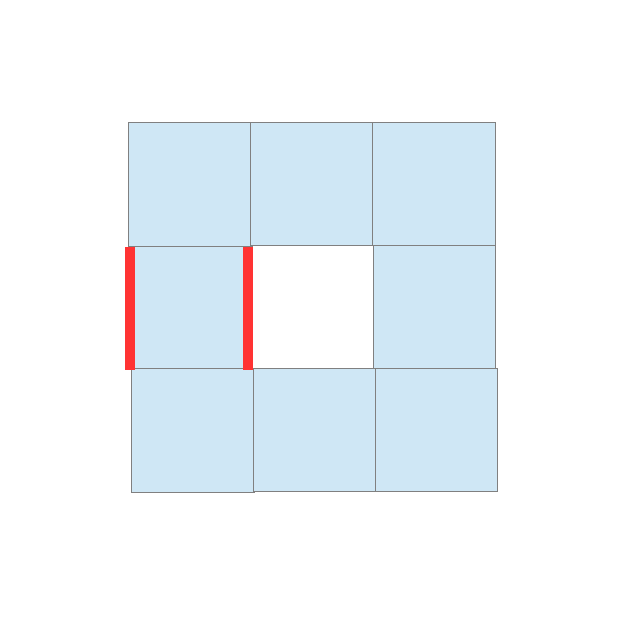}
\end{center}
This duality is a discrete version of Ampere's law. Let us suppose that the annulus above is a cross-section of an insulating domain. The hole therein is part of a conducting domain (that close up to a solid torus in 3d). Ampere's law states that magnetomotive forces on cycles surrounding the branch of a conductor s equal to the current passing through this conductor. It can be interpreted exactly as homology--cohomology duality.
\item Homology over $\mathbb{Z}_2$ is easy to define. In a standard lecture in algebraic topology we always start from homology defined over \emph{integer} coefficients. Due to the so called \emph{universal coefficient theorem for (co)homology}, those are the ones which carry the most information (see the comment above). The reason why we have chosen to use $\mathbb{Z}_2$ homology (or in general, a \emph{field} homology) is that they are needed for \emph{persistent homology} to be well defined. We will talk about persistent homology in the next lecture. 

I want to point out that the fact that we are using $\mathbb{Z}_2$ homology here does not mean that the (co)homology over integers is not used in applied topology. On the contrary, in some applications they are explicitly needed.  
%

%%
%\item The basic property of chains and boundaries, $\partial \partial = 0$ is very closely related with an idea of regular CW-complexes. CW complexes are generalizations of simplicies and cubes. Their basic property is that the closure of every cell is homeomorphic to a sphere of a certain dimension. For instance the CW decomposition of a $S^n$ with one $0-$cell and one $n-$cell is not regular. Regular CW complexes have a lot of nice properties. In particular, For every cell $C$ and for every two cells $D_1$ and $D_2$ in the boundary of $C$, if only $D_1 \cap D_2 \neq \emptyset$, then there exist unique element $E$ in the boundary of both $D_1$ and $D_2$. Moreover, it is very easy to assign incidence coefficients for regular CW-complexes. We will talk more about this when talking about applications of rectangular CW-complexes in a mesh generation. 
\end{enumerate}

\chapter{Persistent homology}

\begin{figure}[h!tb]
\centering
\includegraphics[scale=0.5]{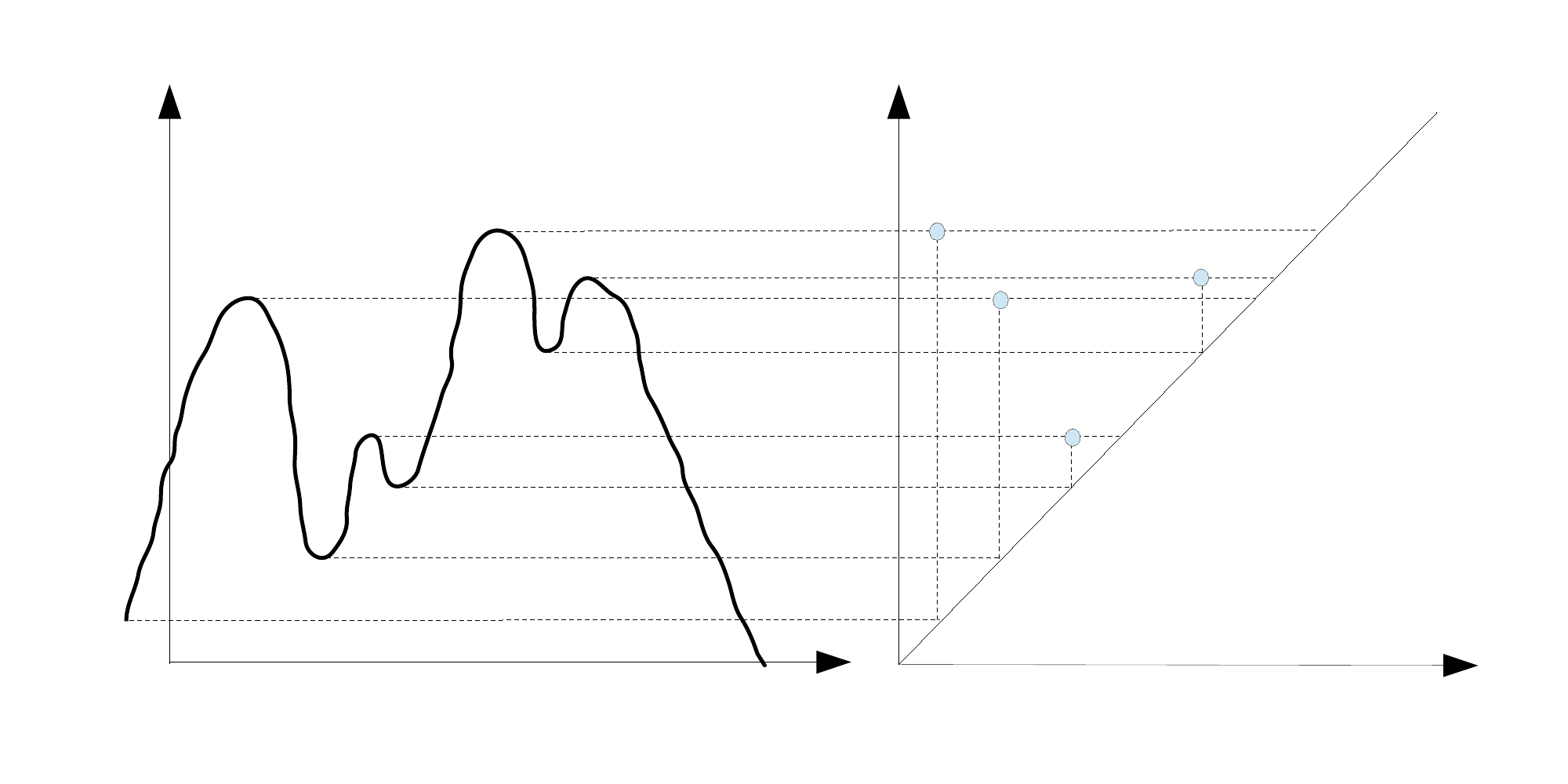}
\caption{Illustration of a basic idea of persistent homology. Let us consider the plot of the function on the left. Let us imagine that this is a surface of a one dimensional terrain and that there is rain falling over that terrain. We assume that the amount of water that falls at every point is the same no matter where the point is. When the plot of the function is floated, we check how many summits we see. In that case, obviously the lowest passes will be floated first. We \emph{pair} each summit with the pass the floating of which causes that we stop to see the summit. We draw the differences of the heights in the diagram on the right. In terms of functions those pairings give us an idea how a minimal perturbation of a function (in the terms of maximum norm) causes the minimum and the corresponding maximum to disappear. Those points, also called intervals, are closely related with the concept of persistent homology.}
\label{fig:persistence}
\end{figure}

So far we have been talking about standard homology. This topic has been known in mathematics since the time of Henry Poincare. 
One can think about persistent homology as a version of a homology theory where the complex under consideration is not static, i.e. contains all the simplices which were meant to be there. But what happens if new cells are being constantly added to the complex? In this new setup, we are likely to have Betti numbers appearing and disappearing when new simplices appear. To summarize this whole process we will require a notion of ''time slots'' in which Betti numbers are active. Let us consider Figure~\ref{fig:persistence} as an example.

We will also look at persistent homology from the perspective of point clouds. In this case we will not keep the single, unique distance in the Rips complex, or radius of the ball in the Cech complex fixed. In contrast, we will let it grow, and we will use persistent homology to see how the number of holes of different dimensions are evolving once the radius is growing. Doing so we will obtain a multi-scale summary of the data. 

\begin{figure}[h!tb]
\centering
\includegraphics[scale=0.5]{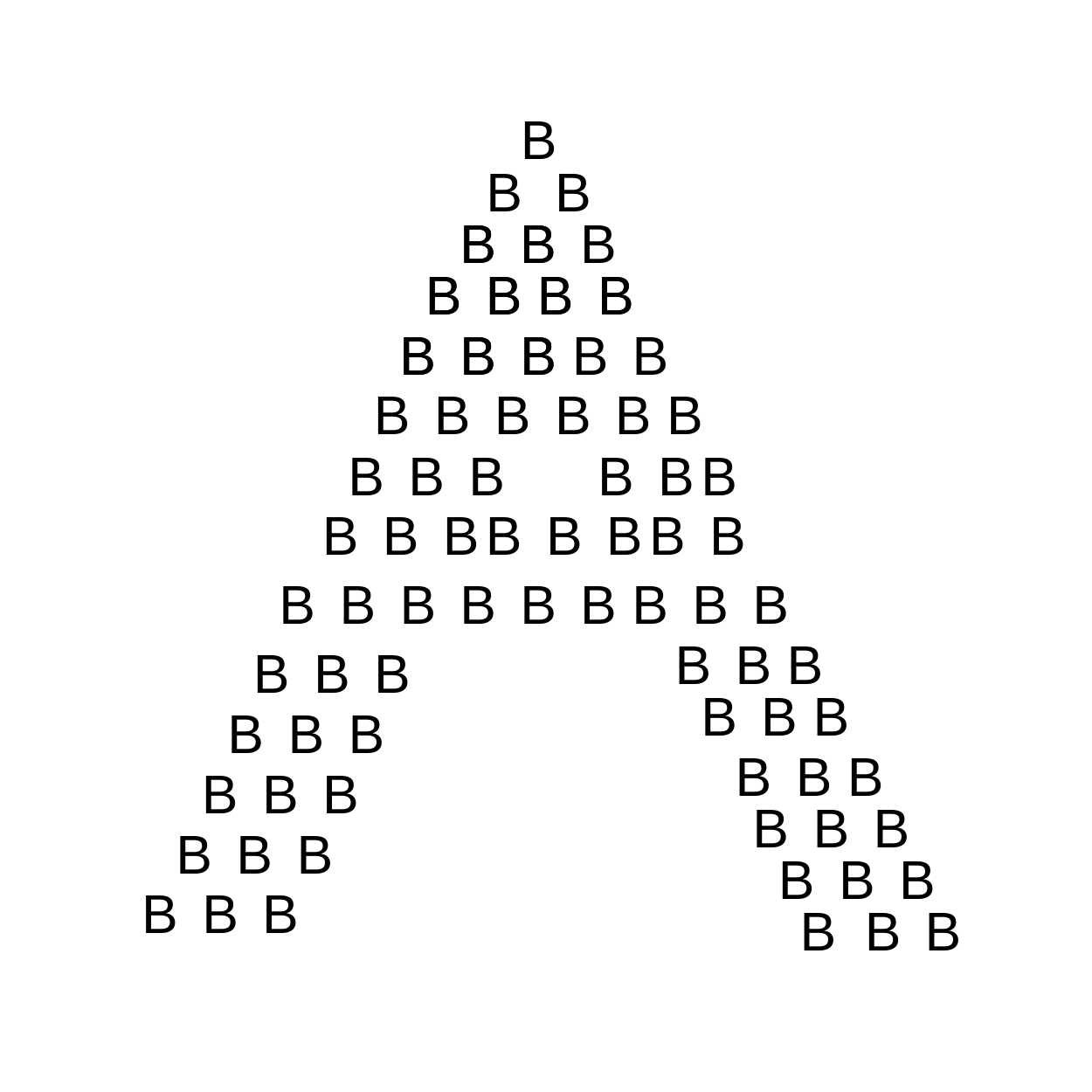}
\caption{Scale matters!}
\label{fig:scale_matters}
\end{figure}

You may ask, why we need a multiscale summary? The reason for that is because often the interpretation of the data depends on the scale, or granularity, you use to look at the data. Given this, different interpretations may be equally correct. Figure~\ref{fig:scale_matters} shows a standard example of a case where information in the picture cannot be summarized in a single spatial scale.

Let us have another look at the example in Figure~\ref{fig:persistence}. In this case, we are tracking the connected components of the sublevelsets of a function $S_a = \{x \in \mathbb{R} | f(x) \leq a\}$

\begin{figure}[h!tb]
\centering
\includegraphics[scale=0.5]{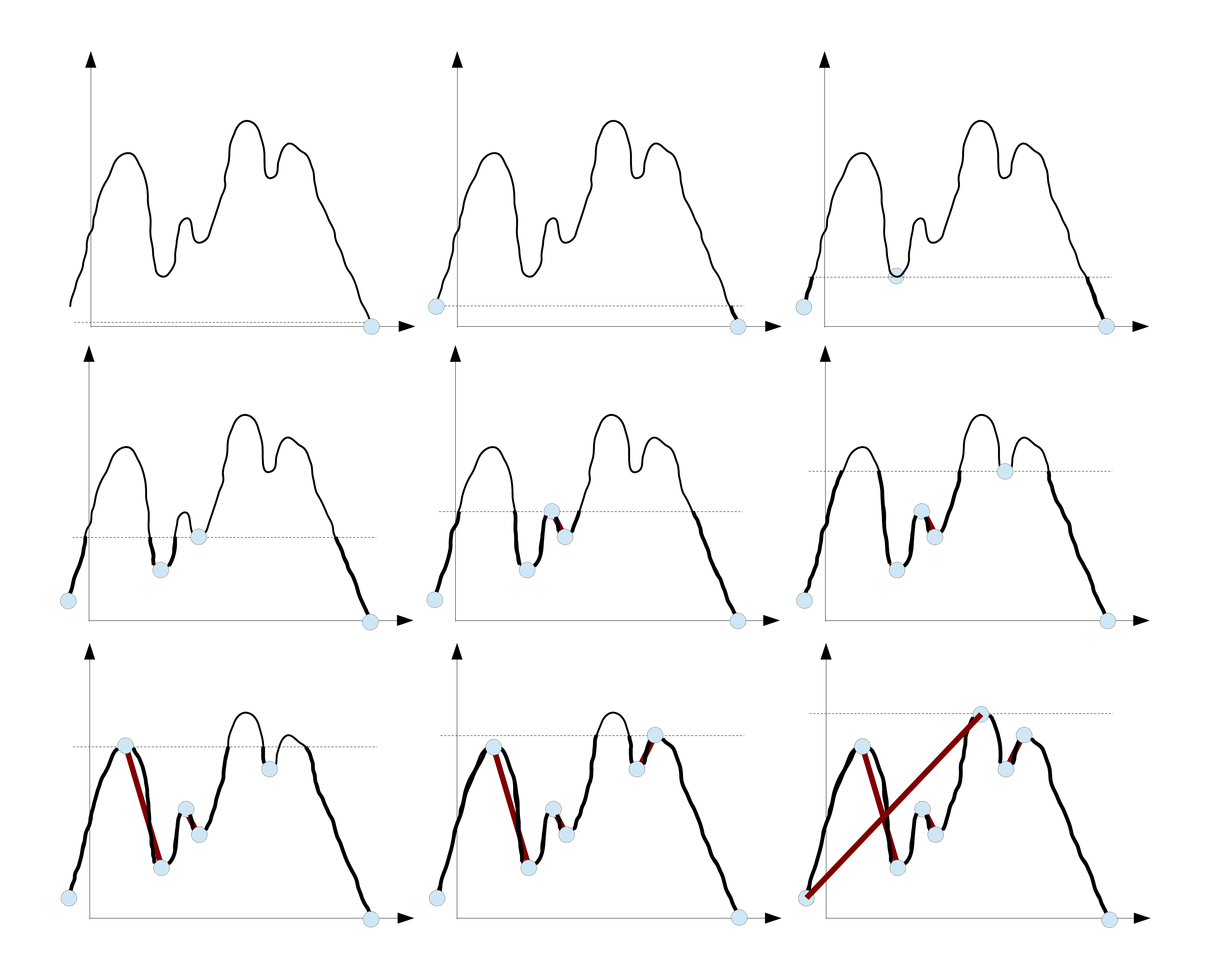}
\caption{The evolution of connected components. Explicit matchings between minima and corresponding saddles are given.}
\label{fig:persistence2}
\end{figure}

In the example in Figure~\ref{fig:persistence2} we see an important convention which is often refereed to as the \emph{elder rule}. In that case, when two connected components are merged, then the one which is \emph{youngest} die. The reason for that is because we want to keep track of the topological features that lives the longest. 

What we saw in Figure~\ref{fig:persistence} and Figure~\ref{fig:persistence2} is exactly what persistent homology is about in dimension $0$. Clearly, in the case of presented functions $f: \mathbb{R} \rightarrow \mathbb{R}$ we cannot have more complicated homology than the one in dimension $0$. But, this can certainly happens for functions $f : \mathbb{R}^n \rightarrow \mathbb{R}$. Let us specify that.

Let $\mathcal{K}$ be a simplicial complex. Let us have $f : \mathcal{K} \rightarrow \mathbb{R}$, a filtering function. Let us recall that this implies, that for every $a \in \mathbb{R}$, $f^{-1}(a)$ is a subcomplex of $\mathcal{K}$.

Please note that Vietoris-Rips and Cech complexes come naturally with a filtration on them which comes from a changing parameter.

In this case a filtration of a complex is a nested sequence of subcomplexes:
\[
\emptyset = \mathcal{K}_0 \subset \mathcal{K}_1 \subset \ldots \subset \mathcal{K}_n = \mathcal{K}
\]
Given a filtration, we can define a filtering function, by setting $f(a) = min_{i\in \{0,\ldots,n\}} a \in \mathcal{K}_i$. A filtering function on a finite complex also defines a filtration of this complex in a natural way.

Given the sequence of complexes one included into the following one, we have a natural sequence of homomorphisms in homology:
\[
0 = H_p(\mathcal{K}_0) \rightarrow H_p(\mathcal{K}_1) \rightarrow \ldots \rightarrow H_p(\mathcal{K}_n) = H_p(\mathcal{K})
\]

\begin{figure}[h!tb]
\centering
\includegraphics[scale=0.45]{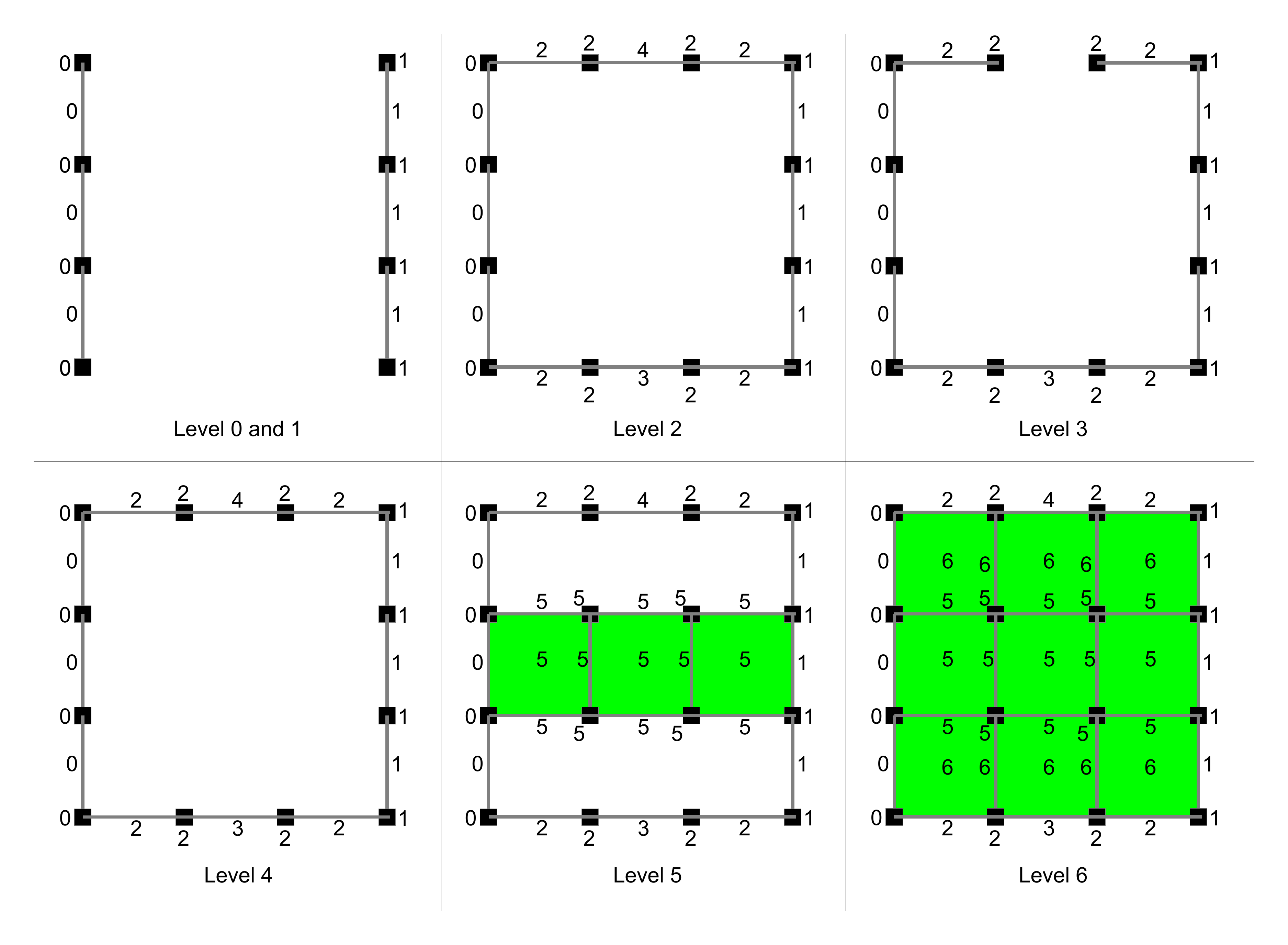}
\caption{The evolution of connected components. In the first picture (level 0 and 1) we have two connected components being born. First in level $0$, second one in level $1$. In he level 2 there is no change in topology. In level $3$ one of the connected component dies. In that case, due to the elder rule, the connected component which was born in the time $1$ dies. Its lifespan is therefore $[1,3]$. In level $4$ the number of connected components remains unchanged, but we have a new $1-$dimensional cycle is created. In level $5$ another $1-$dimensional cycle is created. In level $6$ both one dimensional cycles are killed, and their life span is $[4,6]$ and $[5,6]$.}
\label{fig:persistence3}
\end{figure}

At each step, when we go from $\mathcal{K}_i$ into $\mathcal{K}_{i+1}$, some of the homology classes may become trivial, and some new ones may appear. It turns out that, given such a sequence, we can choose a basis which is compatible for all the steps of the filtration. An algorithm to compute persistent homology is a constructive proof that such a basis can be picked up. Given such a basis we can track the lifespan of a homology classes. When we first see a class, we say that the class is \emph{born}. When the class became trivial, or identical to another class born earlier, we say that the class \emph{dies}. Let us have a look at the example from Figure~\ref{fig:persistence3}.

Given this intuition we are now able to provide a formal definition of persistence. 

\begin{definition}
Let $f_p^{i,j} : H_p(\mathcal{K}_i) \rightarrow H_p(\mathcal{K}_j)$ be a homomorphism induced by inclusion. The p-th persistent homology groups are the images of $f_p^{i,j}$ for $0\ \leq i \leq j \leq n$.
\end{definition}

\begin{figure}[h!tb]
\centering
\includegraphics[scale=0.5]{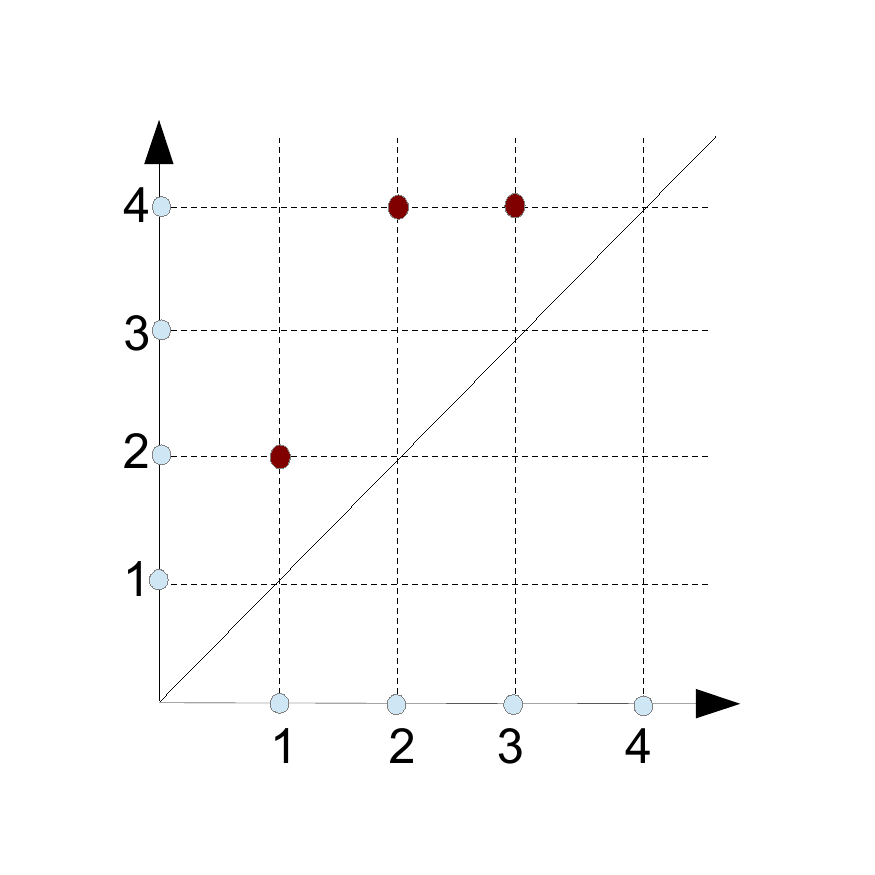}
\caption{Persistence diagram of intervals $[1,2]$, $[2,4]$ and $[3,4]$.}
\label{fig:persistencediagram}
\end{figure}

\begin{figure}[h!tb]
\centering
\includegraphics[scale=0.5]{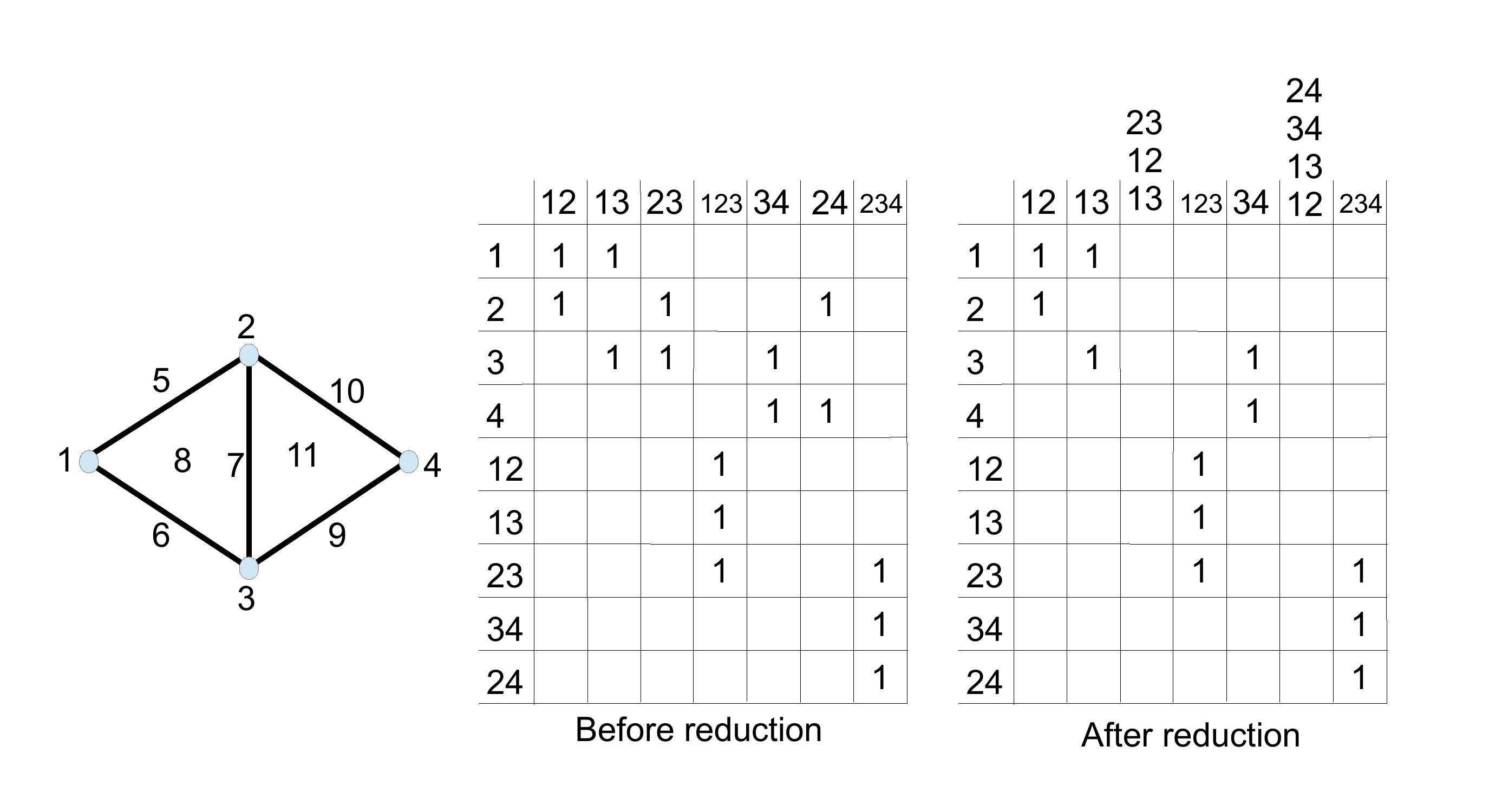}
\caption{Example of the matrix reduction algorithm used to obtain persistence. 
The filtration of vertices of the considered complex is $1$, $2$, $3$ and $4$. We use this filtration to name the edges and vertices. Numbers by edges and vertices denote their filtration. 
The first conflict appears on column $23$. It is resolved by adding column $12$ and $13$ to column $23$. In this way, an empty column is created. It corresponds to a cycle in the subcomplex. The next conflict appears in column $24$ and it is resolved by adding columns $13$, $34$ and $12$ to column $24$. On the right, the reduced matrix is presented. Let us go through it from left to right to read off the persistence. Note that not all columns are presented there (for sake of space we have dropped the zero columns representing the boundary of vertices). Column $12$ has the lowest one $2$. That means, that whatever was created in the vertex $2$ is killed by the edge $12$. That pairing gives rise to the interval $[2,5]$. Analogously column $13$ gives rise to the interval $[3,6]$. Column $23$ is empty. It means, that a $1-$dimensional cycle is created by adding $23$. Column $123$ kills the cycle created in column $23$ and gives rise to the interval $[7,8]$. Column $34$ gives rise to the interval $4,9$ in dimension $0$. Column $24$ is empty and therefore gives rise to the cycle supported in $12+13+34+24$. Column $234$ kills that cycle and give rise to the interval $[10,11]$ in dimension $1$.}
\label{fig:persistenceReductionExample}
\end{figure}

A \emph{lifespan} of a homology class is encoded by a persistence interval $[b,d]$, such that $b < d$. The collection of all persistence intervals is usually encoded as so called \emph{persistence diagram}. It is a finite set of points. For instance, for the intervals $[1,2]$, $[2,4]$ and $[3,4]$ the corresponding diagram is depicted in Figure~\ref{fig:persistencediagram}.
Let us discuss now the algorithms to compute persistent homology. It turns out that it is analogous to the algorithm to compute standard homology we already know. Let us consider the example presented in Figure~\ref{fig:persistenceReductionExample} for further details. 
Note that pairings, which are topological invariants, are directly implied by lowest ones. It requires a proof to show that those are canonical. For that one, consult the book by Edelsbrunner and Harer, page 154.
\section{Comments}
\begin{enumerate}
\item By using ideas from rigorous numerics one can compute persistence not only of a finite complex, but also for a continuous function defined on a compact subset of $\mathbb{R}^n$.
\end{enumerate}

\section{Standard metrics and stability results.}

When analysing noisy data we always need to make sure that the result for noise-less data is not far from the results obtained for noisy data, i.e. that the method is stable. One of main reasons why persistent homology is useful in applied science is precisely this property: persistence is stable, it is robust.
Roughly it means that small change in the filtering function induce small changes in the persistence diagram. In order to make this statement precise, we need to introduce a metric in a space of functions, and more importantly, in the space of persistence diagrams.
In the space of functions we will use standard $L^{\infty}$ metric. Let us define now a metric in a space of persistent diagrams.

Recall that a persistence diagram is a finite multiset of points in the plane $\mathbb{R}^2$ (with points at infinity). For convince, for this finite multiset of points we are adding the points in the diagonal, each with infinite multiplicity. They are needed for the technical reasons. They represent points that are born and die in the same time. Think of them as virtual pairs of particle and anti-particle. You can get them by infinitesimally small perturbation of the filtration. 

Let $X$ and $Y$ be two such persistence diagrams. Let us consider a bijections $b : X \rightarrow Y$. The \emph{Bottleneck} distance is defined in the following way:
\[
W_{\infty}(X,Y) = inf_{b : X \rightarrow Y} sup_{x \in X} || x - b(x) ||
\] 
In other words, we consider all possible bijections between (infinite) multisets of points $X$ and $Y$ and we are searching for the one that minimize the maximal distance between a point $x$ and a point $b(x)$ matched with $x$ via this bijection. This concept is closely related with the earth mover distance. It only cares about the maximal distance. There is another distance which is sensitive to all changes. It is a q-Wasserstein distance and it is defined in the following way:
\[
W_q(X,Y) = (  inf_{b : X \rightarrow Y}  \sum_{x \in X} || x - b(x) ||^q  )^{\frac{1}{q}}
\]

\begin{figure}[h!tb]
\centering
\includegraphics[scale=0.5]{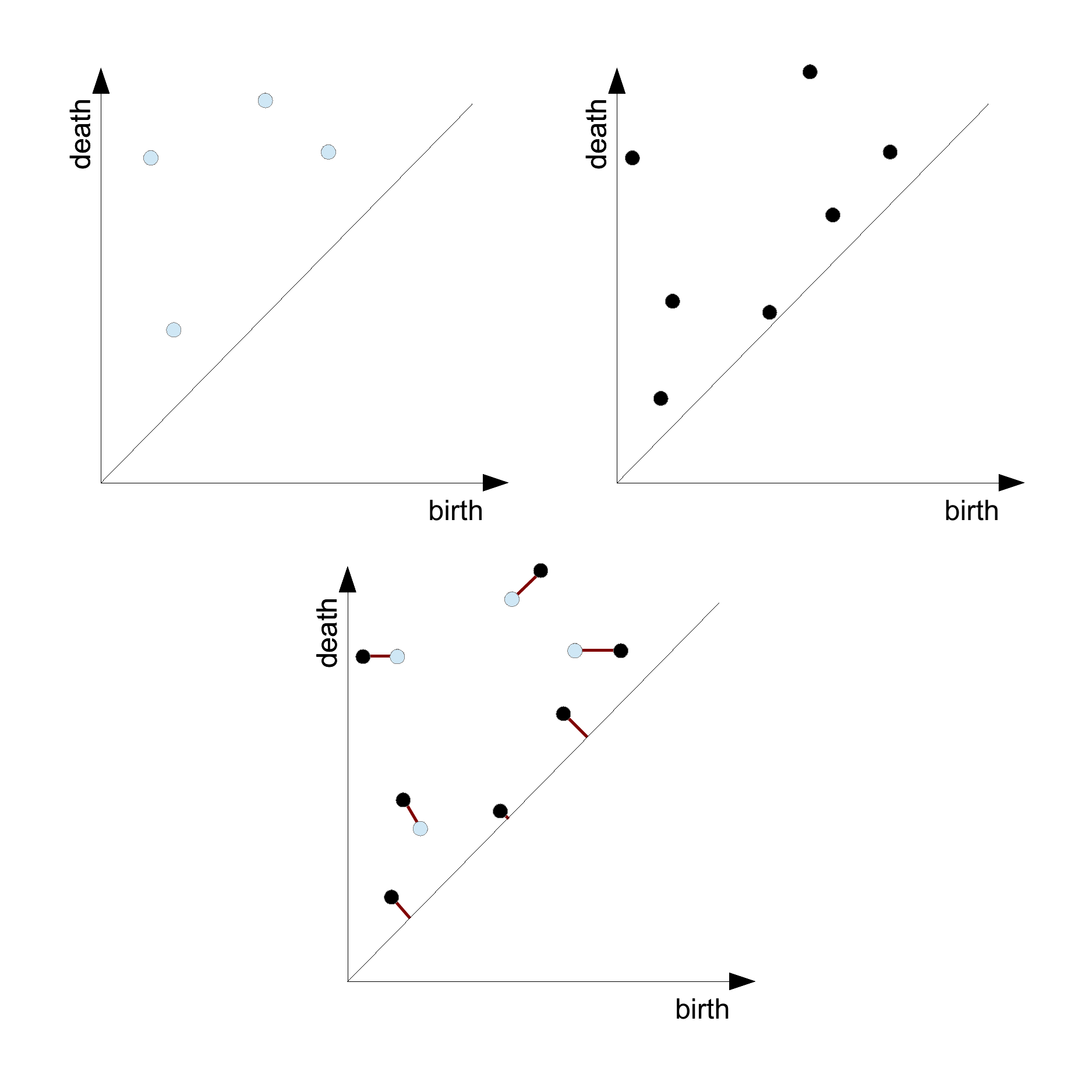}
\caption{Example of a matching between persistence diagrams. For illustration we put the persistence diagram from the top left and the top right into one diagram in the bottom, and we make a matching therein.}
\label{fig:bottelneckDistance}
\end{figure}

Please consult Figure~\ref{fig:bottelneckDistance} for an example of a matching. 

In this section we will provide a stability theorem for the Bottleneck distance. There exists stability theorems for the Wasserstein distances, but they are conceptually more complicated. This stability theorem is due to David Cohen-Steiner, Herbert Edelsbrunner and John Harer~\cite{stability_original}. There are more general stability theorems that work using very mild condition~\cite{stability_general} - 
for a case of Rips complexes with similarity measure.

\begin{theorem}
Let $\mathcal{K}$ be a cell complex and $f,g : \mathcal{K} \rightarrow \mathbb{R}$ be two filtering functions. For each dimension $p$ the bottleneck distance between the diagram of $\mathcal{K}$ with a function $f$ (denoted as $Diag(\mathcal{K},f)$) and the diagram of $\mathcal{K}$ with a function $g$ (denoted as $Diag(\mathcal{K},g)$) satisfies:
\[
W_{\infty}( Diag(\mathcal{K},f) , Diag(\mathcal{K},g) ) \leq || f-g ||_{\infty}
\]
\end{theorem}

In other words, when perturbing the filtering function by $\epsilon$, the points in the diagram will not move further away than $\epsilon$. This is a property which makes persistence useful in applied science. If time permits we will show how it allows us to compute persistence of a continuous function.

Let us put this theorem to the test! We already know how to generate triangulations with Gudhi. Let us generate one, and add a filtration to it:

For the purpose of this exercise we will pick one of the triangulations from the Section
%~\ref{sec:complexes} 
in which we have discussed simplicial complexes and assign a filtration (composed for instance from an increasing sequence of integers) to it:
\begin{lstlisting}
import numpy as np 
import gudhi as gd
import matplotlib
#Create a simplex tree 
#Simplices can be inserted one by one 
#Vertices are indexed by integers 
#Notice that inserting an edge automatically insert its vertices 
(if they were #not already in the complex) 
st = gd.SimplexTree() 
st.insert([1,4,8],0)
st.insert([1,2,8],1) 
st.insert([2,6,8],2) 
st.insert([2,3,6],3)  
st.insert([3,4,6],4)  
st.insert([1,3,4],5)  
st.insert([4,5,9],6)  
st.insert([4,8,9],7)  
st.insert([7,8,9],8)  
st.insert([6,7,8],9)  
st.insert([5,6,7],10)
st.insert([4,5,6],11)    
st.insert([1,2,5],12)  
st.insert([2,5,9],13)  
st.insert([2,3,9],14)  
st.insert([3,7,9],15)  
st.insert([1,3,7],16)  
st.insert([1,5,7],17)  

diagram = st.persistence(persistence_dim_max=True)

plt = gd.plot_persistence_diagram(diagram)
plt.show()
\end{lstlisting}

The numbers after comma (ranging from $0$ to $17$) indicate the filtration value of top dimensional simplices. This is propagated to lower dimensional simplices via so called lower star filtration (i.e. a lower dimensional simplex get a filtration value equal to minimum of filtration values of adjusted top dimensional simplices). 

Now, to illustrate stability theorems, let us perturb the filtration and visualize it, and compute the Bottleneck distances between diagrams. Those bottleneck distances will be bounded by the magnitude of noise we add to the filtration. Below we give some hint on how to use a random number generator to perturb the filtration. 

\begin{lstlisting}
import random as rd
rd.seed()
rd.random() 
\end{lstlisting}

Also remember that in order to compute the Bottleneck distance between two diagrams use:
\begin{lstlisting}
gd.bottleneck_distance(diagram,diagram1)
\end{lstlisting}

Please do it on your own! If there is any problem, you can find the whole code below (with the white ink, sorry... :) You can also find it in the attached directory tree. 
{\color{white}
\begin{lstlisting}
import random as rd
import numpy as np 
import gudhi as gd
#Create a simplex tree 
#Simplices can be inserted one by one 
#Vertices are indexed by integers 
#Notice that inserting an edge automatically insert its vertices (if they were #not already in the complex) 
st1 = gd.SimplexTree() 
st1.insert([1,4,8],0+rd.random())
st1.insert([1,2,8],1+rd.random()) 
st1.insert([2,6,8],2+rd.random()) 
st1.insert([2,3,6],3+rd.random())  
st1.insert([3,4,6],4+rd.random())  
st1.insert([1,3,4],5+rd.random())  
st1.insert([4,5,9],6+rd.random())  
st1.insert([4,8,9],7+rd.random())  
st1.insert([7,8,9],8+rd.random())  
st1.insert([6,7,8],9+rd.random())  
st1.insert([5,6,7],10+rd.random())
st1.insert([4,5,6],11+rd.random())    
st1.insert([1,2,5],12+rd.random())  
st1.insert([2,5,9],13+rd.random())  
st1.insert([2,3,9],14+rd.random())  
st1.insert([3,7,9],15+rd.random())  
st1.insert([1,3,7],16+rd.random())  
st1.insert([1,5,7],17+rd.random())  

diagram1 = st1.persistence(persistence_dim_max=True)

gd.plot_persistence_diagram(diagram1)
plt.show()

diagram = st.persistence_intervals_in_dimension(1)
diagram1 = st1.persistence_intervals_in_dimension(1)
gd.bottleneck_distance(diagram,diagram1)
\end{lstlisting} 
}

\chapter{Experiments with Persistent homology}

In this section we are considering a number of exercises involving persistent homology computations. Please go through them and let me know if you have any questions or problems. 

\section{Open and closed proteins}
This example was entirely designed by Bertrand Michel and is available at his tutorial in here: \url{http://bertrand.michel.perso.math.cnrs.fr/Enseignements/TDA/Tuto-Part2.html}. It is based heavily on the work of Peter Bubenik. Big thanks to both of them. 

In ths example we wil work on the data also considered in the paper \url
{https://arxiv.org/pdf/1412.1394}

The data required for this exercise is available here: \url{https://www.researchgate.net/publication/301543862_corr}

The purpose of this exercise is to understand if persistent homology information can determine the folding state of a protein. Please load the data, unpack them and put them all to the \emph{data} folder. 
For further details, please consult the paper. Here are the steps we are gong to perform in the calculations:
\begin{enumerate}
\item Read the appropriate correlation matrices from csv files (we will use Pandas for that). 
\item Transform a correlation matrix to a pseudo--distance matrix by transforming each entry $a_{ij}$ into $1-|a_{ij}|$.
\item Compute the Rips complexes of the obtained distance matrices (note that in this case, we are constructing a Rips complex of a pseudo--metric space which is not a Euclidean space. Such situations are very typical in this type of bio--oriented research). 
\item Compute persistent homology of the obtained complexes. 
\item Compute all-to-all distance matrix between the obtained persistence diagrams (do it dimension--by--dimension). For that purpose we will use the Bottleneck distance implemented in Gudhi.
\item Use a Multidimensional Scaling method implemented in the scikit-learn library to find the two dimensional projection of the data. 
\item As you can see, in this case we get sort of a separation, but not a linear one. For further studies of what can be done with those methods, please consult the original paper \url{https://arxiv.org/pdf/1412.1394}.
\end{enumerate}

\begin{lstlisting}
import numpy as np
import pandas as pd
import pickle as pickle
import gudhi as gd
from pylab import *
from mpl_toolkits.mplot3d import Axes3D
from IPython.display import Image
from sklearn import manifold
import seaborn as sns

#Here is the list of files to import
files_list = [
'data/1anf.corr_1.txt',
'data/1ez9.corr_1.txt',
'data/1fqa.corr_2.txt',
'data/1fqb.corr_3.txt',
'data/1fqc.corr_2.txt',
'data/1fqd.corr_3.txt',
'data/1jw4.corr_4.txt',
'data/1jw5.corr_5.txt',
'data/1lls.corr_6.txt',
'data/1mpd.corr_4.txt',
'data/1omp.corr_7.txt',
'data/3hpi.corr_5.txt',
'data/3mbp.corr_6.txt',
'data/4mbp.corr_7.txt']
#Read the files:
corr_list = [pd.read_csv(u , header=None,delim_whitespace=True) 
for u in files_list]
#And change correlation matrix do a distance matrix:
dist_list = [1- np.abs(c) for c in corr_list]

#Compute persistence in dimension 1 for all the files. 
#Visualize the persistence diagrams for some of them
persistence = []
for i in range(0,len(dist_list)):
	rips_complex = gd.RipsComplex(distance_matrix=dist_list[i].values,max_edge_length=0.8)                  
	simplex_tree = rips_complex.create_simplex_tree(max_dimension=2)
	simplex_tree.persistence()
	persistence.append( simplex_tree.persistence_intervals_in_dimension(0) )
	
#And compute all-to-all bottleneck distances. Note that this part 
#will take a few seconds:
dist_mat = []
for i in range(0,len(persistence)):
	row = []
	for j in range(0,len(persistence)):
		row.append( gd.bottleneck_distance(persistence[i], persistence[j]) )	
	dist_mat.append(row)	
                           
#We will now use a dimension reduction method to 
#visualize a configuration in  R^2  which almost 
#matches with the matrix of bottleneck distances. 
#For that purpose we will apply a Multidimensional 
#Scaling method implemented in the scikit-learn library.

mds = manifold.MDS(n_components=2, max_iter=3000, eps=1e-9,dissimilarity="precomputed", n_jobs=1)
pos = mds.fit(dist_mat).embedding_      

plt.scatter(pos[0:7,0], pos[0:7, 1], color='red', label="closed")
plt.scatter(pos[7:len(dist_mat),0], pos[7:len(dist_mat), 1], color='blue', label="open")
plt.legend( loc=2, borderaxespad=1)
plt.show()

#repeat this for diagram in dimension 0.                         
\end{lstlisting}
%mat_dist = dist_list[0]
%mat_dist.head()
%rips_complex_ref = gd.RipsComplex(distance_matrix=mat_dist.values,max_edge_length=0.8) 
%simplex_tree_ref = rips_complex_ref.create_simplex_tree(max_dimension=2)
%pers = simplex_tree_ref.persistence()

In one of the next sections we will be discussing persistence representations, and one example we will consider there uses permutation test. Feel free to return to this example and use a permutation test to distinguish two types of proteins.

\section{Classification of smart phone's sensor data} 
In this exercise we will gather accelerometer data of various activities using our mobile phones. For that we need an app to record data on our phones. The one I am using is an android app ''Physical Toolbox". You can find it on a play store \url{https://play.google.com/store/apps/details?id=com.chrystianvieyra.physicstoolboxsuite&hl=en_US}. 

Due to the hardware restrictions I was not able to test in practice, the i-phone analogue, but I think that one of those may work for you if you are an apple user:
\begin{enumerate}
\item VibSensor,\\
 \url{https://itunes.apple.com/us/app/vibsensor-accelerometer-recorder-vibration-analysis/id932854520?mt=8}

\item Physical toolbox -- presumably restricted version of the android app we have discussed above\\ \url{https://itunes.apple.com/us/app/physics-toolbox-sensor-suite/id1128914250?mt=8}\\ and\\ \url{https://itunes.apple.com/us/app/physics-toolbox-accelerometer/id1008160133?mt=8}

\item Accelerometer app\\ \url{https://itunes.apple.com/us/app/accelerometer/id499629589?mt=8}
\end{enumerate}

With this, we need a linear accelerometer, and start to recording at the right moment. The aim is to record various activities, like walking on stairs, normal walking, push ups, jogging, and whatever else you want. Remember to stop recording as soon as you stop the activity. Given the data, the app will store it in a csv file, which we will use for a later analysis.

%\begin{lstlisting}
%import numpy as np
%import pandas as pd
%import pickle as pickle
%import gudhi as gd
%from pylab import *
%import seaborn as sns
%from mpl_toolkits.mplot3d import Axes3D
%from IPython.display import Image
%from sklearn.neighbors.kde import KernelDensity
%import matplotlib
%
%f = open("data_acc","rb")
%data = pickle.load(f)
%f.close()
%
%data_A = data[0]
%data_B = data[1] 
%data_C = data[2]
%label = data[3]
%print(label)
%
%print(np.shape(data_A))
%
%fig = plt.figure()
%ax = fig.add_subplot(111, projection='3d')	
%plt.plot(data_A_sample [:,0],data_A_sample [:,1],data_A_sample [:,2] )
%ax.scatter(data_A_sample [:,0],data_A_sample [:,1],data_A_sample [:,2] )
%plt.show()
%
%Rips_complex_sample = gd.RipsComplex(points = data_A_sample,max_edge_length=0.8 )
%Rips_simplex_tree_sample = Rips_complex_sample.create_simplex_tree(max_dimension=3) 
%
%diag_Rips = Rips_simplex_tree_sample.persistence()
%diag_Rips
%
%diag_Rips_0 = Rips_simplex_tree_sample.persistence_intervals_in_dimension(0)
%
%gd.plot_persistence_diagram(diag_Rips)
%
%\end{lstlisting}
%
%Now we will do the same... but differently! 
Please grab your phones and record accelerometer data when different activities are performed. At the end, export your file, name it with the name of activity, and store it. You shall get file the header of which look like this:
\begin{center}
\includegraphics[scale=0.6]{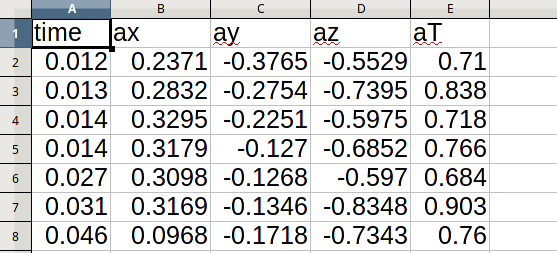}
\end{center}
Remove the first row and the first and last column, and read it. 
\begin{lstlisting}
import numpy as np
import pandas as pd
import pickle as pickle
import gudhi as gd
from pylab import *
import seaborn as sns
from mpl_toolkits.mplot3d import Axes3D
from IPython.display import Image
from sklearn.neighbors.kde import KernelDensity
import matplotlib


data_pd = pd.read_csv('activities/walk.csv',decimal=".",delimiter=",")
data = data_pd.values

data = np.delete(data, (0), axis=1)
data = np.delete(data, (3), axis=1)

#now we can visualize the data:
fig = plt.figure()
ax = fig.add_subplot(111, projection='3d')	
#plt.plot(data [:,0],data [:,1],data [:,2] )
ax.scatter(data[:,0],data[:,1],data[:,2] )
plt.show()

#And now we can compute persistent homology of the data and compare 
#(visually at the moment) different activities. 


Rips_complex_sample = gd.RipsComplex(points = data,max_edge_length=0.6 )
Rips_simplex_tree_sample = Rips_complex_sample.create_simplex_tree(max_dimension=2) 

diag_Rips = Rips_simplex_tree_sample.persistence()
#diag_Rips
#diag_Rips_0 = Rips_simplex_tree_sample.persistence_intervals_in_dimension(0)
plt = gd.plot_persistence_diagram(diag_Rips)
plt.show()
\end{lstlisting}

Note that depending on if you have python 2 or 3, you may get some problem with reading the csv file. This is something we need to work out when doing this exercise.

Check if persistent homology give you a way to cluster the activities. 

Here are a few examples of my activities, their point clouds, and the persistence diagrams obtained for the parameters of the Rips construction as in the example above. 

\begin{tabular}{|c|c|c|}
\hline
 walk & \includegraphics[scale=0.2]{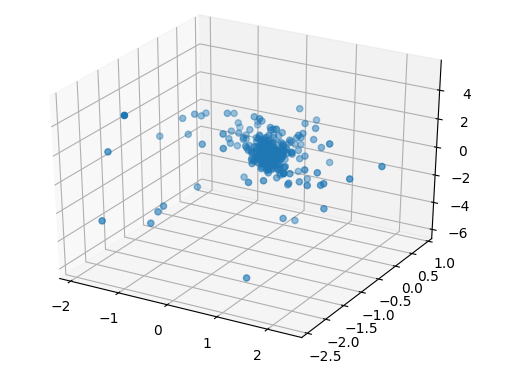} & \includegraphics[scale=0.2]{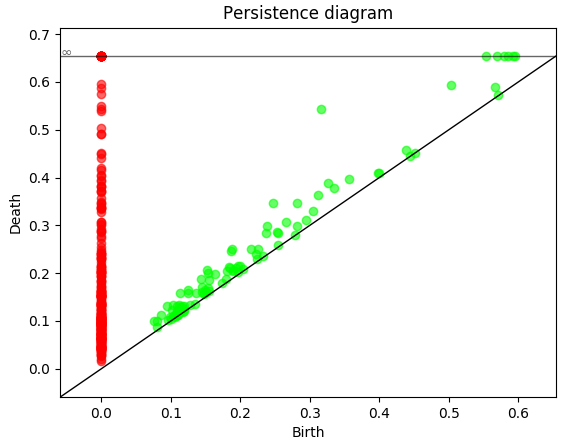}\\
\hline
squats & \includegraphics[scale=0.2]{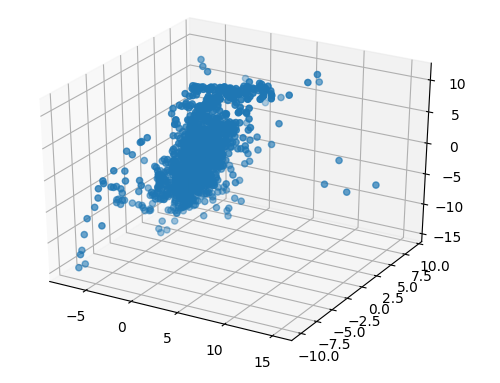} & \includegraphics[scale=0.2]{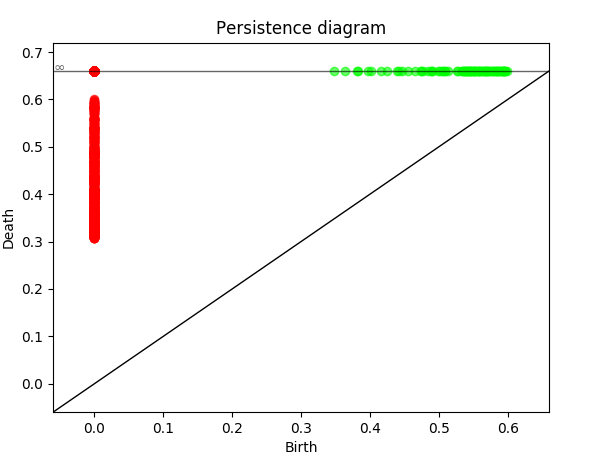}\\
\hline
jumping & \includegraphics[scale=0.2]{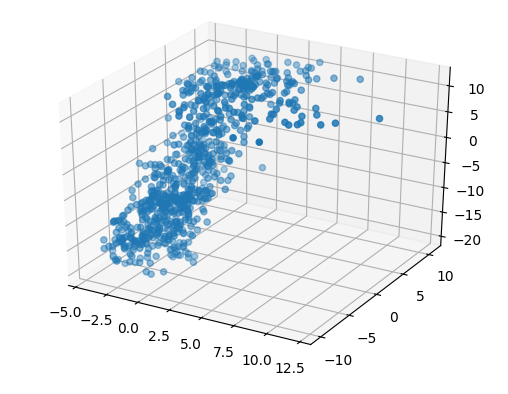} & \includegraphics[scale=0.2]{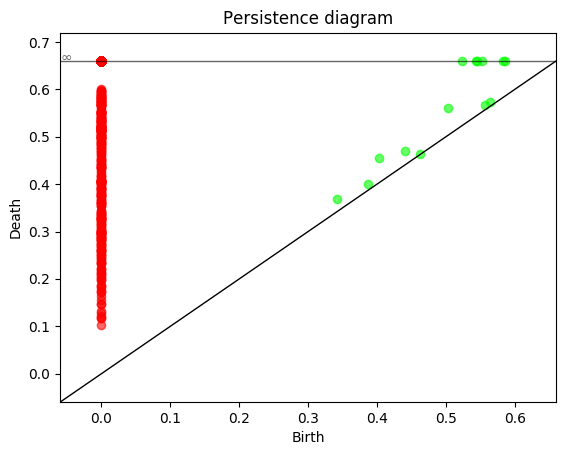}\\  
\hline
downstairs & \includegraphics[scale=0.2]{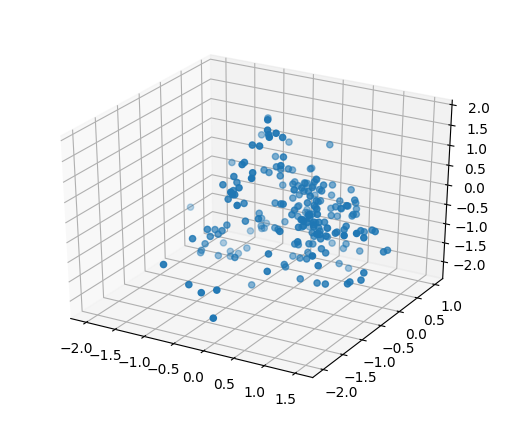} & \includegraphics[scale=0.2]{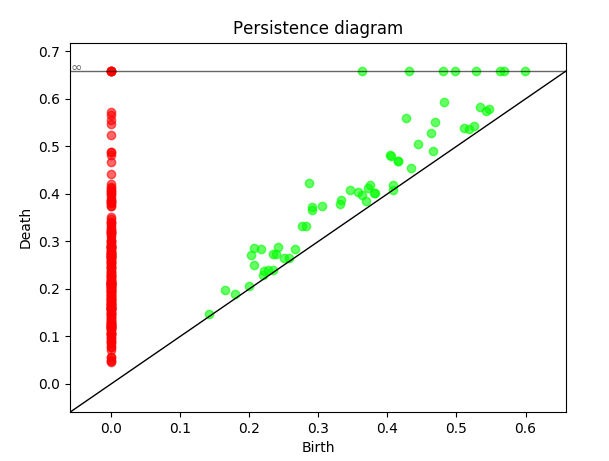}\\  
\hline          
upstairs & \includegraphics[scale=0.2]{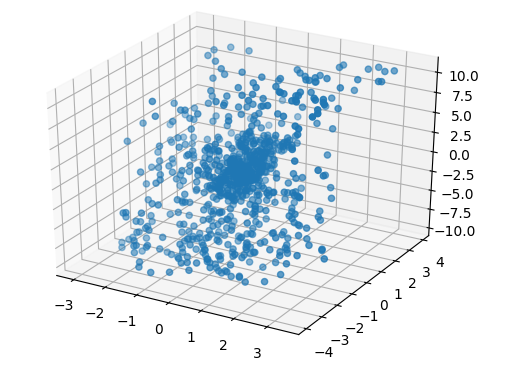} & \includegraphics[scale=0.2]{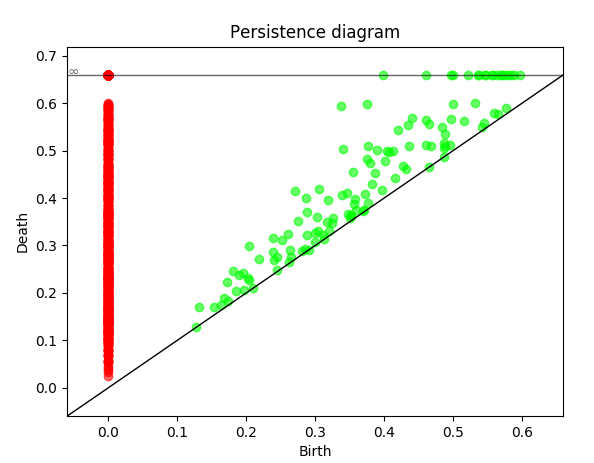}\\                
\hline
\end{tabular}

Please play with different sensors. You can get really nice patterns from theq magnetometer. Run it, rotate your phone, and see what you get!

%\begin{tabular}{|c|c|c|}
%\hline
% walk & \includegraphics[scale=0.2]{walk.png} & \includegraphics[scale=0.2]{walk_diag.png}\\
%\hline
%squats & \includegraphics[scale=0.2]{squats.png} & \includegraphics[scale=0.2]{squats_diag.png}\\
%\hline
%jumping & \includegraphics[scale=0.2]{jumping.png} & \includegraphics[scale=0.2]{jumping_diag.png}\\  
%\hline
%downstairs & \includegraphics[scale=0.2]{downstairs.png} & \includegraphics[scale=0.2]{downstairs_diag.png}\\  
%\hline          
%upstairs & \includegraphics[scale=0.2]{upstairs.png} & \includegraphics[scale=0.2]{upstairs_diag.png}\\                
%\hline
%\end{tabular}

Check it up for your case! Think of the characteristics of persistence diagrams that discriminate those activities. Think of alternative (to persistent homology) characteristics of point clouds that could differentiate the type of activity. Think about it next time you give an app permission to access your phone sensor. And share your ideas and observations!

\section{Back to the point clouds}
%In the Section~\ref{sec:complexes_based_on_point_clouds} 
So far we have generated a number of point clouds. Computing their persistent homology is the first task we can do in this section. In fact, computations of persistent homology is an optional part in all the codes available 
%in the Section~\ref{sec:complexes_based_on_point_clouds}. 
therein. The point that is missing is to visualize persistence diagrams. This can be achieved using the following code:

\begin{lstlisting}
import matplotlib
#suppose that over here the simplex tree has been generated.
plt = gd.plot_persistence_barcode(pers)
plt.show()
\end{lstlisting}
for persistence barcodes or

\begin{lstlisting}
import matplotlib
#suppose that over here the simplex tree has been generated.
plt = gd.plot_persistence_diagram(pers)
plt.show()
\end{lstlisting}
for persistence diagramss

Please carry on the experimentation with all the point clouds and check if the results are as predicted. 

Try to do the sampling a few times. Plot the persistence diagrams. Are they similar? Verify this by computing the distances between them. For that purpose you can use either python-based Wasserstein distance, or C++ based persistence representations. 
If you take a reasonable number of points sampled from the set, you will see almost exactly the same persistence diagram. There is in fact a \emph{stability result} saying that if the so called \emph{Hausdorff distnace} between point clouds	 is bounded by $\epsilon$, then the persistence diagrams of that point cloud are bounded by the same $\epsilon$,	 subject to a few mild assumptions. 

\section{Random cubical complexes and generalized percolation}
In this exercise we will create a number of cubical complexes according to the following model: A top dimensional cube is present in the complex with probability $p \in [0,1]$. Pick a few values of $p$, and generate a few instances of cubical complexes (choose the dimension and size as you wish, but be reasonable as you do not want to blow up your memory). Compute persistent homology of those complexes. What can you observe? Can you track for evidences of phase transitions happening in the system? How this experiment is related to percolation theory.

Use the code below as a starting point in your experiments:

\begin{lstlisting}
import matplotlib.pyplot as plt
import numpy as np
import random as rd
from mpl_toolkits.mplot3d import Axes3D
import csv
from sklearn import decomposition
import math
import gudhi as gd
from PIL import Image

#in this example we are considering two dimensional complexes
N = 100

#this is where you set the probability:
p = 0.6

bitmap = []
for i in range(0,N*N):	
	x = rd.uniform(0, 1)
	if ( x < p ): bitmap.append(1)
	else: bitmap.append(0)
				
bcc = gd.CubicalComplex(top_dimensional_cells = bitmap, dimensions=[N,N])
diag = bcc.persistence()
#now we can check how many generators in dimension 0 and in dimension 1 we have:
dim0 = bcc.persistence_intervals_in_dimension(0)
dim1 = bcc.persistence_intervals_in_dimension(1)

print "Here is the first Betti number: ", len(dim0)
print "Here is the second Betti number: ," len(dim1)
\end{lstlisting}

You can also modify this exercise to create cubical complexes such that cubes get not only filtration $0$ or $1$, but the whole spectrum $[0,1]$. Another interesting experiment aims in picking a grid $0 \leq p_1 < p_2 < \ldots < p_n \leq 1$, computing the Betti numbers for each probability, averaging them for each probability, and plotting the results. 

\section{Ssid walks}
If during your ssid--walk you have made a cycle, you can now go back to the data and check if you can see the dominant interval in one dimensional persistent homology. Observe the persistence of this interval compares to the persistence of other (shorter) intervals that come from the noise.

\chapter{Persistence representations.}
In this section we will take a look at various alternative ways of representing persistence diagrams which is suitable for statistical operations. For many years scientists have tried to provide a sound way of doing statistics on diagrams. The most obvious approach which has been used already many time is to compute some function of a diagram like:
\begin{enumerate}
\item Average length of an interval.
\item Median of interval's length.
\item Average of squares of length.
\item Sum of length's squares.
\item Length of maximal interval.
\item ...
\end{enumerate}
All of this is done either to do a hypothesis testing, or to project the data from the space of persistence diagrams down into $\mathbb{R}$, where averaging is much easier to do. 

\begin{figure}[h!tb]
\centering
\includegraphics[scale=0.8]{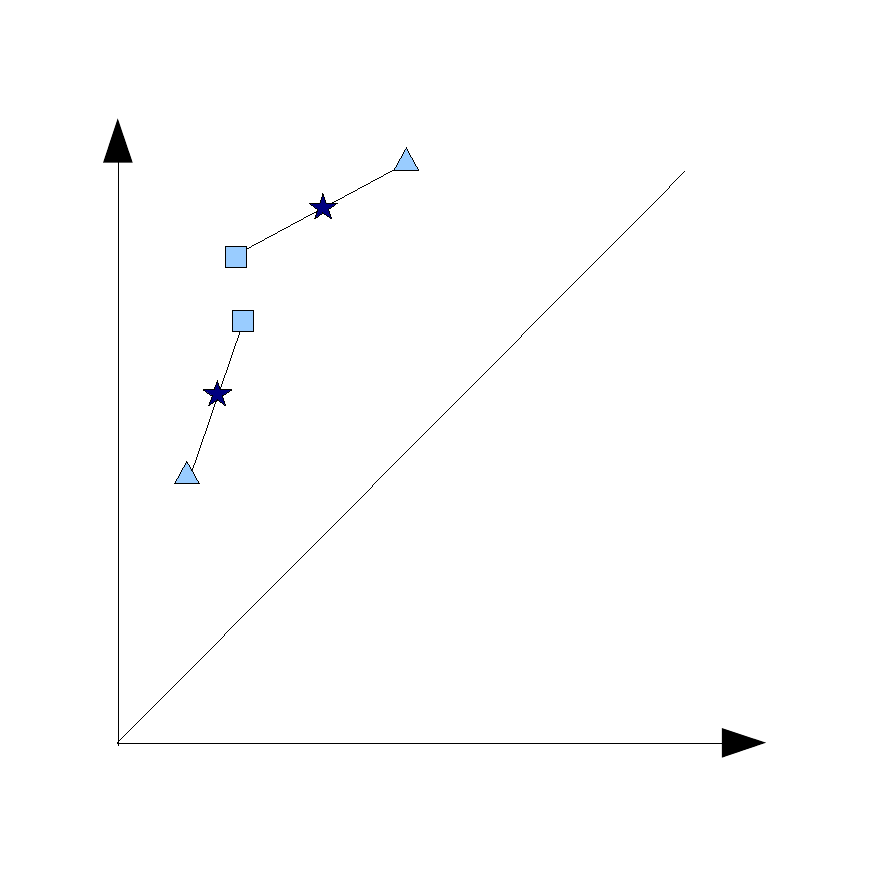}
\caption{Frechet mean of two diagrams. Elements of the diagram $D_1$ are marked with squares. Elements of the diagram $D_2$ are denoted with triangles. Elements of an averaged diagram are marked with stars.}
\label{fig:FrechetMean}
\end{figure}

Let us discuss the problem with averages. Why can one not define an average on diagrams? There exists a concept of a Frechet mean for a metric space. It can be used to define an average of two diagrams in the following way. Take two diagrams $D_1$ and $D_2$. Compute a minimal matching between them. For every $x \in D_1$ matched with $y \in D_2$, take a midpoint of a line segment between $x$ and $y$ as an element of the averaged interval. See Figure~\ref{fig:FrechetMean}

\begin{figure}[h!tb]
\centering
\includegraphics[scale=0.8]{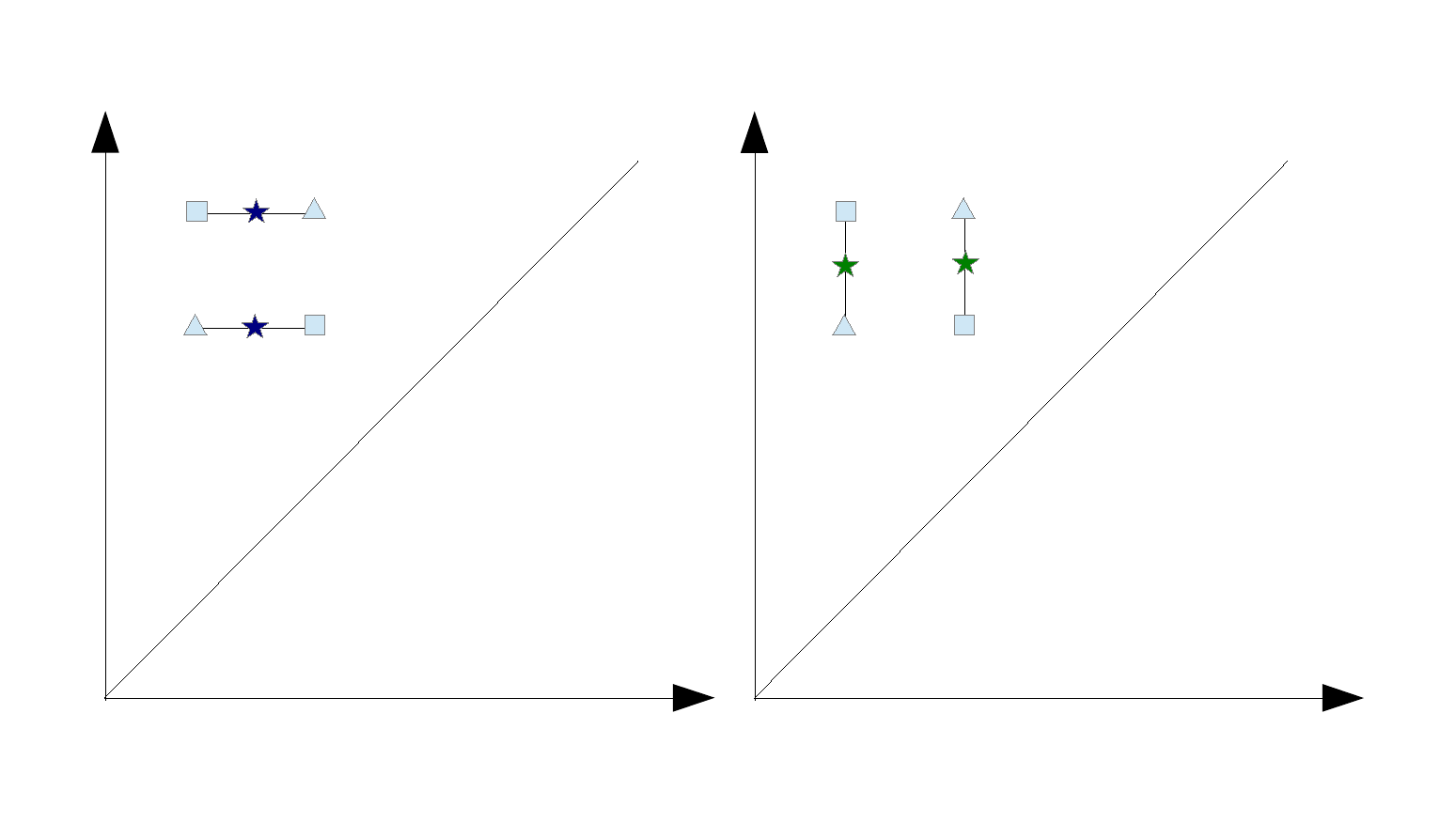}
\caption{Problem with Frechet mean. In this case we have two possible matching between diagrams. Using one of them gives us one possible average. Using another, gives an average which can be arbitrary far from the first one.}
\label{fig:FrechetMeans2}
\end{figure}

But there is a problem with the concept of Frechet mean. There are situations when it is not unique. Consult Figure~\ref{fig:FrechetMeans2} for an example.

I like to think about persistent diagrams as I think of natural numbers. In both cases, there is no clear way to define averages. In the case of natural numbers a solution is to start using rational numbers. What is the corresponding rational number in the case of persistence diagrams? For that aim we construct an embedding into various functional spaces. We will present a few of them. 

\section{Persistence Landscapes}
One option was proposed by Peter Bubenik\cite{landscapes_original} and implemented by me~\cite{landscaspes_algo}. This is the idea of persistence landscapes. 

Let us fix first the multiset of persistence intervals $\{(b_i , d_i)\}_{i=1}^n$. For every interval, let us define a function:
\begin{equation}
 f_{(b,d)} =
  \begin{cases}
   0 & \text{if } x \not \in (b,d) \\
   x - b & \text{if } x \in (b , \frac{b+d}{2}] \\
   -x + d       & \text{if } x \in (\frac{b+d}{2} , d) 
  \end{cases}
\end{equation}

Then a \emph{persistence landscape} is a set of functions $\lambda_k : R \rightarrow R$ such that $\lambda_k(x) =$ k-th largest value of $\{f(b_i,d_i)(x)\}_{i=1}^n$. This may be a little hard to digest at first. Let us take a look at Figure~\ref{fig:persistenceLandscapeIllustration} to get the idea.

\begin{figure}[h!tb]
\centering
\includegraphics[scale=0.7]{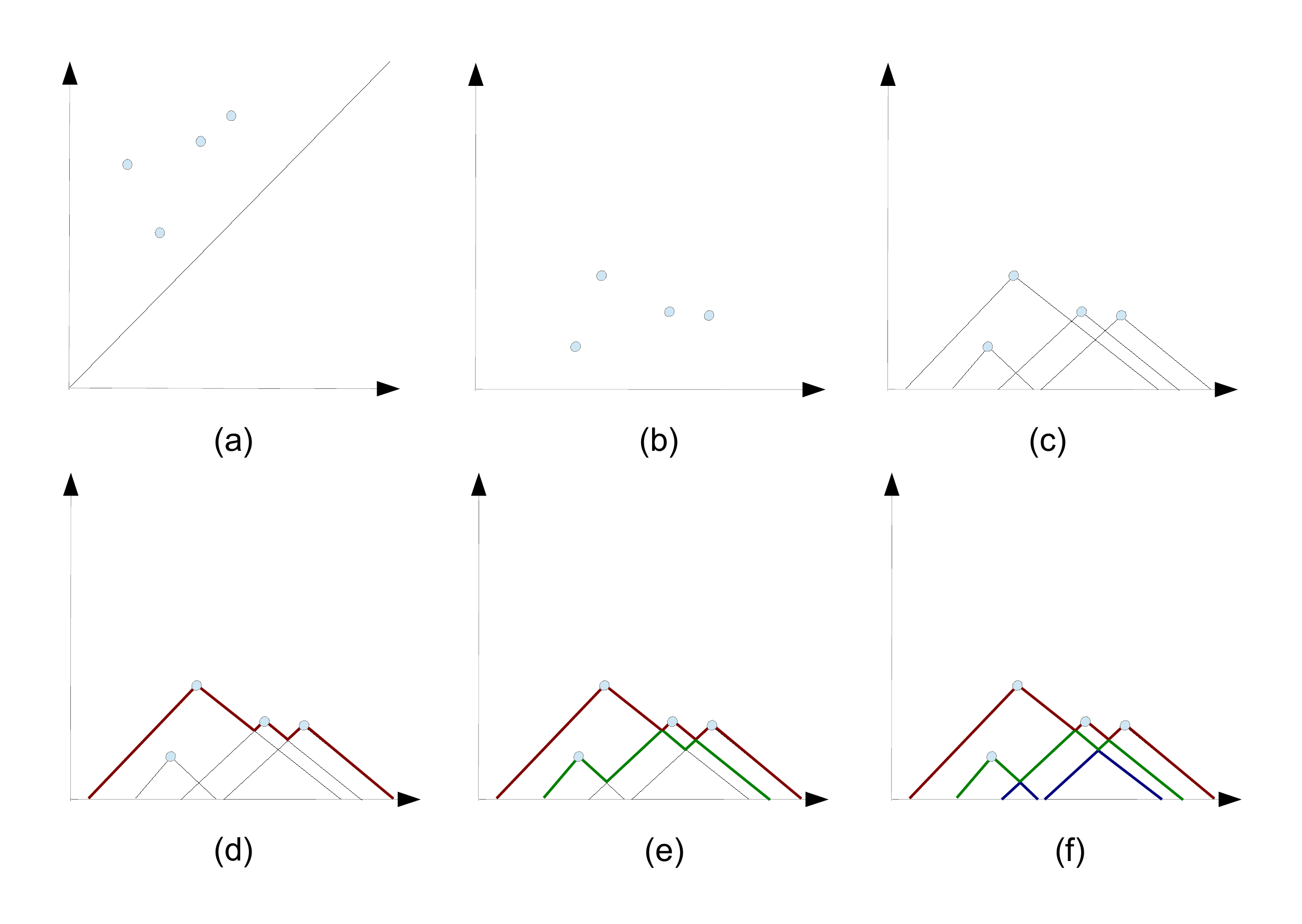}
\caption{An idea of a persistence landscape. First, we ''lie down'' the persistence diagram (b). Formally this corresponds to moving from $(birth,death)$ coordinates into $(middlife, half\ life)$ ones. Then, for the image of every interval $(b,d)$ in the new coordinates, we draw a plot of the function $f_{(b,d)}$ as in (c). Then, the first landscape function, $\lambda_0$ is depicted in Figure (d), the $\lambda_2$ in (e) and $\lambda_3$ in (f).}
\label{fig:persistenceLandscapeIllustration}
\end{figure}

Persistence landscapes are an equivalent way of representing persistence diagrams. There are algorithms to get one representation from anther. Also, given the intuition from Figure~\ref{fig:persistenceLandscapeIllustration}, it is clear that $\lambda_{n+1} \leq \lambda_n$.

Lets talk a little bit more about the idea of this representation. What if, at a given $x \in \mathbb{R}$, we have nonzero landscapes $\lambda_1,\ldots,\lambda_n$ and $\lambda_{n+1} = 0$? Well, it means that at level $x$, our filtered complex has the rank of the homology in the considered dimension equal $n$. Moreover, the minimal amount of perturbation we need to make to kill at least one homology generator is $\lambda_n(x)$.

The obvious question is -- how does introducing persistence landscapes help us to solve the problems we have with defining an average? Well, it does. An average of two functions $\lambda_i : \mathbb{R} \rightarrow \mathbb{R}$ and $\lambda'_i : \mathbb{R} \rightarrow \mathbb{R}$ is simply a point-wise average of them. So, the average is well defined, it is unique. The price to pay is that average of two landscapes is a partially linear function which typically does not come from any persistent diagram. This is something that make some people worry. In my opinion it is a very typical situation that in order to define an average of two nice objects, we need to allow ourselves a little more freedom, and move to a larger space (exactly like in case of integer and rational numbers). Only then one can state a definition that make sense. I think it is the case for persistence landscapes. 

But lifting the persistence diagrams to the $L^p( \mathbb{R} \times \mathbb{N} \rightarrow \mathbb{R} )$ space, for $p \in [1,\infty]$, gives us more. There is a natural distance function in this space: $d(f,g) = (\sum_{i \in \mathbb{N}} \int_{\mathbb{R}}|f_n(x) - g_n(x)|^p)^{\frac{1}{p}}$. Moreover those distances are stable. 

The persistence landscape toolbox and the Gudhi implementation of persistence landscapes have the following functionalities:
\begin{enumerate}
\item Computations of a distance matrix.
\item Various plotting procedures.
\item Computations of averages and standard deviation.
\end{enumerate}

%
%\item \emph{Classifiers} -- Suppose we are given a $n$ different classes $C_1,\ldots,C_n$ of persistence diagrams. Suppose also, that a new data are coming and we want to find a class among $C_1,\ldots,C_n$ that fits the data best. For that purpose, we can use a very simple nearest neighbor classifier implemented in the PLT. The idea is simple. For each class $C_i$, compute its average $\hat{C_i}$. Then, when a new diagram $N$ (or a set of graded persistence diagrams) came in, one simply compute all the distances from $N$ to $\hat{C_1},\ldots,\hat{C_n}$. The result of classification is the $i$, such that $\hat{C_i}$ is closest to $N$.

\section{Persistence Heat Maps}
Many groups at different time have come up with an idea of building a functional descriptor of a persistence diagram by placing a kernel (typically Gaussian kernel) on each point of a diagram. In order to make this construction stable some type of \emph{weighting} is necessary: otherwise on persistence points of very low persistence one should place the same kernel as the points of high persistence. As the first ones can appear and disappear in any number, even when a small amount of noise is introduced, they can change the resulting function by an arbitrary factor. To make this descriptor stable two strategies have been implemented:
\begin{enumerate}
\item Weight the kernel placed on a point $(b,d)$ by the total persistence of the point $b-d$. This strategy has been used in so called Persistence Weighted Gausian Kernel proposed by Yasuaki Hiraoka and collaborators~\cite{pwgk} as well as Persistence Images~\cite{persistence_images}.
\item Weight all the kernels by the same number, but for any kernel placed in the point $(b,d)$ place the symmetric kernel at the point $(d,b)$ with the weight $-1$. This strategy has been proposed in the Persistence Stable Space Kernel~\cite{pssk}.
\end{enumerate}
An example of a kernel with no weight is presented in Figure~\ref{fig:persistenceLandscapeIllustration}.

\begin{figure}[h!tb]
\centering
\includegraphics[scale=0.5]{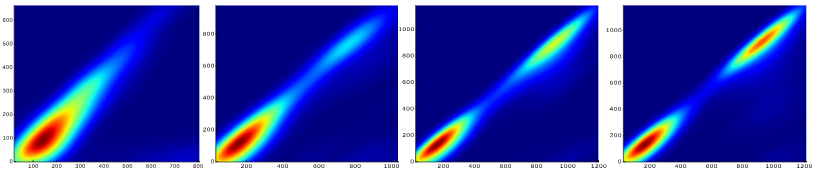}
\caption{Persistence images obtained from diagrams describing neuronal morphologies, see~\cite{topaz} for details.}
\label{fig:persistenceLandscapeIllustration}
\end{figure}

As much as Persistence Landscapes, various persistence heat maps enable all the main operations required in statistics and machine learning. 

\section{Persistence representations in Gudhi}

All those persistence representations and more have been implemented in the package \emph{Persistence Representations} available in Gudhi. At the moment unfortunately they are not available in python, but this will change soon.

In the next section we present an exercise we can do to familiarise ourselves with the concept. Unfortunately we do not yet have the persistence landscapes available in python, so to do that we need to get our hands dirty with C++. We are here to help you with that!

\section{Exercise: detecting the dimension of a random point cloud examination of their persistence diagrams.} 
At the moment the exercise requires working on the C++ level, since some functionalities related to averaging various representations of persistence has not been cytonized yet. For that purpose, please download the Gudhi library from \url{http://gudhi.gforge.inria.fr/}, and subsequently use the attached 

\url{https://www.dropbox.com/s/4js8g9k0evnzd1o/study_dimensions_with_persistence.zip?dl=0}

C++ file which need to be added to the Gudhi library and compiled with it. Please consult the linked file for further technical details. 
For the purpose of the exercise, please generate $N$ random point clouds for two different dimensions $d_1$ and $d_2$. Then generate a random point cloud in $\mathbb{R}^{d_1},\mathbb{R}^{d_2}$.  Using a permutation test, please check if there is a statistical difference between Persistence Landscapes of different dimensions. Any necessary details about this task will be given on demand, so please ask!

\section{Permutation tests}
\emph{Permutation test} -- Suppose we are given two sets $A$, $B$ of persistence diagrams and we would like to show that they are essentially different. For simplicity let us assume that $\card{A} = \card{B}$. If we have some assumptions, we can use a t-test (implemented in the PLT). If not, we can always use the permutation test. It works as follow:
	\begin{enumerate}
	\item[STEP 1:] Compute averages $\hat{A} = $ average of landscapes constructed based on the diagrams from set A, $\hat{B} = $ average of landscapes constructed based on the diagrams from set B. Let $D = $distance between $\hat{A}$ and $\hat{B}$. Let $z = 0$.
	\item[STEP 2:] Put all persistence diagrams from $A$ and $B$ in a set $C$. Shuffle the set $C$, divide it into two new sets $E$ and $F$. Compute averages $\hat{E}$, $\hat{F}$ and a distance $\hat{D}$ between them. Every time $\hat{D} > D$, increment the counter $z$.
	\item[STEP 3:] Repeat step 2 $N$ times (typically 10000-1000000 times). 
	\end{enumerate}
The only output from this procedure we care about is the counter $z$. Given it, we compute the ration $\frac{z}{N}$. If it is small, that means that the clusters are well separated. If it is large, then just the opposite. In statistics the magic value below which the clusters are considered to be separated well is $\frac{5}{100}$. Also note that here, efficient computations of averages and distances is crucial.

\section{Comments}
The idea of persistence landscapes is in essence similar to the idea of \emph{size functions} introduced by the group of Massimo Ferri, Patrizio Frosini and Claudia Land from Bologna in early '90. Long before the concept of persistence was introduced, they had persistence in dimension zero. In order to measure it, they have been using size functions. 

\chapter{Sliding window embedding.}
In the last sections I tried to convinced you that topology may be a good tool to look at in the applied sciences. Typically the inputs are in a form of point clouds, cubical or simplicial complexes or continuous functions. It turns out that topology may be useful to analyse time series. In particular it is a handy tool to verify (semi)periodicity in data\footnote{I will remain vague here what exactly I mean by periodicity and semi-periodcity.}. In that case, the input is a time series, i.e. a sequence of measurements, typically taken over constant time intervals. In this section we will give some raw intuition on what are the characteristics of (semi)-periodicity of time series that can be captured with persistent homology. For a further, more formal, treatment please consult~\cite{perea_harer}.

\begin{figure}[h!tb]
\centering
\includegraphics[scale=0.2]{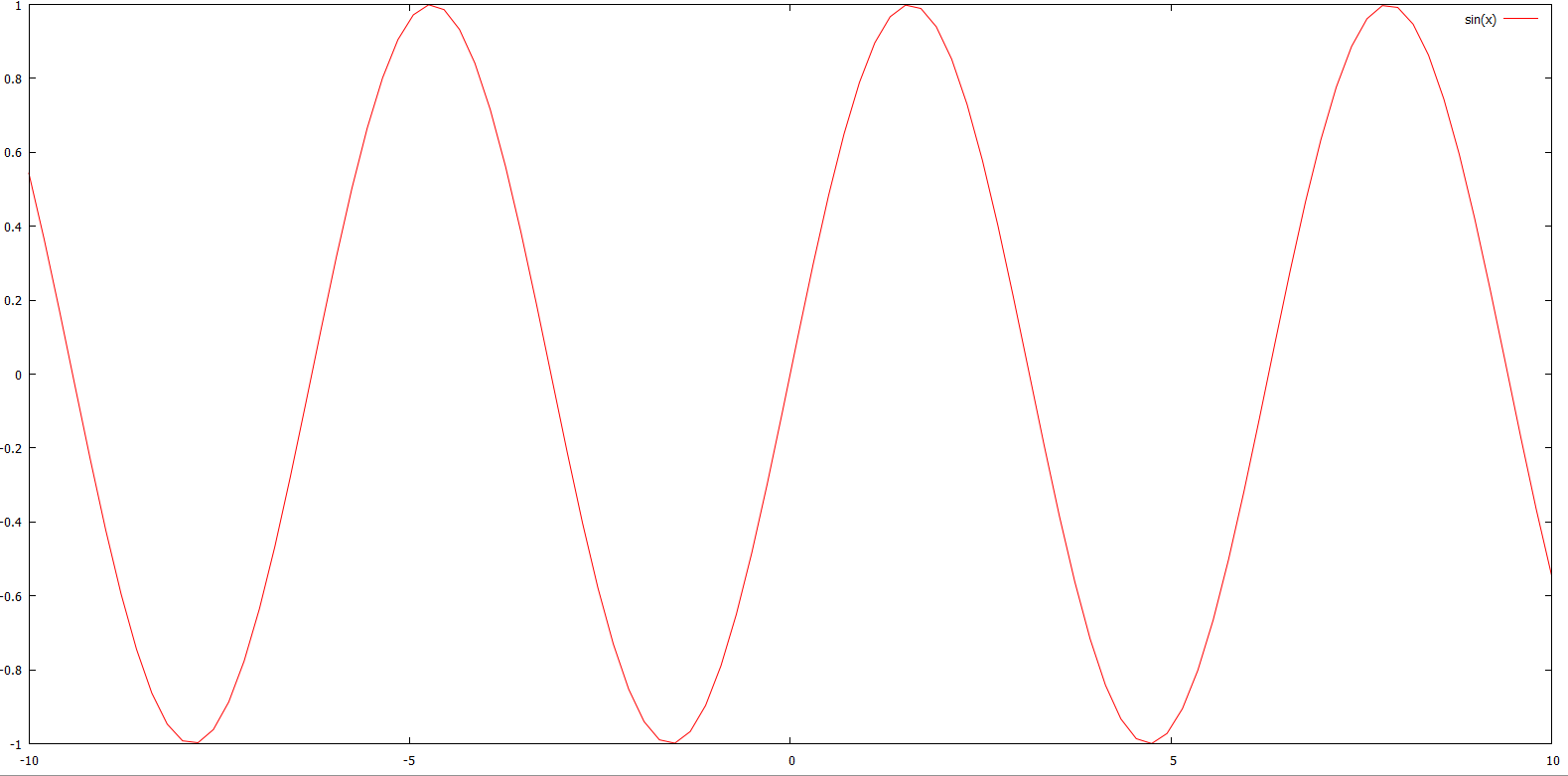}
\caption{sin(x).}
\label{fig:sinus}
\end{figure}

Let us start with a motivating example, a platonic idea of a periodic function. The function $sin(x)$, Figure~\ref{fig:sinus}.

\begin{figure}[h!tb]
\centering
\includegraphics[scale=0.5]{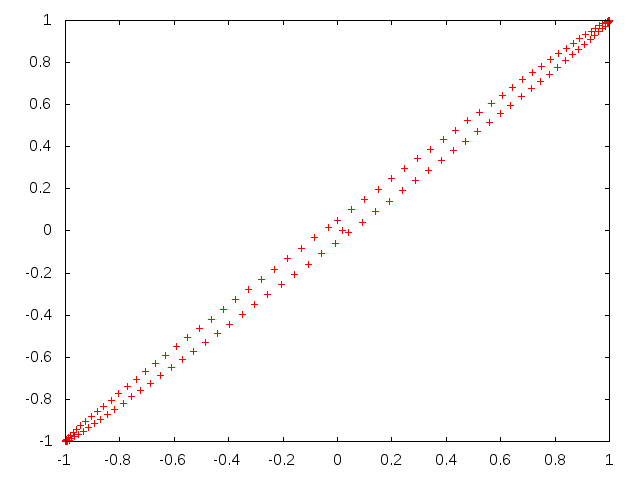}
\label{fig:embeddig_of_sin}
\end{figure}

Let us start with a very simple experiment. You can use python, C++ or even exel to perform it. Let us generate a subdivision of an interval $[0,2\pi]$ into a number of points $0,\epsilon,2\epsilon,\ldots,2\pi$. Given that, let us create a point cloud by computing values $sin(x_i), sin(x_{i+1})$, where $x_i$ and $x_{i+1}$ are all the constitutive grid elements. By doing so we get:
\[
sin(0), sin(\epsilon)\\
sin(\epsilon), sin(2\epsilon)\\
\ldots
sin(n\epsilon), sin((n+1)\epsilon)
\ldots
\]
Let us then draw this point cloud. What do we see? We see some embedding of an ellipse. A quite flat one, but an ellipse, see Figure~\ref{fig:embeddig_of_sin}.
It turns out that the point cloud get more \emph{round} if the length of the sliding window embedding is comparable to the period of the function~\cite{perea_harer}.

This is the general idea of a sliding window embedding. We group $N$ ($2$ in the example above) values of the time series into a single point. By doing so we transform a time series into a point cloud\footnote{Note that we are losing the information about time when performing this construction -- the order of points indicate the direction of time, but it is to some extent disregarded in this construction.}. 
Given this, we can measure the periodicity by checking how round this point cloud is -- in a sense it give us a sort of periodicity (or  cyclicity to be precise) measure. 
Instead of giving a proof of this statement, we will present an intuition. Suppose we are dealing with
continuous, smooth functions $f$. We sample the points $p_1,\ldots,p_n$ to obtain a time series. For the sake
of argument let us assume that the period of $f$ is $\epsilon$. Then from smoothness of $f$ a point $f(x_i),\ldots,f(x_{i+N})$ will
be close to a point $f(x_i+\epsilon),\ldots,f(x_{i+N}+\epsilon)$. Most likely it will not be the same point,
but due to the continuity of $f$ it will be close. It will also be far away from the intermediate points in between $f(x_i),\ldots,f(x_{i+N})$ and $f(x_i+\epsilon),\ldots,f(x_{i+N}+\epsilon)$. Given this, starting from
the point $f(x_i),\ldots,f(x_{i+N})$, moving forward we move away from it, and then come back when approaching
the point $f(x_i+\epsilon),\ldots,f(x_{i+N}+\epsilon)$. This induce a circle topology which can be detected
with persistent homology. Given a periodic, or semi periodic signal, we can use the lifespan of the
corresponding persistence generators as a score of how periodic the signal is. If it is small, we have very
little evidence to support such a claim.

But, from the picture we made we can clearly see that is for platonic ideas of a periodic function the ellipse we get does not look very convincing. The reason for this is the embedding we have chosen somehow random parameters and an embedding into a low dimensional space (embedding to $\mathbb{R}^2$). 

Note that everything I wrote above is true under the assumption that a signal is periodic. There are known cases when a signal is not periodic (even chaotic), but we see a clear cycle in persistent homology. An example of this situation is given below:

Let us consider a function $f(x) = sin(x)*w(x)$, where $w(x)$ is a random function constant on the intervals $[i\pi,(i+1)\pi]$ and equal to $1$ or $-1$ on those intervals (probability of each of them is $\frac{1}{2})$. Once we take sufficiently large domain in which the function $w(x)$ has both values, we will get almost exactly the same point cloud as from the sliding window embedding of a function $g(x) = sin(x)$. Yet, the function $f$ is clearly not periodic. There are ways of detecting this, but they will not be discussed in this lecture. 

As an exercise, we will first generate the point cloud coming from the function $sin(x)$. Please do it yourself and if needed, consult the code provided below.

\begin{lstlisting}
import numpy as np
import gudhi as gd
import math
#construction of the grid points x, and the values therein, y = sin(x). 
arg = np.linspace(0,10*math.pi,1000)
values = np.sin(arg)
#Size of the sliding window:
N = 200;
swe = []
for i in range(0,len(values)-N):
	point = []
	for j in range(0,N):
		point.append( values[i+j] );	
	swe.append( point );
#Now we have the point cloud, and we can compute the persistent homology 
#in dimension 1:


rips_complex = gd.RipsComplex(points=swe,max_edge_length=20)
simplex_tree = rips_complex.create_simplex_tree(max_dimension=2)
simplex_tree.persistence()
persistence = simplex_tree.persistence_intervals_in_dimension(1)	
persistence
\end{lstlisting}

In the code above please experiment with $N$ (the size of the sliding window), max\_edge\_length (note that for higher dimensions it should be increased to glue the cycle). Please convince yourself, that we see a dominant cycle in persistence, and that it comes from the fact that we are dealing with a periodic function here. For that compare the persistence of the cycle you get with the persistence of other classes (in dimension $0$). Please be careful, since for some parameters the program may consume all your ram and force you to restart your computers.

Having this part of the exercise done, we shall add a little bit of noise to the function, and check if the sliding window embedding still successfully detects a cycle coming from the periodicity of the underlying function. For that purpose, when constructing sliding window embedding add a minimal random perturbation to the points. Note that in this case you are likely to get a lot of ''noisy'' circles, therefore use the gudhi tools to visualize them.

\begin{lstlisting}
import numpy as np
import gudhi as gd
import random as rd
import matplotlib

import math
#construction of the grid points x, and the values therein, y = sin(x). 
arg = np.linspace(0,10*math.pi,1000)
values = np.sin(arg)
#Size of the sliding window:
N = 200;
swe = []
rd.seed()
for i in range(0,len(values)-N):
	point = []
	for j in range(0,N):
		point.append( values[i+j]+rd.random()*1 );	
	swe.append( point );
#Now we have the point cloud, and we can compute the persistent homology 
#in dimension 1:


rips_complex = gd.RipsComplex(points=swe,max_edge_length=20)
simplex_tree = rips_complex.create_simplex_tree(max_dimension=2)
persistence = simplex_tree.persistence()
#persistence = simplex_tree.persistence_intervals_in_dimension(1)	
plt = gd.plot_persistence_diagram(persistence)
plt.show()

\end{lstlisting}

Given this additional experience we can play now with some real periodic, or semi periodic data. Let us have a look at the webpage:\\ \url{https://datamarket.com/data/set/2324/daily-minimum-temperatures-in-melbourne-australia-1981-1990#!ds=2324&display=line}. You can get from there the spreadsheet with daily minimum temperatures from Melburne in the years 1981-1990. Those data should be almost periodic (almost, since it is not expected that temperatures yesterday, today and tomorrow will be exactly the same as last year those days). Let us try to use the tools we have just learned to see if a sliding window embedding generates a reasonably long living persistent homology class in dimension $1$. 
Note that for some reason the last line of the csv files you have imported looks like three positions, and pandas are having problems with it. Please remove the first and the last line before going to execute the code below. Also some of the temperature readings have a character ? in it. Please edit the csv files and remove this characters, since it will cause problems when reading the file. 

Watch out, since this example tends to consume a lot of RAM! If you do not have too much of it in your machine, may want to restrict the data to some smaller time interval.

\begin{lstlisting}
import numpy as np
import gudhi as gd
import random as rd
import matplotlib
import pandas as pd
import math

data_pd = pd.read_csv('semi_periodic/daily-minimum-temperatures-in-me.csv')
#We can now remove the first column:
data = data_pd.values
data = np.delete(data, (0), axis=1)

#convert it from multi dimensional array to one dimensional array:
data = data[:,0]

#in order to make the computations feasible, we should restrict the data a bit.
#experiment with various ranges, do not make it too large unless you want to restart
#your computer :)
#data = data[0:700]
#data = data[700:1500]
data = data[500:1500]

values = data.astype( np.float )
	
#Size of the sliding window. We will try to set it close to 
#the predicted value of period, which is in this case, the 
#number of days in a year. 
N = 100;
swe = []
for i in range(0,len(values)-N):
	point = []
	for j in range(0,N):
		point.append( values[i+j] );	
	swe.append( point );
#Now we have the point cloud, and we can compute the persistent homology in dimension 1:


rips_complex = gd.RipsComplex(points=swe,max_edge_length=60)
simplex_tree = rips_complex.create_simplex_tree(max_dimension=2)
persistence = simplex_tree.persistence()
simplex_tree.persistence_intervals_in_dimension(1)	

plt = gd.plot_persistence_diagram(persistence)
plt.show()

\end{lstlisting}

This is a typical output we can get. Note the dominant persistence interval in dimension $1$:
\begin{center}
\includegraphics[scale=1]{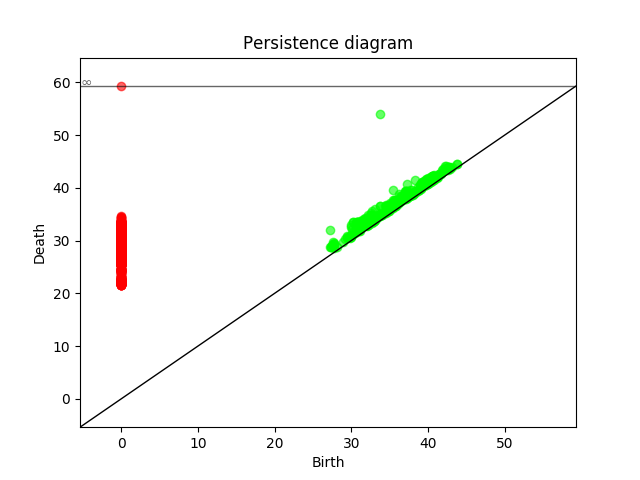}
\end{center}
Again, be careful not to pick up to large range. You do not want to kill your computer! :)

That is all for today. Thank you very much for taking this tutorial. Hope you liked it and that you have learned some new stuff you will be able to use later in your work. If you have any questions please let me know on p.t.dlotko@swansea.ac.uk

%\begin{center}
%\Huge{Thank you!}
%\vskip 1 cm
%\includegraphics[scale=0.8]{images/thats_all_folks.jpg}
%\end{center}

\Large{
Please, do me one favour. Once you have completed the tutorial, please let me know (with anonymous email if you wish):
\begin{enumerate}
\item What you liked?
\item What you disliked?
\item What you think can be improved or added to the tutorial?
\end{enumerate}
I will be very grateful for your feedback!}

\bibliographystyle{plain}
\bibliography{my_bib}
\end{document}